\documentclass[prx,aps,twocolumn,footinbib,superscriptaddress,floatfix]{revtex4-1}

\usepackage{CJK}
\usepackage[colorlinks,bookmarks=false,citecolor=blue,linkcolor=red,urlcolor=blue]{hyperref}
\usepackage{color} 
\usepackage{graphicx}
\usepackage{float}
\usepackage{multirow}
\usepackage{amsmath}
\usepackage{bm}
\usepackage{bbold}
\usepackage{amsfonts}
\usepackage{braket}
\usepackage{subfigure}
\usepackage{cleveref}
\usepackage{comment}
\usepackage{chemfig}
\usepackage{dcolumn}
\usepackage{amssymb}
\usepackage{adjustbox} 
\usepackage{microtype}
\usepackage{xfrac}
\usepackage{array}
\usepackage[dvipsnames]{xcolor}
\usepackage{feynmp-auto}

\newcolumntype{P}[1]{>{\centering\arraybackslash}p{#1}}

\begin{document}

\title{$q$-state Potts ice
}
\author{Mark Potts}
\affiliation{Max Planck Institute for the Physics of Complex Systems, N\"{o}thnitzer Str. 38, Dresden 01187, Germany}
\author{S.A. Parameswaran}
\affiliation{Rudolf Peierls Centre for Theoretical Physics, Parks Road, Oxford OX1 3PU, United Kingdom}
\affiliation{Hertford College, Catte Street, Oxford OX1 3BW, United Kingdom}
\author{Roderich Moessner}
\affiliation{Max Planck Institute for the Physics of Complex Systems, N\"{o}thnitzer Str. 38, Dresden 01187, Germany}
\affiliation{Hertford College, Catte Street, Oxford OX1 3BW, United Kingdom}

\begin{abstract}
Classical and quantum spin ice  arguably provide the simplest route towards spin liquids and their emergent gauge fields. $q-$state Potts ice models have been constructed that generalize spin ice, hosting multiple emergent $\text{U}(1)$ gauge fields and excitations charged under non-trivial combinations of these fields. We present a general treatment of classical $q-$state Potts ices relating their properties to the $\mathfrak{su}(q)$ Lie algebras, and demonstrate how the properties of charged excitations in the classical model can be related to this symmetry group. We also introduce quantum generalizations of the Potts ice models, and demonstrate how charge flavor changing interactions unique to $q>2$ models dominate their low energy physics. We further show how symmetries inherited from the $\mathfrak{su}(q)$ can lead to flux vacuum frustration, greatly modifying the dynamical properties of charged excitations. 
\end{abstract}

\maketitle
The field theories  that arise as long-wavelength, low-energy descriptions of condensed matter systems, whether classical or quantum, often offer much greater variety than those that are {\it experimentally} relevant in the high-energy setting. The loosening of the straitjacket of Poincar\'e invariance means that models originally devised as special limits or conceptual toy models in the latter context can acquire a new significance in the former, as a means to understand distinct phases of matter. Perhaps the best known cases involve  the Landau paradigm of symmetry-breaking:  for example, liquid crystals and quantum magnets exemplify a far wider array of broken symmetries than the relatively austere set relevant to the Standard Model. 

More recently, emergent gauge theories have provided ways to describe and distinguish phases in systems that evade symmetry breaking down to very low temperatures. Introducing gauge fields can provide an elegant field-theoretic encoding of local energetic constraints, or enable a dual description in more convenient variables. Two instances where this approach has found experimental validation are the application of Chern-Simons theories to capture the charge-flux duality in the fractional quantum Hall effect \cite{PhysRevLett.48.1559,PhysRevLett.62.82,PhysRevB.44.5246,doi:10.1142/S0217979292000037}, and the use of gauge constraints to implement the ``ice rule''  at the heart of classical and quantum spin ice compounds in the  rare earth pyrochlores \cite{doi:10.1126/science.1064761,PhysRevLett.79.2554,doi:10.1126/science.1178868,PhysRevX.12.021015,j451-ztvr,Sibille2020}. Quantum spin ice offers great promise as a tabletop realization of the strongly-coupled regime of quantum electrodynamics, since its low energy description involves an emergent fluctuating compact Maxwellian $U(1)$ gauge field, coupled to gapped magnetic and electric charges \cite{Balents2010,Savary_2017,Knolle19,RevModPhys.89.025003,PhysRevLett.108.037202,PhysRevB.69.064404,Gingras_2014,doi:10.1126/science.1064761,Castelnovo2008}. Apart from this motivation, already in the classical limit spin ices furnish perhaps the best understood examples of how local energetics can be rewritten in terms of a Gauss-law constraint: enforcing the constraint leads to a disordered ground state manifold characterized by dipolar ``Coulomb-phase'' correlations, whereas discrete excitations that violate it can be viewed as monopole sources of both the emergent and physical magnetic field. 

Given these examples, it is natural to ask if emergent gauge structures beyond the $U(1)$ relevant to spin ice can be identified: initially theoretically, in simple lattice models, and eventually also in experiments on real  materials. Pyrochlore magnets offer  a partial answer in the affirmative: for example, the ground state manifold of the classical Heisenberg pyrochlore antiferromagnet  is described by three emergent $U(1)$ gauge fields each satisfying a Gauss-law constraint, although there the spectrum of excitations out of the ground-state manifold is massless \cite{PhysRevLett.80.2929}. A similar three-$U(1)$ structure also arises for the 4-state Pyrochlore Potts model (or ``color ice''), but here the excitations out of the ground state manifold are again point defects, with the added twist that they are `bions' charged under two of the three emergent $U(1)$ fields \cite{PhysRevB.52.6628,PhysRevB.86.054411,PhysRevLett.112.020601,PhysRevB.94.174417,Lozano-Gómez2024}. Three-color ice models have also been constructed \cite{https://doi.org/10.1155/2013/836168} or encountered in spin models on certain tri-coordinated lattices \cite{PhysRevLett.120.117202,PhysRevB.99.104433,PhysRevB.103.144414,Zhang2025}. However, a systematic construction of color ice models, and the relationship between the lattice symmetries, emergent gauge structure, and the spectrum of elementary charges, has not previously been attempted. Absent that, the fate of such models under the addition of quantum dynamics is largely unchartered territory. 

In this work, we provide such a systematic treatment of $q$-state Potts ices, both from a classical and a quantum perspective. We first identify an intimate connection to the properties of the Lie groups $\mathfrak{su}(q)$. Specifically, we describe the extensively degenerate gapped ground states of the classical q-state Potts ices, and relate fluctuations between these in terms of gapped excitations that we term ‘rootons’, whose charges take on the values of the roots of $\mathfrak{su}(q)$, and which interact
entropically via an effective Coulomb interaction. The corresponding abelian gauge group, consisting of $(q-1)$ $U(1)$ factors, is understood as being generated by the Cartan sub-algebra of $\mathfrak{su}(q)$, and the theory is thus an abelian projection of SU$(q)$ gauge theory.

Having established these aspects of the classical ices, we then extend the study of $q$-state Potts ices to include  quantum fluctuations perturbatively. This imbues the gauge field with dynamics, introducing $(q-1)$ gapless and uncharged `photon' modes, as well as gapped visons: topological defects in the emergent gauge fields. Interactions between charged particles may proceed analogously as in QED via exchange of virtual photons, or via direct flavor changing three-field interactions existing only in the $q>2$ theories, which are found to dominate at low energies. Such interactions are understood to be remnants of off-diagonal gluon-gluon interactions in the abelian projected theory, but also have an appealing interpretation under the `standard model' dynamics familiar from spin ice \cite{Ryzhkin2005,Jaubert-dynamics,doi:10.1126/science.add1644}. We also consider the issue of other non-confining vacua  (i.e., distinct `Coulomb phases'), by using exactly solvable models to link the symmetry of the Cartan sub-algebra to the flux assignment in the ground state for $q>2$.

Our work provides a bridge between the standard construction of $\mathfrak{su}(q)$ lattice gauge theories, which typically involve complex multi-spin interactions and an explicitly imposed gauge constraint, and their realizations as emergent descriptions of physically-realistic spin models on lattices. This echoes recent themes in quantum simulation   of $\mathfrak{su}(N)$ magnetism and gauge dynamics in a range of platforms ranging from ultracold atoms, to Rydberg arrays, to superconducting qubit devices. 

This paper is organized as follows. First, in section \ref{sec:Exposition}, we provide a simple construction of the emergent gauge theory for classical $q$-state Potts ices with explicit links to the properties of $\mathfrak{su}(q)$. This is then consolidated with Monte Carlo numerical work on various classical models in section \ref{sec:Numerics}. Our analysis of the quantum extension of the Potts ice models is given in section \ref{sec:Quantum}, employing Gauge Mean Field Theoretic (GMFT) methods, and constructing exactly solvable models in the vein of the toric code to explore physical phenomena present in these models that are not observed in quantum spin ice --- specifically, the influence of the flavor-changing interactions on the mean-field phase diagram of the $q=3,4$ models, a brief discussion of the dynamical gauge theory, and an examination of flux-frustrated vacua occurring in $q>2$ models.

\section{Classical $q$-color Potts Ices}\label{sec:Exposition}

\subsection{Construction of $q$-state Potts ices}

\begin{figure*}
    \centering
    \subfigure[]{
    \includegraphics[width=0.3\linewidth]{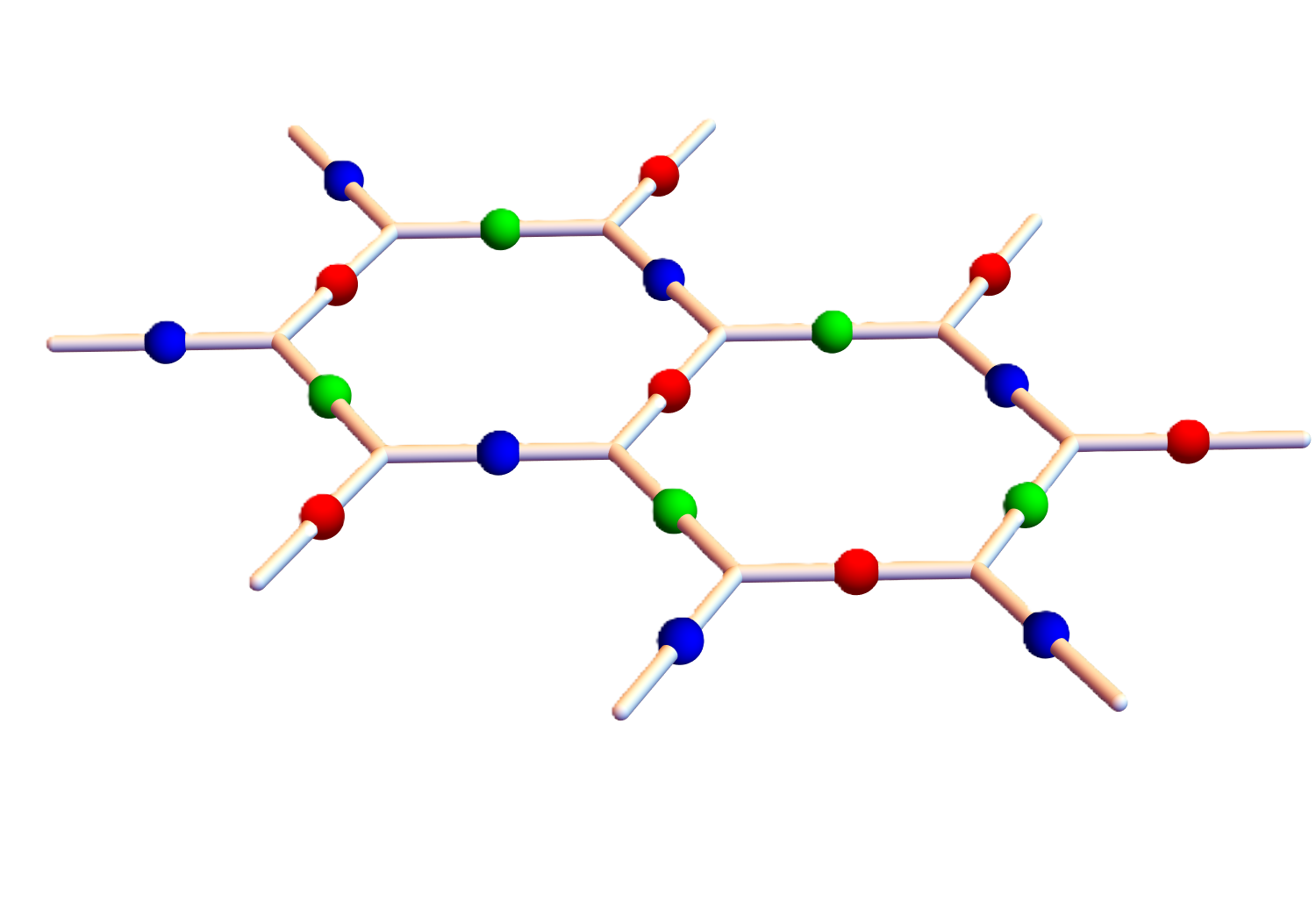}
    }
    \subfigure[]{
    \includegraphics[width=0.3\linewidth]{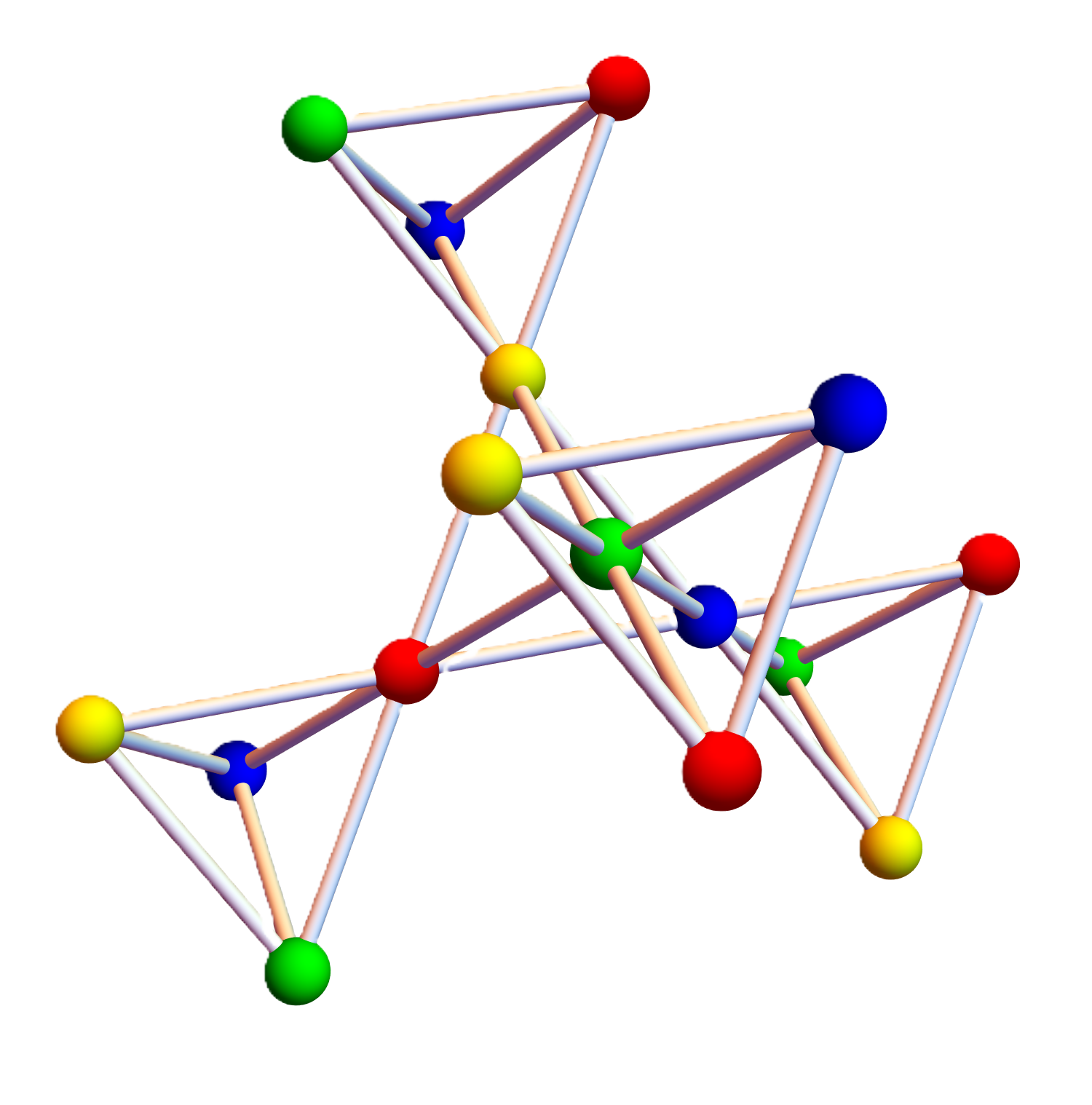}
    }
    \subfigure[]{
    \includegraphics[width=0.3\linewidth]{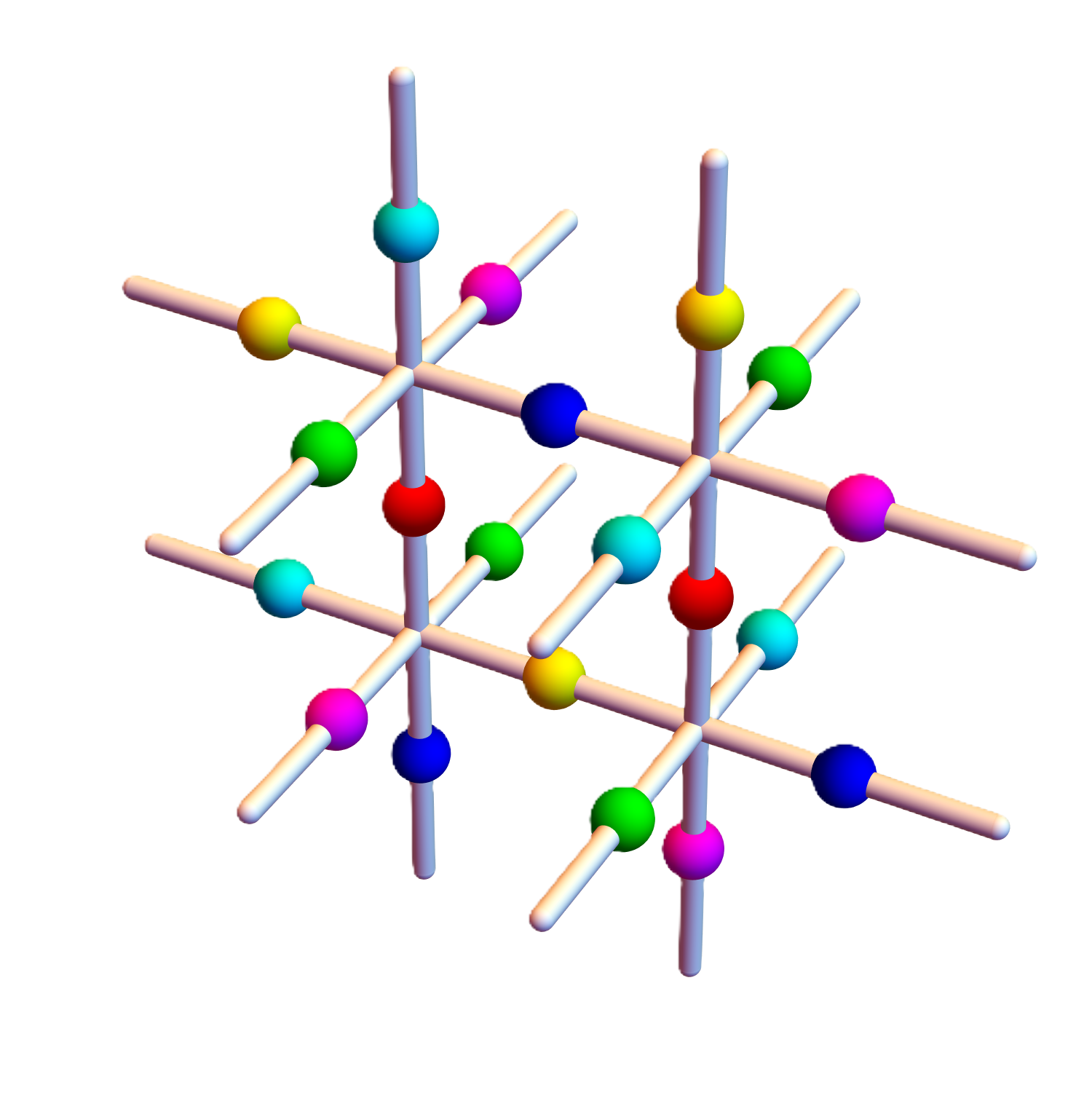}
    }
    \caption{Various realizations of $q$-state Antiferromagnetic Potts model ground states on lattices in two and three dimensions: $q=3$ model on the edges of the Honeycomb lattice ((a));  $q=4$ model on the vertices of the pyrochlore lattice (line-graph of diamond lattice) ((b)); $q=6$ model on the edges of the cubic lattice ((c)). In all cases $z=1$, so ground states consists of all edges neighboring each vertex (or each vertex of each cell for the line graph, as in (b)) having a distinct color.}
    \label{fig:Color_Ice_models}
\end{figure*}

We consider quite general $q$-color antiferromagnetic Potts models, with variables defined equivalently on the edges of some bipartite lattice, or the vertices of its line-graph (i.e. a lattice formed by placing vertices on the edges of a parent lattice, and connecting those vertices whose parent edges met at an original vertex). The coordination number of this lattice is assumed to be $\mathcal{N}=zq$, for integer $z$, such that when we place an antiferromagnetic Hamiltonian on the lattice:
\begin{equation}
    H_{\text{cl}}= J_Q \sum_{\langle r,s\rangle}\delta_{c_rc_s}, \label{eq:Classical_Hamiltonian}
\end{equation}
the ground state of the system consists of coloring the links of the lattice such that there are an equal number $z$ of each of the $q$ colors meeting at a vertex. Here $\langle r,s \rangle$ implies that $r$ and $s$ are edges connected by a vertex, or equivalently connected vertices on the line-graph. $c_r$ denotes the color degree of freedom on link $r$. Several examples of such models for various values of $q$ are shown in Fig. \ref{fig:Color_Ice_models} in ground states. 

The equal color number condition serves as the analogue of the two-in-two-out `ice-rule' of classical spin ice. Excitations from the ground state can be constructed by changing a single color variable, creating two local ice-rule defects, each of energy $J_Q$. These excitations are fractionalised, in the sense that they act as independent particles that can be moved independently along `worms' of alternating colors. Creating a pair of defects by flipping a red edge to green creates a `red-green' and a `green-red' type particle, which can be moved by further flipping loops of alternating red-green edges without incurring additional energy costs.

This is all very similar to the situation in classical spin ice, and this correspondence can be made concrete by also reinterpreting this physics in terms of an emergent electric field. The generalization we will require, as already shown for $q=4$ \cite{PhysRevB.86.054411}, is to introduce $N_F$-component electric fields $E^{i}_{c_r}$, assigning an as yet unspecified fixed vector to each of the $q$ colors. To capture the behavior of the defect excitations, one can then define $N_F$-component charges at each vertex (or cell of the line graph) $C$ as the lattice divergence of these fields in the following way:
\begin{equation}
	Q^i_C = \eta_C \sum_{r \in C} E^i_{c_r}, \label{eq:Charge_definition}
\end{equation}
where, making use of the fact the lattice is bipartite, we define $\eta_C=1$ on sublattice A, and $-1$ on sublattice B.

For ground states of Eq. (\ref{eq:Classical_Hamiltonian}) to correspond to divergence-free field configurations, it must be the case that one can rewrite $H_{\text{cl}}$ in terms of the $Q^i_C$ as:
\begin{equation}
	H_{\text{cl}}= K \sum_{C,i} \left( Q_C^{i} \right)^2,
\end{equation}
up to some global constant. Equating the two expressions for the classical Hamiltonian, and requiring that an equal-colors configuration corresponds to $Q_C^i=0$, one obtains the following conditions for the field variables:
\begin{align}
	2K \sum_i E^i_{c_r}E^i_{c_s} &= J_{Q}\delta_{c_r,c_s} +const. , \label{eqMan:Inner_product_condition}\\
	\sum_i (E^i_c)^2 &=\rho^2 \ \forall c \ , \label{eqMan:Equal_length_condition}\\
    \sum_cE^i_c &=0 \ \forall_i \ . \label{eqMan:Divergence_free_condition}
\end{align}
The unique solution to these conditions is that the fields $E^i_c$, treated as $N_F$-dimensional vectors, describe the vertices of a $(q-1)$-simplex. The inner product between any two field vectors of differing color is $\mathbf{E}_{c_1}\cdot\mathbf{E}_{c_2}=-\frac{\rho^2}{q-1}$, and each vector has the same length $\rho^2$. This assignment respects the $S_q$ symmetry between the different colors, and crucially the sum over all color vectors vanishes, so each ground state is automatically a divergence-free field configuration. This construction is consistent with the treatments of the $q=4$ model on pyrochlore \cite{PhysRevB.86.054411,PhysRevLett.112.020601}, and is consistent with results for a similar `tripole' model on the kagome lattice \cite{https://doi.org/10.1155/2013/836168}, but generalizes these to all $q$. 

Note that the construction does not uniquely fix the magnitude of the field vectors $\kappa$; for definiteness we choose to set $K=J_Q/2$, and $\rho^2=(q-1)/q$. 

That the required field values form a simplex has an important consequence. If we define now the diagonal matrices $E^i_{c_r,c_s}=E^i_{c_r}\delta_{c_r,c_s}$, then it can be verified that these matrices form a basis for the set of $q \times q$ traceless diagonal matrices, and hence are a parameterisation of a maximally commuting sub-algebra (Cartan sub-algebra) of $\mathfrak{su}(q)$. Just as the two-color model (spin ice) can be written as a nearest neighbor AFM model coupling the $z-$components of $\mathfrak{su}(2)$ degrees of freedom, the general $q$-Potts ices are equivalent to nearest neighbor AFM models coupling all of the many mutually commuting `$z-$components' of $\mathfrak{su}(q)$ degrees of freedom. 

This connection between Potts ice models and $\mathfrak{su}(q)$ is highlighted in the specific case of the $q=4$ model on the square lattice by Kondev and Henley \cite{PhysRevB.52.6628} in the context of a height model formulation of the problem. Our construction cleanly demonstrates that this relationship exists for arbitrary $q$, and does so more directly than through the conformal field theory considerations discussed in that work.

The relationship to $\mathfrak{su}(q)$ is not circumstantial; it can be used to cleanly describe the properties of the charged excitations of the model. If the diagonal matrices $E^i_{c_r,c_s}$ are the elements of the Cartan of $\mathfrak{su}(q)$, then the remaining non-commuting generators of the algebra-- which generate color swaps --can be denoted as $R^{\alpha}$. Each of these acts of a basis state of a given color (say red) and exchanges it with another (say green), and annihilates all other basis states. The $R^{\alpha}$ thus parameterize the natural fluctuations of the model, with pairs of excitations created by open worms of $R^{\alpha}(R^{\alpha})^{\dagger}$ operators.

The index $\alpha$ is used to parameterize each color pairing, and is more precisely the root vector associated with the operator $R^{\alpha}$, defined through the fundamental commutation relation: $[E^i,R^{\alpha}]=\alpha^iR^{\alpha}$. 

From this relation, it follows that the action of $R^{\alpha}$ (and thus the effect of changing one color variable starting from a ground state configuration) is to raise/lower the electric field at a given site by an amount $\alpha$, and thus the charge defects created by $R^{\alpha}$ acting on link $r$ at the adjoining cells $C,C'$ have vector charges $\pm \alpha^i$. The charged excitations of the $q$-color Potts ices thus all have $\mathfrak{su}(q)$ root valued charges.

In \cite{PhysRevB.86.054411}, the charged excitations are termed `bions', because, for a certain choice of basis, the roots of $\mathfrak{su}(4)$ can be written in a way such that each excitation is charged under precisely two of the three emergent fields with unit magnitude charge. Such a basis choice can only be made for certain values of $q$, and so to keep our nomenclature as general as possible, we will refer to all divergence-free violations as `rooton' particles.

Note that there exists also the arbitrary choice of coordinate axes for the electric fields $E^i_c$-- the orientation of the simplex in root space is not physically meaningful-- and so all observables must be $\text{SO}(q-1)$ singlets when expressed in terms of the emergent fields.

\subsection{Correlations in the ground and excited states}

As with classical Ising spin ice, the presence of emergent Coulomb physics is manifest in the ground state correlation functions of the classical Potts ices, and in entropic interactions between charges.

We define a color structure factor as follows for general $q$-Potts ice:
\begin{equation}
    \tilde{S}_{rs}=\left\langle \sum_{\mu}C^{\mu}_rC^{\mu}_s\right\rangle -\frac{1}{q},
\end{equation}
where $C^{\mu}_r$ is an operator that equals $1$ when the color on site $r$ is $c_{\mu}$, and zero otherwise. Expressed in terms of the emergent electric fields, the structure factor becomes:
\begin{equation}
    \tilde{S}_{r,s}= \left\langle \sum_iE^i_{c_r}E^i_{c_s}\right\rangle.
\end{equation}
We obtain $(q-1)$ copies of the correlations of divergence-free electric fields. The field flavors are not however independent degrees of freedom, so it must be verified numerically that there are no non-trivial correlations between the fields that prevent the system from realizing a Coulomb phase. This has been previously verified for the $q=4$ models on pyrochlore and the square lattice \cite{PhysRevB.86.054411,PhysRevB.52.6628}, and in a later subsection we will present corroborating calculations for the $q=6$ model on the cubic lattice (or its line graph the `octachlore' lattice).

If we do assume that the different field flavors are statistically independent in the bulk when averaging over all ground state configurations, then an effective bulk free energy is of the form:
\begin{equation}
    F_{\text{eff}}= -TS = U_0 \int \text{d}^D\mathbf{r} \sum_i |\mathbf{E}^i(\mathbf{r})|^2,
\end{equation}
where $U_0$ is an unknown constant. Evaluating the field correlations over divergence free configurations, the resulting $k-$space correlations take the same general form as in Ising spin ice:
\begin{equation}
    \langle E_a^i(\mathbf{k})E^j_b(\mathbf{-k)}\rangle =\frac{1}{U_0}\delta^{ij}\left(\delta_{ab}-\frac{k_ak_b}{k^2}\right),
\end{equation}
where the indices $a,b$ run over the $D$ spatial dimensions. The components parallel to the electric field are projected out, resulting in characteristic pinch point singularities in the correlation function.

If the divergence free condition is violated at some fixed set of locations $\mathbf{r}_n$, with root charges $\alpha_n$, then the resulting free energy, when integrating over the allowed field configurations, takes the form of an entropic interaction between these violations:
\begin{equation}
    U_{\text{S}}(\{\mathbf{r_n}\}) \propto \sum_{n,m\neq n}\frac{\sum_i\alpha^i_n\alpha^i_m}{|\mathbf{r}_n-\mathbf{r}_m|} \ .
\end{equation}

This generalizes the form of the entropic interactions found for the four-color Potts ices \cite{PhysRevB.86.054411,PhysRevLett.112.020601}. Most importantly, the interaction strength between two rootons is proportional to the inner products between those roots. The Killing form inner product $(\alpha,\beta)$ between two roots is related to this simple inner product for our choice of normalization by $\sum_i\alpha^i\beta^i=2q(\alpha,\beta)$. Making use of the properties of this Killing form inner product for $\mathfrak{su}(q)$ Lie algebras \cite{Groups_Fulton-Harris}, one obtains that the simple inner product $\sum_i\alpha^i\beta^i$ can have magnitude $2,1$, or $0$, regardless of the value of $q$.

Each $R^{\alpha}$ operator is associated to a pair of colors; the interaction strength $\sum_i\alpha^i\beta^i=2$ when the interacting rootons share both of these colors. For pairs sharing only one color, $\sum_i\alpha^i\beta^i=1$, and rootons with no colors in common do not interact directly.

\section{Monte Carlo Results for classical models} \label{sec:Numerics}

\begin{figure}
    \centering
    \subfigure[]{
    \includegraphics[width=1.0\linewidth]{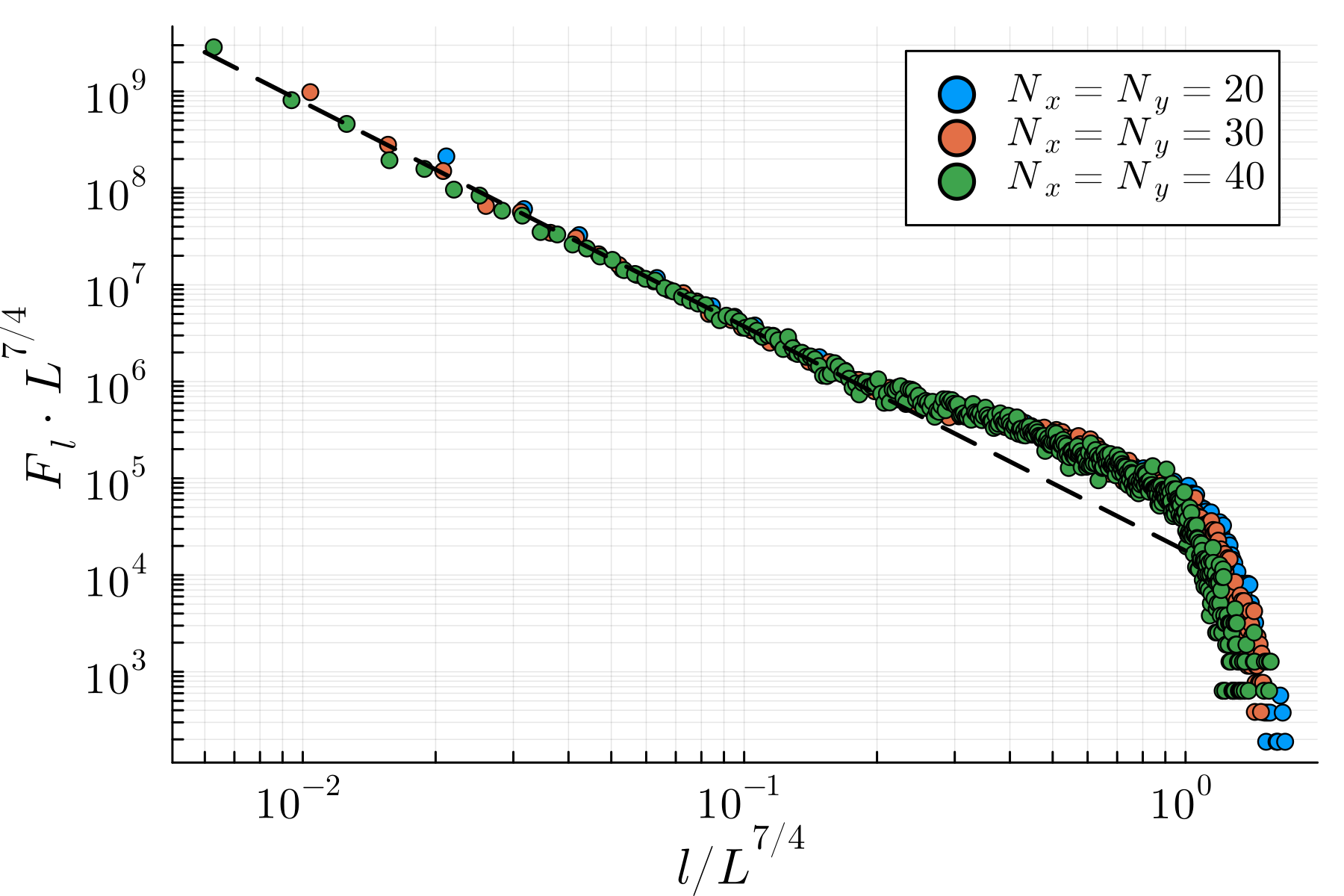}
    }
    \subfigure[]{
    \includegraphics[width=1.0\linewidth]{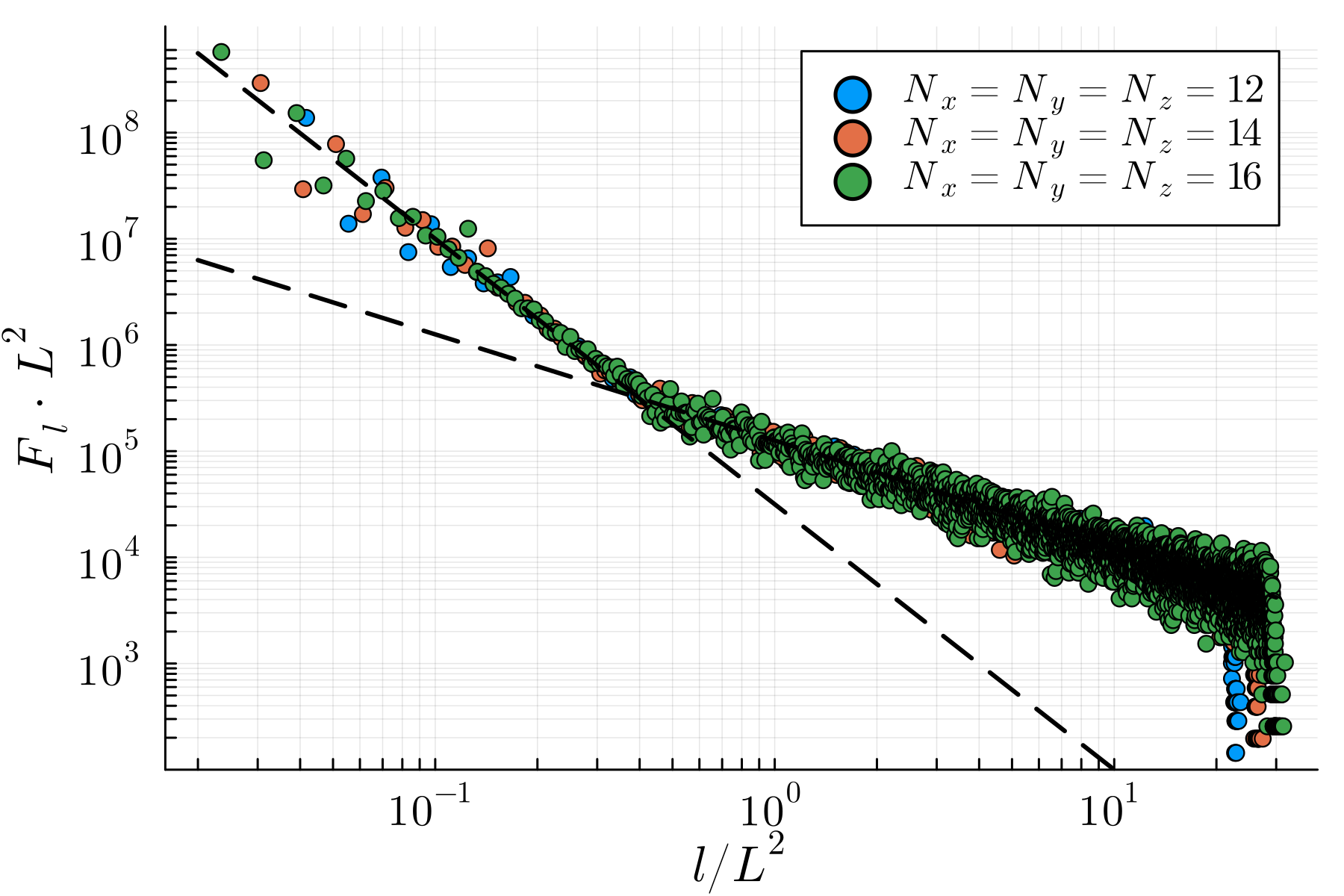}
    }
    \subfigure[]{
    \includegraphics[width=1.0\linewidth]{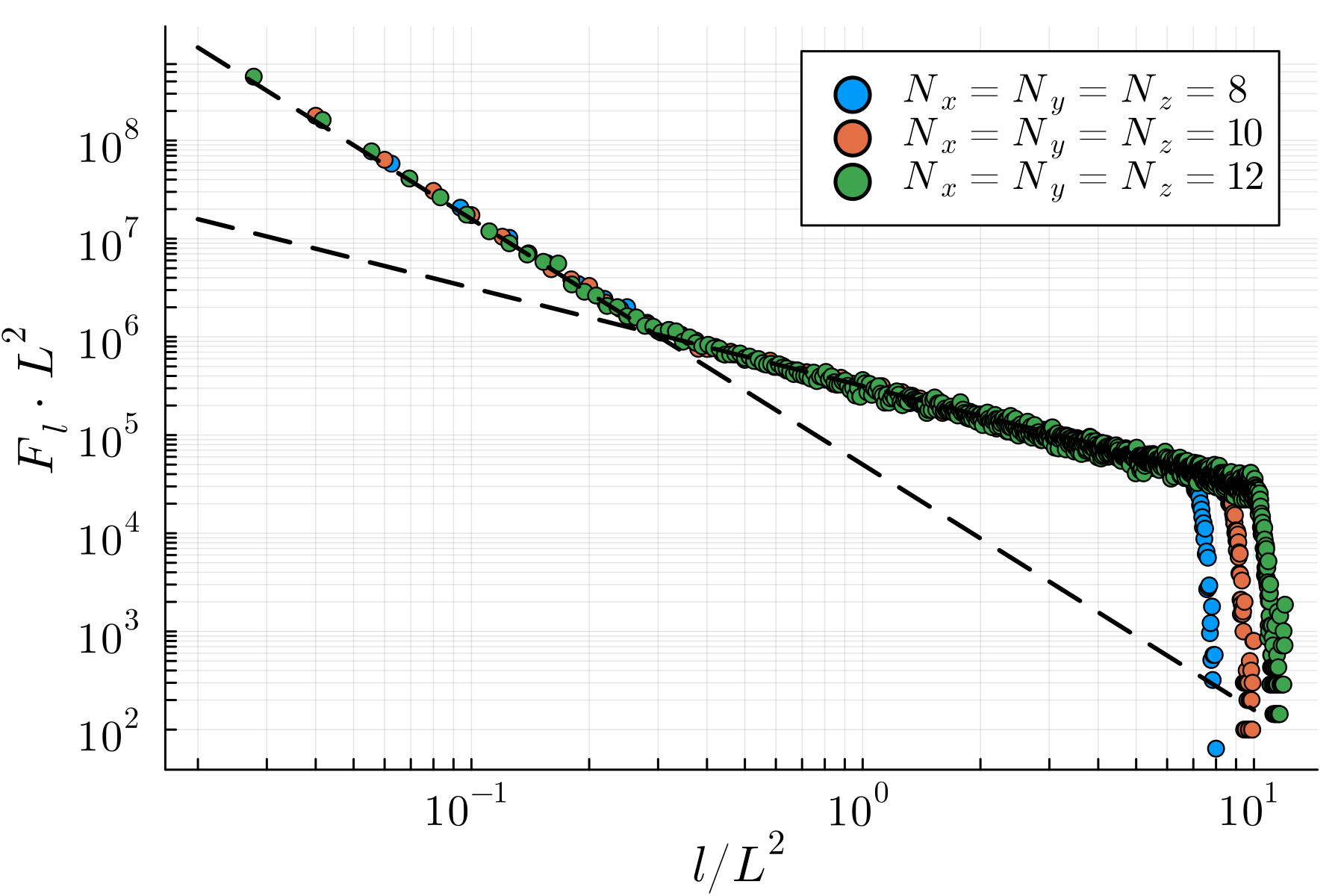}
    }
    \caption{Worm length statistics for the $q=4$ models on (a) the square lattice and (b) the pyrochlore lattice, and (c) the $q=6$ model on the cubic lattice. All three exhibit the same behavior observed in the Coulomb phases of classical $q=2$ spin ice in two and three dimensions respectively \cite{PhysRevLett.107.177202}. The two dimensional $q=4$ model is described by a single power law with a non-universal exponent $n\sim-2.32$, whilst the three dimensional models possess identical statistics: short worm frequency obeys a power law with exponent $n_1=-2.5$, whilst longer worms fall in prevalence with power $n_2=-1.0$.}
    \label{fig:Loop_Statistics}
\end{figure}

To support our analysis of the classical spin ices, we present in this section the results of Monte Carlo simulations of the $q=4$ Potts ices on the square lattice and on the pyrochlore lattice, and the $q=6$ Potts ice on the cubic lattice. The $q=4$ models have been studied numerically in previous works \cite{PhysRevB.52.6628,PhysRevB.86.054411,PhysRevLett.112.020601}; we demonstrate that the $q=6$ model possesses the same characteristic signatures of Coulomb physics, supporting our claim that emergent Coulomb physics is ubiquitous in Potts ice models.

\subsection{Worm length statistics}

\begin{figure}
    \centering
    \includegraphics[width=0.75\linewidth]{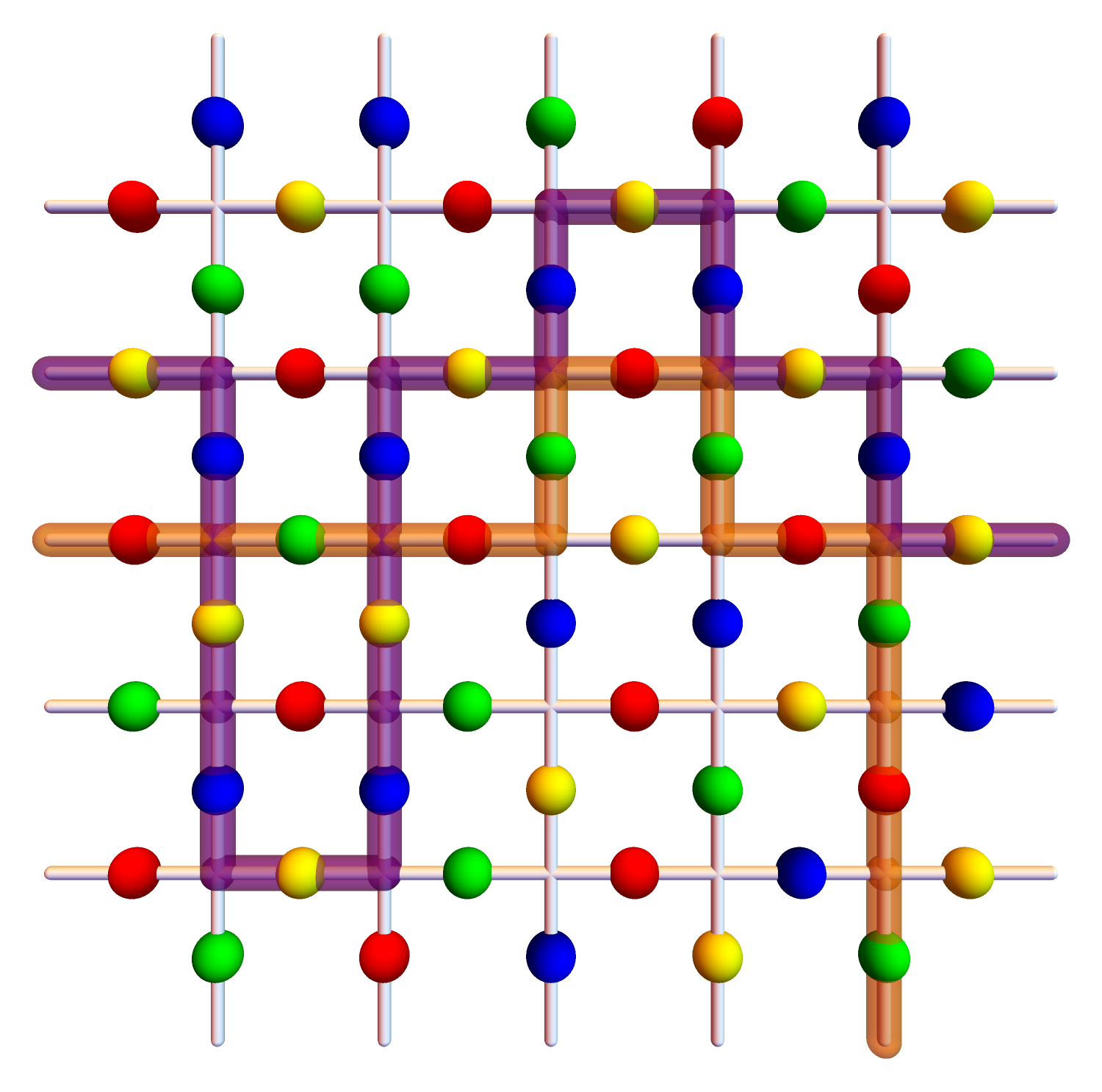}
    \caption{Ground state of $q=4$ model on the square lattice with worms of different flavors highlighted}
    \label{fig:Worm_diagram}
\end{figure}

The statistics of `loops' of like spins, and `worms' of alternating spin orientation (see Fig. \ref{fig:Worm_diagram}) have been studied in spin ice models, and shown to be characteristic of random walks in two and three dimensions, consistent with the principle that the Coulomb phase in spin ice corresponds to deconfinement of the electric field \cite{PhysRevLett.107.177202}. 

We can similarly analyze the statistics of worms of alternating colors in various Potts ice models, and ascertain if these conform to the expected behavior in a Coulomb phase. In two dimensions, the distribution of worm lengths is predicted to follow a power law with non-universal exponent. In three dimensions, there is a distinction between the distribution of short worms, and long worms which wrap around the periodic boundaries of the system. Short worm lengths are distributed as a power law with exponent $-2.5$ in three dimensions, whilst the longer worm distribution falls away more slowly with exponent $-1$ \cite{PhysRevLett.107.177202}. The distinction between the two behaviors lies in the finite probability of random walks in three dimensions not to recur.

Each flavor of worm is generated by loops of $R^{\alpha}$ and $R^{-\alpha}=(R^{\alpha})^{\dagger}$ operators. We note that for $z=1$, the assignment of worms to a given color configuration is unique, whilst for $z\ge2$ this is not the case, as mutliple worms of the same flavor intersect at each vertex, as is the case for spin ice.

Worm length statistics for the $q=4$ models and the $q=6$ model on the cubic lattice is presented in Figure \ref{fig:Loop_Statistics}. The expected power law behavior is observed in all cases, with deviations found only for worms of length comparable to the system size. 

These statistics are obtained through sampling ground state configurations obtained via Markov chain Monte Carlo, wherein the allowed updates are the flipping of all colors along a randomly selected worm. Worm lengths are sampled from all worms of all flavors for a given ground state configuration, not just from the worm flipped in the Markov update.

The ergodicity of these updates in two dimensions for $z=1$ are not trivial, as the worms are self avoiding, and indeed these updates are not ergodic for the $q=3$ model on the Honeycomb lattice \cite{10.21468/SciPostPhys.7.3.032,10.21468/SciPostPhys.10.2.042}, but have shown to be so for the $q=4$ square lattice model \cite{J_K_Burton_Jr_1997,Ferreira1999}.

\subsection{Entropic Coulomb interactions}

To complement the establishment of Coulomb physics evidenced by the statistics of worm lengths, we also present numerical data for states with two excitations, wherein we observe an effective entropic force between these particles consistent with an effective Coulomb interaction of the correct form for two or three dimensions.

The entropic interaction between rooton-antirooton pairs is probed by sampling their separations whilst they undergo dynamics generalized from the `standard model' dynamics of spinons in spin ice \cite{Ryzhkin2005,Jaubert-dynamics,doi:10.1126/science.add1644}. Under such dynamics, updates to the color configurations are allowed if they would incur no change in energy. For $z=1$ models, this almost confines the rooton and antirooton to a single worm of alternating colors, however these dynamics allow for neighboring pairs to change their flavor by changing the color on the bond between the defects, thus allowing the excitations to change worms at a collision.

If the two particles have an effective entropic interaction $U(r)$, then the distribution of their separations will go as $P(r) \propto N_r\exp(-\beta U(r))$. $N_r$ here is the discrete `measure' enumerating the total number of ways a separation of $r$ can be achieved holding one rooton fixed. Thus in two and three dimensions, we expect for a Coulomb effective interaction:
\begin{align}
    P_{2D}(r) &\propto N_r\exp(-\kappa\ln(r))=\frac{N_r}{r^{\kappa}}, \\
    P_{3D}(r) &\propto N_r\exp\left(-\frac{\kappa'}{r}\right).
\end{align}

\begin{figure}
    \centering
    \subfigure[]{
    \includegraphics[width=0.88\linewidth]{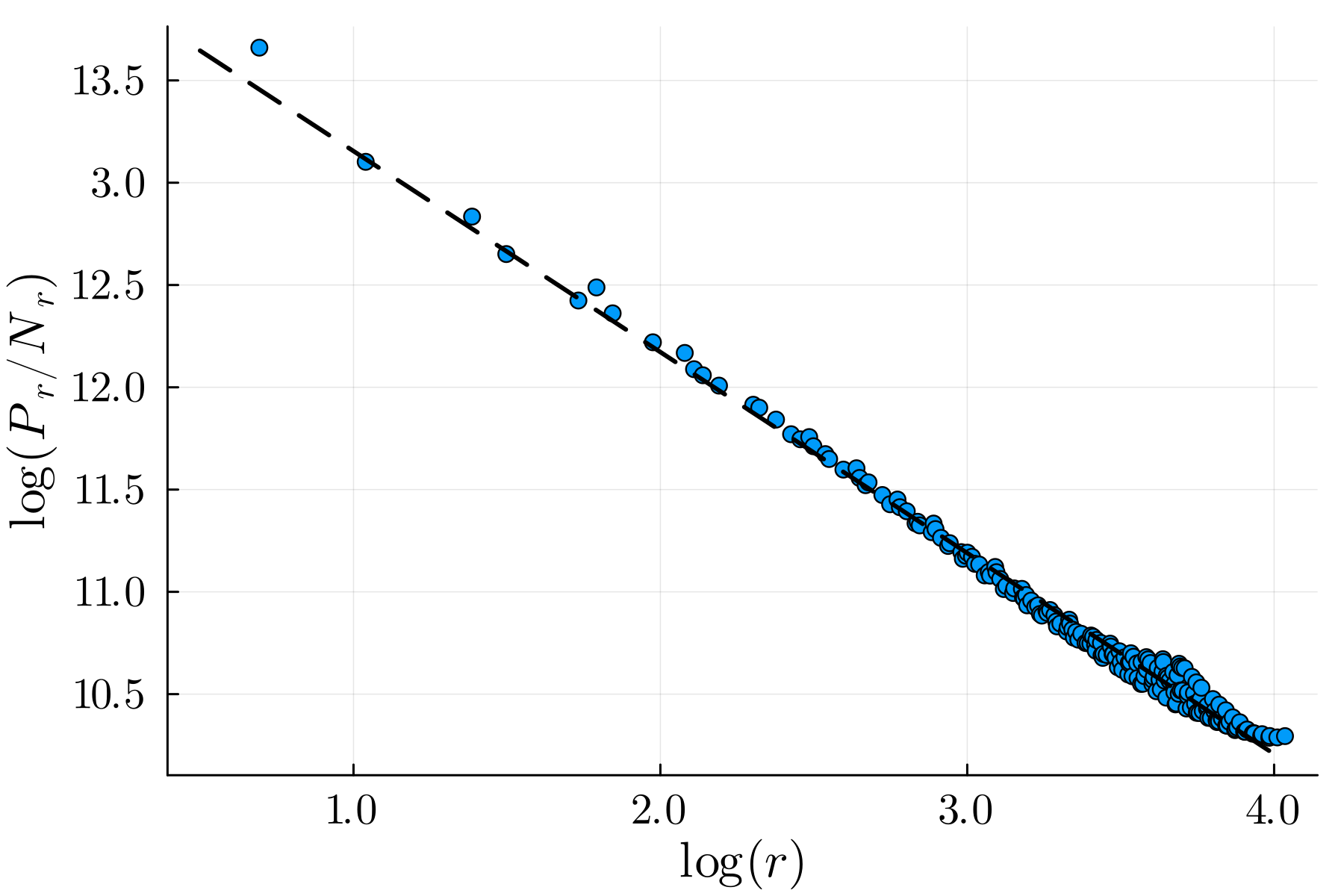}
    }
    \subfigure[]{
    \includegraphics[width=0.88\linewidth]{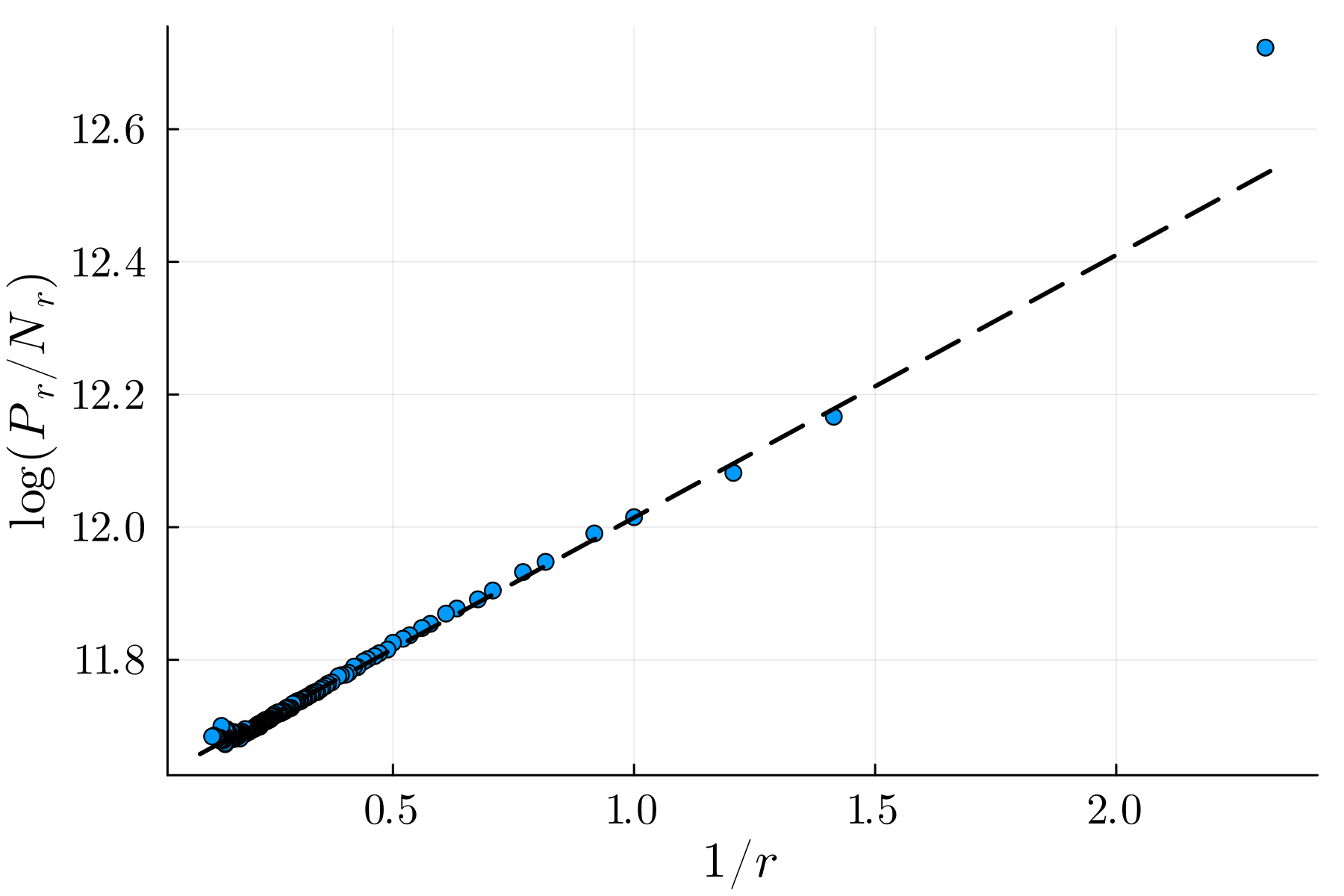}
    }
    \subfigure[]{
    \includegraphics[width=0.88\linewidth]{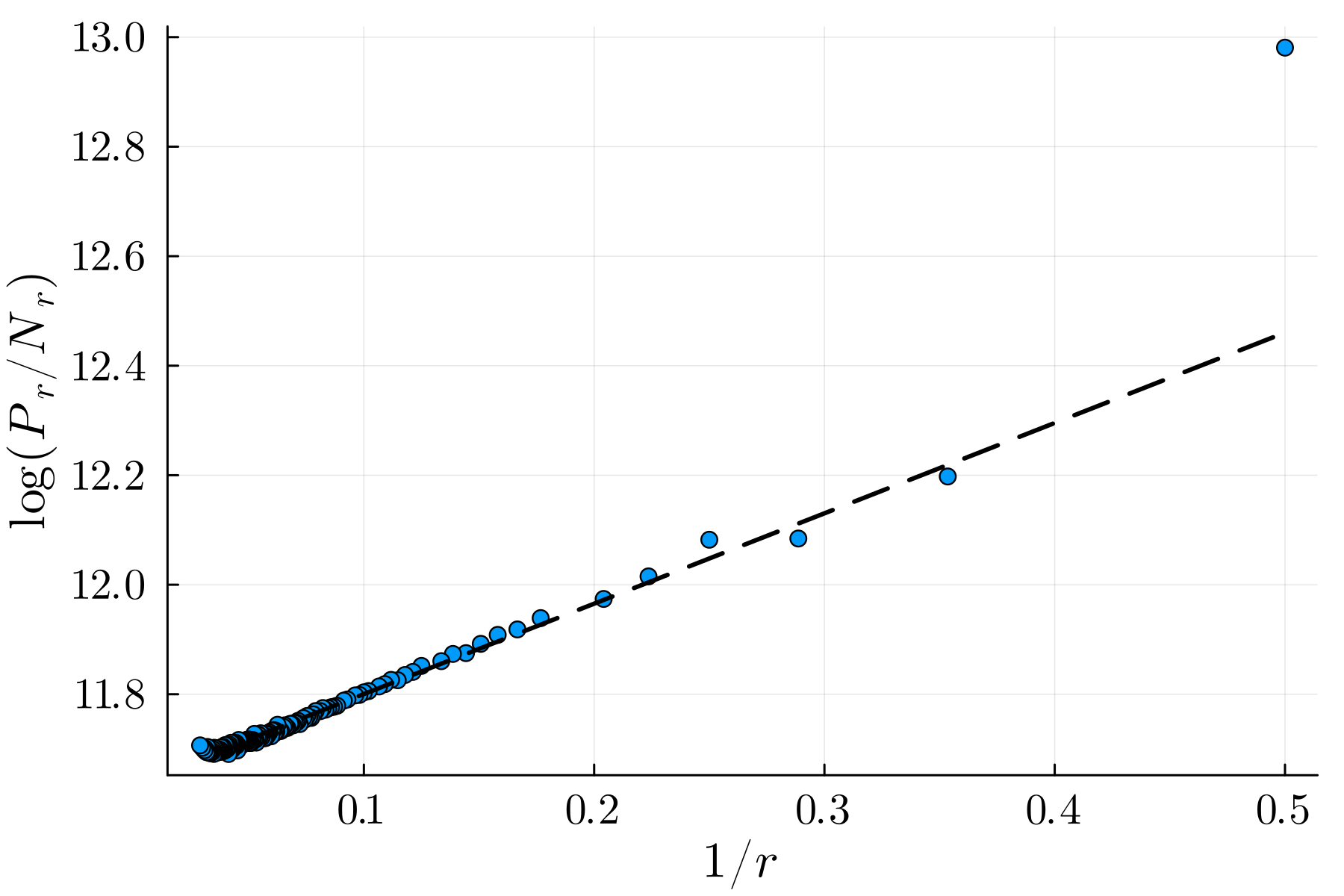}
    }
    \caption{Logarithms of separation frequency divided by the discrete measure $N_r$ displayed as functions of ((a)) $\log(r)$ or ((b) and (c)) $1/r$ for rooton-antirooton dynamics. We find that the effective interaction between particle-antiparticle pairs obeys the form expected of Coulomb interactions in two ((a)) and three ((b) and (c)) dimensions, excepting an over representation of states where the excitations sit on neighboring sites.}
    \label{fig:Coulomb_interaction}
\end{figure}

Separation data for rooton-antirooton pairs under generalized standard model dynamics is presented in Fig. \ref{fig:Coulomb_interaction} for the $q=4$ models on the square and pyrochlore lattice, and for the $q=6$ cubic lattice. We find that the separation data are consistent with the expected power law in two dimensions and a $1/r$ potential in three dimensions. Critically again, the $q=6$ model behaves identically to the $q=4$ model in three dimensions, supporting our claim that emergent Coulomb fields are a generic feature of all $q$-state Potts ices.

\section{Quantum $q$-color Potts Ices} \label{sec:Quantum}

The $q$-state classical Potts ices have been shown to be effectively described by emergent Coulomb physics, with fields $E^i_{c_r,c_s}$ taken from the generators of the Cartan sub-algebra of $\mathfrak{su}(q)$. We now extend these models by introducing quantum fluctuations, and explore in some detail the constraints the Cartan structure of the emergent gauge fields imposes on the phenomenology of the theory. 

In analogy with typical Hamiltonians for quantum spin ice, which are constructed from nearest neighbor quadratic spin operators such as $S^+_iS^-_j$, we append similar nearest neighbor fluctuations of the form $R^{\alpha}_rR^{-\alpha}_s$ to Eq. (\ref{eq:Classical_Hamiltonian}):
\begin{equation}
	H = \frac{J_Q}{2} \sum_{C}\sum_{i}(Q^i_{C})^2 - J_{\pm} \sum_{\langle r,s \rangle}\sum_{\alpha>0} R^{\alpha}_{r}R^{-\alpha}_{s} + \text{h.c}. .\label{eq:Quantum_Hamiltonian}
\end{equation}
We note here that the form of the Hamiltonian chosen possesses a global symmetry which is related to the conservation of charge per sublattice. The elements of the global symmetry group are concretely parameterized as:
\begin{equation}
    U_{\theta}=\exp\left(i\sum_{r,i} \theta^iE^i_r\right).
\end{equation}
This symmetry group is compact $\text{U}(1)^{\otimes(q-1)}\bigr|_{\mathbb{Z}_2\times S_{q}}$; the manifold upon which the $\text{U}(1)$ factors are compactified has the $\mathbb{Z}_{2}\times S_{q}$ point group symmetry of the root lattice of $\mathfrak{su}(q)$. 

In the limit $J_{\pm} \ll J_{Q}$, the quantum fluctuations can be applied perturbatively to the classical degenerate state space, wherein the lowest order terms are ring exchange operators around primitive plaquettes. With the introduction of a chemical potential term, the resulting effective Hamiltonian can be deformed into a Rokhsar Kivelson form \cite{PhysRevLett.61.2376}, giving:
\begin{footnotesize}
\begin{multline}
  H= \sum_{\text{plaquette}}  -K\left(\Ket{
\adjustbox{valign=c,raise=1.0ex,scale=0.36}{
\chemfig{*6(-=-=-=)} 
};\textcolor{red}{\bullet}\textcolor{ForestGreen}{\bullet}
}
\Bra{
\adjustbox{valign=c,raise=1.0ex,scale=0.35}{
\chemfig{*6(=-=-=-)} 
};\textcolor{red}{\bullet}\textcolor{ForestGreen}{\bullet}
} + \text{h.c.}\right) \\ + \mu\left(\Ket{
\adjustbox{valign=c,raise=1.0ex,scale=0.36}{
\chemfig{*6(=-=-=-)} 
};\textcolor{red}{\bullet}\textcolor{ForestGreen}{\bullet}
}\Bra{
\adjustbox{valign=c,raise=1.0ex,scale=0.36}{
\chemfig{*6(=-=-=-)} 
};\textcolor{red}{\bullet}\textcolor{ForestGreen}{\bullet}
}+\Ket{
\adjustbox{valign=c,raise=1.0ex,scale=0.36}{
\chemfig{*6(-=-=-=)} 
};\textcolor{red}{\bullet}\textcolor{ForestGreen}{\bullet}
}\Bra{
\adjustbox{valign=c,raise=1.0ex,scale=0.36}{
\chemfig{*6(-=-=-=)} 
};\textcolor{red}{\bullet}\textcolor{ForestGreen}{\bullet}
}\right) \\ + (\text{other color pairs}),
\end{multline}
\end{footnotesize}
where $\Ket{
\adjustbox{valign=c,raise=1.0ex,scale=0.36}{
\chemfig{*6(-=-=-=)} 
};\textcolor{red}{\bullet}\textcolor{ForestGreen}{\bullet}
}$ is a state where the given plaquette has a loop of colors alternating red-green, and $\Ket{
\adjustbox{valign=c,raise=1.0ex,scale=0.36}{
\chemfig{*6(=-=-=-)} 
};\textcolor{red}{\bullet}\textcolor{ForestGreen}{\bullet}
}$ is a state with a loop on the same plaquette, but with red and green interchanged. At $\mu=K$, it can be verified that an equal weighted superposition of classical ground states (within each boundary flux sector with periodic boundaries) is an exact ground state. This is in complete analogy with quantum spin ice, and just as there is there is strong evidence that QSL behavior persists to $\mu=0$ in that model \cite{PhysRevLett.108.067204,PhysRevLett.115.077202,PhysRevLett.120.167202}, we might tentatively expect similar quantum `color liquid' behavior in the Potts ices, at least for small $J_{\pm}$.

To describe such a quantum liquid phase, we must suitably quantize the rooton excitations and the electric fields, which are expected to be the relevant degrees of freedom, and re-express the operators $R^{\alpha}$ in terms of these fields. 

In order to quantize the electric field via the introduction of canonically conjugate gauge field $A^i_r$, the operators $E^i_r$ must be expanded such that, instead of having only $q$ eigenvalues, the electric field can take any value on the root lattice $\Lambda$ of $\mathfrak{su}(q)$, such that the commutation relation $[A^i_r,E^j_s]=i\delta^{ij}_{rs}$ holds. To enable us to project back into the physical Hilbert space via an energetic constraint, the origin of the lattice $\Lambda$ is chosen to lie at the center of a simplex, such that the original $q$ degrees of freedom have electric fields of the smallest magnitude. With this adjustment made, the conjugate gauge field can then be defined, and has eigenvalues that are taken from the Brillouin zone of the root lattice.

To quantize rooton excitations, we define a canonical conjugate to the charge defects $Q^i_C$ to be $\varphi^i_C$. An operator that then creates a rooton excitation is: $(\Phi^{\alpha})^{\dagger}=e^{i\alpha^i\varphi^i}$.

Following the Gauge Mean Field Theory approach as used in quantum spin ice \cite{PhysRevB.86.104412,PhysRevLett.108.037202,PhysRevB.87.205130,PhysRevLett.132.066502,PhysRevB.90.214430}, the $R^{\alpha}_r$ operators can be fractionalized in this expanded Hilbert space into fields creating and destroying charges at cells $C_A$ and $C_B$, and an operator shifting the electric field on link $r$ by root $\alpha$:
\begin{equation}
    R^{\alpha}_{r} = \frac{1}{q}(\Phi^{\alpha}_{C_A})^{\dagger} e^{i\alpha^iA^i_r} \Phi^{\alpha}_{C_B} \ . \label{eq:R_fractionalisation}
\end{equation}

Note that going forward we will routinely employ the Einstein summation convention for sums over the $(q-1)$ root components. With these quantization definitions, and the above fractionalisation of physical degrees of freedom, the theory possesses a gauge symmetry when expressed in the new degrees of freedom, which is of the form:
\begin{align}
    \Phi^{\alpha}_C \rightarrow&e^{i\alpha^i\chi^i_C} \Phi^{\alpha}_C \\ 
    A^i_{r} \rightarrow& A^i_r +\chi^i_{C_A}-\chi^i_{C_B}
\end{align}
The gauge group characterizing these transformations is also $\text{U}(1)^{\otimes(q-1)}\bigr|_{\mathbb{Z}_2\times S_{q}}$. We should be careful to note that whilst the $U(1)^{\otimes(q-1)}$ gauge symmetry under which the fields $\Phi^{\alpha}_C$ transform is a redundancy of our degrees of freedom, the global symmetry under which the $R^{\alpha}_r$ transform is not, and may be spontaneously broken.

As mentioned previously, the emergent gauge structure of the theory is the Cartan sub-group of SU$(q)$, and so the emergent gauge theory encountered here is an abelian projection of the full SU$(q)$ gauge theory. Such a projection can be performed by gapping off-diagonal, or charged, gluons \cite{HOOFT1981455,PhysRevD.57.7467}. In the Potts ices, these gapped, charged, excitations created by off-diagonal elements of $\mathfrak{su}(q)$ are the rootons. The remnants of non-abelian behavior from the larger gauge group should thus be found in their mutual interactions.

If we consider possible combinations of matter operators defined on a single state, the allowed gauge invariant combinations are overall charge neutral. For $q>2$, the set of gauge invariant matter operators at one's disposal is not limited to powers of $(\Phi^{\alpha})^{\dagger}\Phi^{\alpha}$; the root charges are vectors, and any sum of roots that forms a closed polygon corresponds to a neutral operator.

\begin{figure}
    \centering
    \includegraphics[width=1.0\linewidth]{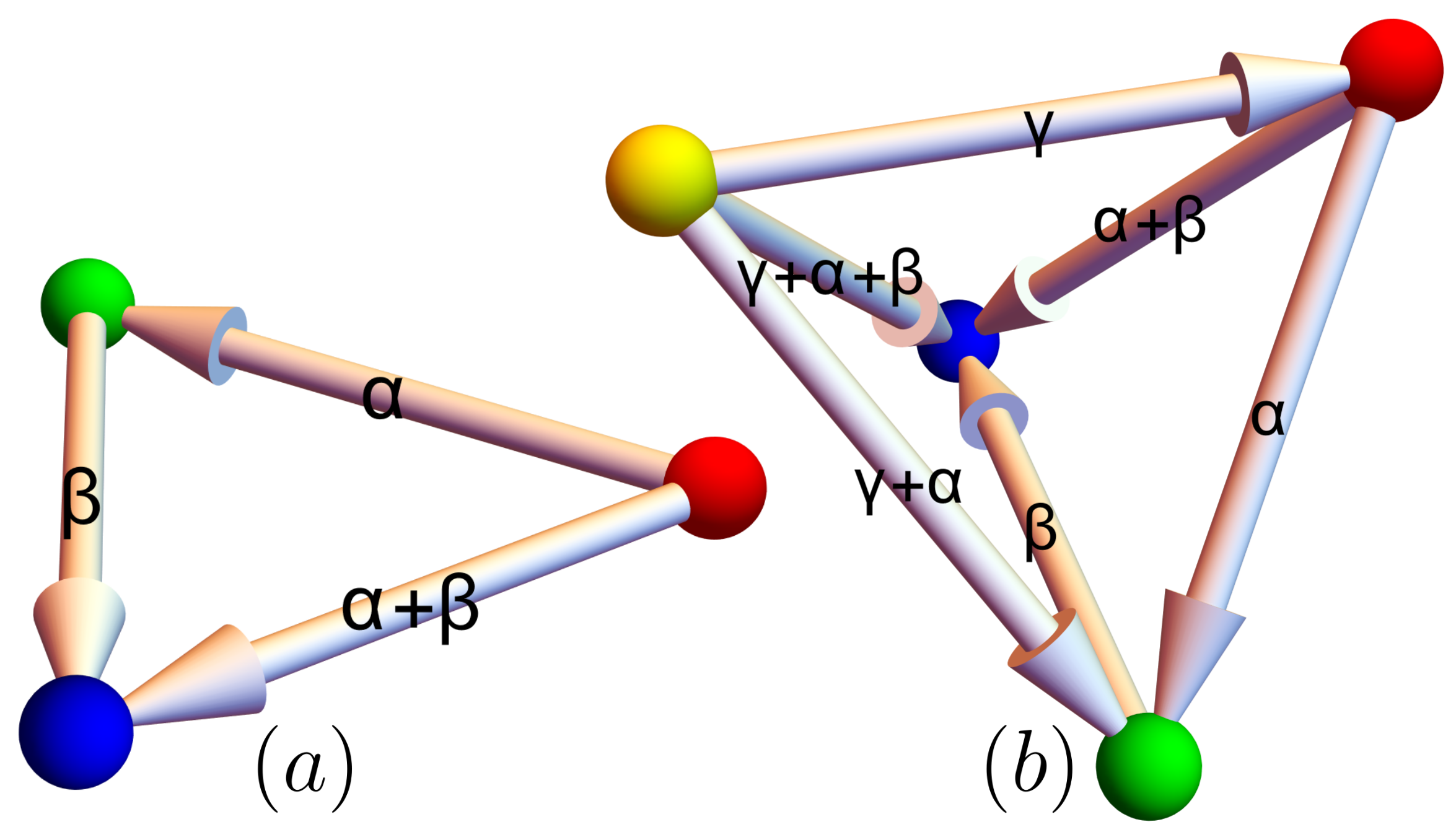}
    \caption{Diagrams of the root systems for (a) $\mathfrak{su}(3)$ and (b) $\mathfrak{su}(4)$, with roots labeled as sums of simple positive roots. Each root is associated with the exchange of two Potts state colors. Each edge of the $(q-1)$-simplex corresponds to a different matter field flavor. Combinations of fields whose root sum is zero are gauge invariant operators.}
    \label{fig:Root_Diagrams}
\end{figure}

The primitive operators generating non-trivial matter interactions are the gauge-invariant three-field operators of the general form $(\Phi^{\alpha+\beta}_r)^\dagger\Phi^{\alpha}_r\Phi^{\beta}_r$. These are associated with each constituent equilateral triangle of the $(q-1)$-simplex, and correspond to flavor changing, particle number non-conserving processes.

Continuing the analogy with SU$(q)$ gauge theory, these operators correspond to three-gluon interaction vertices, but their origin can be also understood microscopically in the lattice model. If a red link is flipped to a green link to create a pair of red-green type rootons, and then subsequently flipped from green to blue, red-blue type rootons result. Thus the creation of a red-green, then a green-blue rooton, must correspond to the creation of a single red-blue rooton, and it is this behavior the three-field operators reflect.

Consulting Fig. \ref{fig:Root_Diagrams}, it can be seen that for $q\ge 4$ one can also construct non-trivial 4-cycles on the simplex corresponding to gauge invariant combinations of four distinct rooton fields. Naturally, one can also construct four-field operators involving pairs of distinct rootons such as $\Phi^{\alpha\dagger}\Phi^{\alpha}\Phi^{\beta\dagger}\Phi^{\beta}$.

The three-field operators are important for a number of reasons. They are relevant operators in the renormalisation group sense, and so their presence greatly influences the critical behavior of the theory. Descending from the non-abelian parts of the SU$(q)$ gauge group, they also induce wholly different, non-QED physics unique to the $q>2$ Potts ices, enriching the quantum Potts ices beyond simply possessing multiple emergent photons.

In the following sub-sections, we shall give an overview of phenomena particular to the $q>2$ quantum Potts ices that follow from the constraints of their Cartan gauge structure.

\subsection{Gauge mean field theory} \label{sec:GMFT}

In this section we  write down an effective action for the quantum liquid phase that may be stabilized by the model for perturbative $J_{\pm}$, and provide a short treatment within the framework of gauge mean field theory. To begin, we consider the model in the absence of matter excitations. For small, perturbative $J_{\pm}$, the effective Hamiltonian for the matter-free sector is generated by operators that consist of rings of $R^{\alpha}_rR^{-\alpha}_s$ around primitive plaquettes $p$. These are operators that flip worms of alternating colors identified in the classical model. Expressed in terms of the gauge field, these plaquette flip operators have the form:

\begin{equation}
    \Pi^{\alpha}_p = \exp\left(i\alpha^i\sum_{l\in \partial p}s_lA^i_l\right).
\end{equation}
Each plaquette should be assigned an orientation defining a positive and negative sense of field circulation. The symbol $s_l$ is then positive if the orientation of edge $l$ (which is chosen to be pointing from sublattice A to sublattice B) is parallel to the circulation sense, and negative otherwise. The expression $\sum_{l \in \partial p} s_lA^i_l$ is the lattice curl, and can be understood to be a lattice magnetic flux density $B_p^i$, which also takes values from the Brillouin zone of the root lattice. 

In addition to the magnetic field terms, an energetic constraint $\frac{U}{2}E^2$ on the electric field is required to return the model to the physical Hilbert space, as alluded to above.

The effective Hamiltonian for the pure gauge sector can then be written in the following form:
\begin{equation}
    H_{\text{eff}} = \frac{U}{2} \sum_{r}(E^i_r)^2-2V\sum_{p,\alpha} \cos(\alpha^i B^i_p) . \label{eq:Gauge_eff_H}
\end{equation}
Just as for quantum spin ice, the effective Hamiltonian closely resembles that of classical electromagnetism, though with the additional complication of compactified $U(1)$ fields, which allow for non-trivial topological configurations of the fields, and in particular topological defects-- visons. In analyzing this Hamiltonian, the standard approach is to identify the energy minima, and expand about these minima for small fluctuations of the gauge field, which is suitable for a deconfined phase where the electric field hosts large fluctuations, and if we only wish to study photon modes and discard the visons. However, with $q>2$, it is not entirely trivial to identify these minima as a function of the $(q-1)$ dimensional Brillouin zone on which the $B^i_p$ are defined. Certainly for positive $V$, the vacuum state is simply that with zero flux through every plaquette, but for negative $V$, there are several non-equivalent flux states that give equal minimal energies per plaquette. This issue of \textit{flux frustration} will be discussed in some detail in subsection \ref{sec:flux}. For now, we  focus on the more familiar situation found for $V>0$.

The Lagrangian density for the pure gauge sector in imaginary time is simply the above Hamiltonian, which can be expressed purely in terms of the gauge field and its time derivative:
\begin{equation}
    S_{\text{EM}}= \int_0^{\beta}\text{d}\tau\sum_{l} \frac{1}{2U}(\dot{A}^i_{l;\tau})^2 -2V \sum_{p,\alpha}\cos\left(\alpha^i\sum_{l \in \partial p}s_lA^i_{l;\tau}\right). \label{eq:EM_action}
\end{equation}
Note that strictly here we can only equate the time derivative of the gauge field with the `transverse' part of the electric fields whose divergence is zero. The longitudinal parts are determined by enforcing the Gauss law when matter excitations are present.

Equation (\ref{eq:EM_action}) describes the propagation of $(q-1)$ species of photon when the cosine is expanded for small fluctuations. Using the property of roots that $\sum_{\alpha} \alpha^i \alpha^j \propto \delta^{ij}$ \cite{Groups_Fulton-Harris}, it can be verified that on expanding Eq. (\ref{eq:EM_action}) to leading order in the $A^i_{l;\tau}$, cross terms between the different photon flavors vanish.

Now turning to the effective theory and action for the matter fields, one must make a decision of how to approximate the rotor operators $\Phi^{\alpha}$ when constructing a field theory. We  follow other authors \cite{PhysRevB.86.104412,PhysRevLett.108.037202,PhysRevB.87.205130,PhysRevLett.132.066502} and approximate these operators as complex scalar fields $\phi^{\alpha}$, whilst introducing suitable Lagrange multiplier terms to the action to return the model to the physical Hilbert space. As for $q=2$ quantum spin ice, one such required term is $\lambda_{C;\tau} (\phi^{\alpha*}_{C;\tau}\phi^{\alpha}_{C;\tau}-1)$ which normalizes the complex scalar fields to one, as required of rotors. However, for $q>2$, the algebra obeyed by the rotor operators has a greater structure, and we must also encode relationships such as $\Phi^{\alpha+\beta*}\Phi^{\alpha}\Phi^{\beta}=1$ so that the theory `knows' about the $\mathfrak{su}(q)$ structure of the excitations. We thus also introduce Lagrange multiplier terms of the form: $g_{C;\tau}(\phi^{\alpha+\beta*}_{C;\tau}\phi^{\alpha}_{C;\tau}\phi^{\beta}_{C;\tau}-1)$. For the sake of symmetry, and to ensure that the action is Hermitian, we include as many such terms as there are primitive equilateral triangles constituting the $q-1$ simplex, plus their complex conjugates.

To properly account for the Gauss law, and include Coulomb interaction effects, a further Lagrange constraint strictly enforcing the Gauss law must also be included for a lattice formulation of the theory. The longitudinal part of the electric field term that did not contribute to the photon dynamics must also be included here. The matter sector action is then:
\begin{widetext}

\begin{multline}
    S_{\text{Mat}} = \int_0^{\beta}\text{d}\tau \left(\sum_{\mathbf{r}_{C}}\left[ iQ^i_{\mathbf{r}_{C};\tau} \partial_{\tau} \varphi^i_{\mathbf{r}_{C};\tau} +\frac{J_Q}{2}(Q^i_{\mathbf{r}_{C};\tau})^2+\eta^i_{\mathbf{r}_{C};\tau}\left\{\sum_{l \in \partial C}^4E^{i\parallel }_{\mathbf{r}_{l};\tau}-Q^i_{\mathbf{r}_{C};\tau}\right\}+\lambda_{\mathbf{r}_{C};\tau} \sum_{\alpha}\left\{\phi^{\alpha*}_{\mathbf{r}_{C};\tau}\phi^{\alpha}_{\mathbf{r}_{C};\tau}-1\right\} \right. \right. \\ \left. \left.+g_{\mathbf{r}_C;\tau} \sum_{\text{3-cycles}}\left\{\phi^{\alpha+\beta*}_{\mathbf{r}_C;\tau}\phi^{\alpha}_{\mathbf{r}_C;\tau}\phi^{\beta}_{\mathbf{r}_C;\tau} +\text{c.c}-2\right\} \right]  +\frac{U}{2}\sum_{\mathbf{r}_l}(E^{i\parallel}_{\mathbf{r}_l;\tau})^2 -\frac{J_{\pm}}{q^2} \sum_{a=A,B}\sum_{\langle C_{a},C_a'\rangle}\sum_{\alpha}  \phi^{\alpha*}_{\mathbf{r}_{C_a};\tau} \phi^{\alpha}_{\mathbf{r}_{C'_a};\tau} e^{-i\eta_{a}(A_{\mathbf{r}_{l};\tau}-A_{\mathbf{r}_{l'};\tau})}\right), \label{eq:Matter_effective_action}
\end{multline}
\end{widetext}
where in the last term, $\langle C_a, C_a'\rangle$ indicates a sum over cells on sublattice $a$ that are nearest neighbor with respect to that sublattice, and $l$ and $l'$ are links which connect $C_a$ and $C'_a$ respectively to the opposite sub-lattice site that sits between them. The sum over 3-cycles should be understood as a sum over all sets of three roots that sum to zero, these each corresponding to one primitive equilateral triangle of the root diagram of $\mathfrak{su}(q)$. Both the fields $\lambda$ and $g$ may be approximated as fixed constants if the constraints they enforce are relaxed to hold only when averaged over space and imaginary time:
\begin{align}
	\frac{1}{\beta N} \sum_{\mathbf{r}_{C};\tau} \langle \phi^{\alpha*}_{\mathbf{r}_{C};\tau}\phi^{\alpha}_{\mathbf{r}_{C};\tau} \rangle =& 1,  \label{eq:Two-body constraint}\\
	\frac{1}{\beta N} \sum_{\mathbf{r}_{C};\tau} \langle \phi^{\alpha+\beta*}_{\mathbf{r}_{C};\tau}\phi^{\alpha}_{\mathbf{r}_{C};\tau}\phi^{\beta}_{\mathbf{r}_{C};\tau} \rangle =& 1. \label{eq:Three-body constraint}
\end{align}

In following sections we will consider effective long-wavelength limits of this action in three spatial dimension, first without a dynamical gauge field (\ref{sec:Landau}), followed by a brief discussion of the theory with gauge field dynamics (\ref{sec:QED}).

Before this however, let us look at the action of Eq.~(\ref{eq:Matter_effective_action}) within the gauge mean field theory approximation by setting $A=\bar{A}=0$, suitable for $V>0$. Remaining couplings to other fields can be removed by first strictly imposing the gauss law, integrating out charges $Q^i_{\mathbf{r}_C;\tau}$ (which requires approximating this field as continuous, which is not suitable when wanting to study the coupling to visons), and then expressing $\alpha^i\partial_{\tau}\varphi^{i}_{\mathbf{r}_C;\tau}$ as $\frac{i}{2}(\phi^{\alpha*}\partial_{\tau}\phi^{\alpha}-\phi^{\alpha}\partial_{\tau}\phi^{\alpha*})$. With the assumptions that violations in the hard rotor constraint are small, the action can be massaged into the form:
\begin{multline}
    S_{\text{GMFT}} \approx \int_0^{\beta} \text{d}\tau \left(\sum_{\mathbf{r}_C} \left[\frac{\kappa}{2} |\partial_{\tau} \phi^{\alpha}_{\mathbf{r}_C;\tau}|^2+\lambda|\phi^{\alpha}_{\mathbf{r}_C;\tau}|^2\right] \right. \\-\frac{J_{\pm}}{q^2}\sum_{a=A,B}\sum_{\langle C_{a},C_a'\rangle}\sum_{\alpha}  \phi^{\alpha*}_{\mathbf{r}_{C_a};\tau} \phi^{\alpha}_{\mathbf{r}_{C'_a};\tau} \\ \left. +g\sum_{\text{3-cycles}}\sum_{\mathbf{r}_C}\left\{\phi^{\alpha+\beta*}_{\mathbf{r}_C;\tau}\phi^{\alpha}_{\mathbf{r}_C;\tau}\phi^{\beta}_{\mathbf{r}_C;\tau} +\text{c.c}\right\} +V(\{\phi^{\alpha}\})\right),
\end{multline}
where $\kappa$ is a kinetic factor dependent on $J_Q$ and the strength of coulomb interactions $U$, and $V(\{\phi^{\alpha}\})$ consists of four-field coulomb interactions. 

GMFT actions are commonly employed to model spinon dynamics in quantum spin ice \cite{PhysRevB.86.104412,PhysRevLett.108.037202,PhysRevB.87.205130,PhysRevLett.132.066502}. However, the GMFT action for $q>2$ is non-trivially modified by the presence of the three-field interaction $g$ introduced by the additional Lagrange constraints, which limits the utility of this approach. Crucially, the three-field interaction is \textit{RG relevant} in three dimensions, meaning that we cannot treat this interaction perturbatively if we wish to calculate bulk observables, including the expectation value of Eq. (\ref{eq:Three-body constraint}) that would self-consistently fix the value of $g$.

\subsection{Continuum GMFT and Landau theory} \label{sec:Landau}

\begin{figure}
    \centering
    \includegraphics[width=0.9\linewidth]{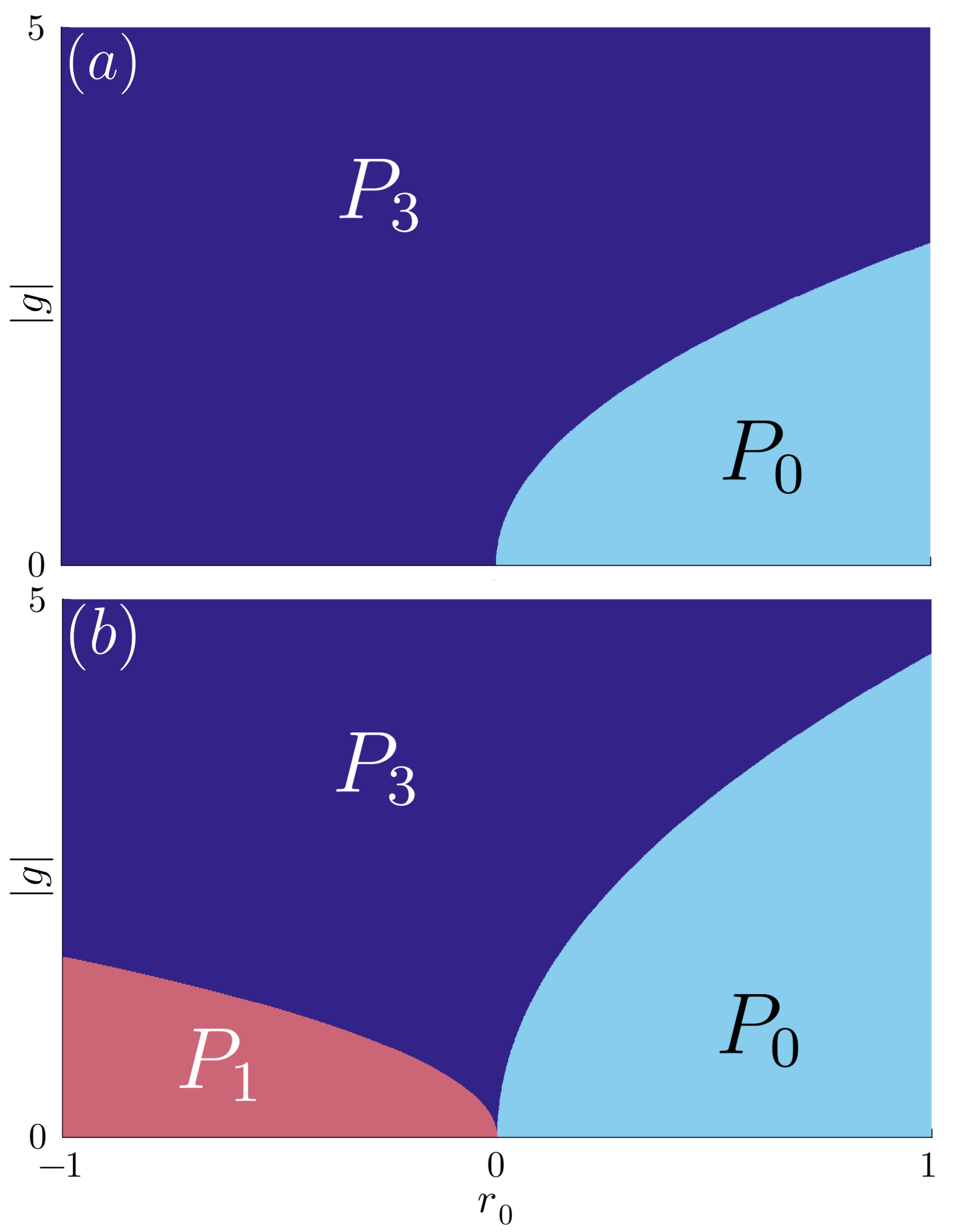}

    \caption{Phase diagrams for the $q=3$ GMFT action in $(3+1)$ dimensions. In (a) $u_1=4 $, $u_{2;a}=0$ and in (b) $u_1=1$, $u_{2;a}=2$. Three distinct phases are present: The gauge symmetric phase $P_0$, a phase $P_1$ where one of the two $\text{U}(1)$ factors in the Cartan is broken, and a fully broken phase $P_3$. All phase boundaries are lines of constant $r_0/g^2$ at the mean field level.}
    \label{fig:Nc3_phase_diagrams}
\end{figure}

\begin{figure*}
    \centering
    \includegraphics[width=1.0\linewidth]{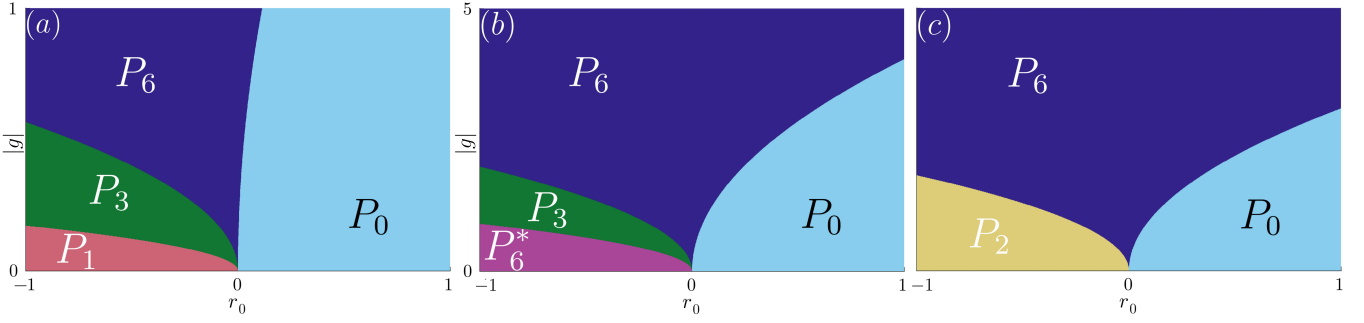}

    \caption{Phase diagrams for the $q=4$ GMFT action in $(3+1)$ dimensions. In (a) $u_1=2$, $u_{2;a}=1.2$, $u_{2;b}=2$, $u_4=0$; in (b) $u_1=3$, $u_{2;a}=2$, $u_{2;b}=1$, $u_{4}=2$; in (c) $u_1=1$, $u_{2;a}=2$, $u_{2;b}=0$, $u_4=0$. Symmetry broken phases $P_n$ are labeled by the number of matter fields that are condensed. Both the $P_6$ and $P_6^*$ phases break the entire Cartan symmetry, however they differ in their behavior under the root point group $\mathbb{Z}_2\times S_{q}$, with $P^*_6$ breaking the $\mathbb{Z}_2$ factor associated with flipping the signs of all root charges.}
    \label{fig:Nc4_phase_diagrams}
\end{figure*}

\begin{figure*}
    \centering
    \includegraphics[width=0.8\linewidth]{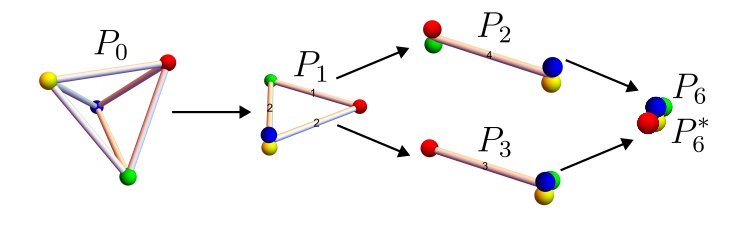}
    
    \caption{Progressive condensation of matter fields shown diagrammatically on the root diagram of $q=4$. Condensation of a given root associates the two Potts colors swapped by that root, and associates matter fields with now identical start and end points on the condensed simplex. For $q=4$, 0,1,2,3, or 6 fields can be condensed simultaneously. The $P_6$ and $P_6^*$ phases remain distinguished by their transformation properties under the point group $\mathbb{Z}_2\times S_{q}$. }
    \label{fig:Nc4_Root_condensation}
\end{figure*}

The RG relevance of the three-field operators poses an obstacle to calculating the effects of quantum fluctuations away from the saddle points of the GMFT action on observables, however we can still identify these saddle points within Landau theory. Throughout this section, we maintain the mean gauge field condition $\bar{A}=0$ relevant for $V>0$.

In identifying saddle points of the GMFT action, we must keep in mind that this complex scalar boson field theory is approximating the true hard rotor theory. This approximation introduces a number of spurious non-gauge phase degrees of freedom that are fixed to zero when the Lagrange constraints $(\Phi^{\alpha+\beta}_r)^\dagger\Phi^{\alpha}_r\Phi^{\beta}_r=1$ are strictly enforced. The GMFT action explcitly has this restriction relaxed, and this will have to be kept in mind.

As $\text{U(1)}$ gauge theories suffer from instanton instabilities \cite{POLYAKOV197582,Polyakov_book} in $(2+1)$ dimensions, we choose to examine the effective theories for $q=3$ and $q=4$ in three spatial dimensions and imaginary time.

A full renormalisation group analysis including single loop corrections for both theories is given in supplementary material \cite{supp}. Corrections to the pure matter theory phase diagrams beyond mean field theory are minor, as the gross behavior is dominated by the two RG relevant couplings $r_0$ for the two-field operator, and $g$ for the three-field operator. As such, we limit our discussion here to the mean field results.

The Landau free energy density for the $q=3$ theory, dropping spatial variations in the fields, and including all RG relevant and marginal operators is:

\begin{multline}
    \mathcal{F} = \sum_{\alpha>0}\left\{r_0|\phi^{\alpha}|^2+\frac{u_1}{4}|\phi^{\alpha}|^4\right\} +u_{2;a}\sum_{\substack{\alpha,\beta \neq \alpha \\ (\alpha,\beta)\neq 0}} |\phi^{\alpha}|^2|\phi^{\beta}|^2 \\ + g\left\{\phi^{\alpha+\beta*}\phi^{\alpha}\phi^{\beta} + \text{h.c.} \right\}
\end{multline}
The $q=3$ theory possess two distinct four-field operators-- one responsible for interactions between rootons of a single flavor (and their antiparticles) and another giving rise to interactions between distinct rootons. Both of these $u_{\mu}$ couplings would be generated by integrating out electromagnetic couplings from a full dynamical theory. 

Phase diagrams for this free energy for fixed values of $u_1$ and $u_{2;a}$ are presented in Fig. \ref{fig:Nc3_phase_diagrams}. Phase boundaries occur along lines of constant $r_0/g^2$, and phases are labeled $P_n$, where $n$ is the number of matter fields that have condensed. Transitions are generically first order, save those through the critical point $g=0,r_0\approx0$.

Rooton condensation proceeds via the Anderson-Higgs mechanism, which gaps some or all of the emergent photon modes dependent on the number of rootons condensed. In the $P_1$ phase, gauge transformations orthogonal to the root vector of the condensed field survive, and only one of the two photons becomes gapped, whilst in the $P_3$ phase, all photons become gapped. 

Condensation of a rooton field $\phi^{\alpha}$ also implies that the corresponding gauge-invariant $\mathfrak{su}(3)$ generator $R^{\alpha}$ acquires a non-zero expectation value, breaking the global $\text{U}(1)^{\otimes(q-1)}\bigr|_{\mathbb{Z}_2\times S_{q}}$ symmetry. Thus each of these Higgs phases is also an ordered phase, with an equal number of goldstone modes to gapped photons.

Turning now to the $q=4$ model, the spatially uniform free energy with all RG relevant and marginal terms is:

\begin{multline}
    \mathcal{F} = \sum_{\alpha>0}\left\{r_0|\phi^{\alpha}|^2+\frac{u_1}{4}|\phi^{\alpha}|^4\right\} +u_{2;a}\sum_{\substack{\alpha,\beta \neq \alpha \\ (\alpha,\beta)\neq 0}} |\phi^{\alpha}|^2|\phi^{\beta}|^2 \\ +u_{2;b}\sum_{\substack{\alpha,\beta \neq \alpha \\ (\alpha,\beta)= 0}} |\phi^{\alpha}|^2|\phi^{\beta}|^2+ g\sum_{\text{3-cycles}}\left\{\phi^{\alpha+\beta*}\phi^{\alpha}\phi^{\beta} + \text{h.c.} \right\} \\ + u_4\sum_{\text{4-cycles}}\left\{\phi^{\alpha+\beta+\gamma*}\phi^{\alpha}\phi^{\beta}\phi^{\gamma} + \text{h.c.} \right\}
\end{multline}
New four-field operators are present in the $q=4$ theory that are absent in the $q=3$ theory. These are the $u_{2;b}$ operator, coupling orthogonal rooton pairs that have no classical entropic interaction, and $u_4$, which is associated with a gauge invariant four-cycle of distinct rooton fields. For the $q=4$ model, there are three such 4-cycles on its simplex, and four triangular 3-cycles contributing to the $g$ term.

There are also many more symmetry broken phases encountered in the $q=4$ theory. A selection of phase diagrams for various fixed values of $u_1,u_{2;a},u_{2;b},$ and $u_4$ are presented in Fig. \ref{fig:Nc4_phase_diagrams}. Whilst most phases are distinguished by the number of condensed matter fields, the $P_6$ and $P_6^*$ phases both condense all fields, but differ in the value of non-gauge phase differences between the complex matter fields. As such non-gauge phases are not physical degrees of freedom in the original rotor model, the $P_6^{*}$ phase exists only in the complex scalar theory, and is entirely off-shell in the rotor theory.

The allowed field condensation patterns for any $q$ can be deduced from the root diagram simplex for that model. This is shown in Fig. \ref{fig:Nc4_Root_condensation} for the $q=4$ model. Starting from the initial simplex, one can progressively condense fields by removing an edge from the graph and identifying its end points. Multiple edges which now possess the same start and end point are also associated, giving a new simplex associated with the partially broken Cartan symmetry. This process of eliminating edges continues until all possible condensations have been identified, up to possible differences in the non-gauge relative phases between the condensed fields, though these are again non-physical for the original rotor degrees of freedom.

\subsection{Coupling to dynamical gauge field for $q=4$} \label{sec:QED}

Whilst a perturbative treatment of the effective field theory for $q>2$ Potts ices is not suitable for the IR due to the relevance of three-field interaction terms, we are able to make a few qualitative descriptions of the full theory with dynamical coupling to the gauge field. Thus, in this section we briefly explore the properties of a modified QED encountered for the $q=4$ model in $(3+1)$ dimensions.

An effective low-energy Lagrangian for such a model can be written down by minimally coupling the relevant GMFT Lagrangian to the $(q-1)$ $\text{U(1)}$ gauge fields, and imbuing these fields with dynamics through the standard Maxwell action. This minimal coupling is achieved through covariant derivatives $D^{\alpha}_{\mu}$ defined with reference to the charge of the relevant field $\alpha$ as:
\begin{equation}
    D^{\alpha}_{\mu}=\partial_{\mu}-i\alpha^iA^i_{\mu}.
\end{equation}
In an effective low-energy theory, the normalization $|\alpha|^2$ of the root charges is an unknown coupling parameter, unlike in a microscopic construction where its value would be 2. The electromagnetic field tensors are defined for each of the $q-1$ gauge field components as:
\begin{equation}
\mathcal{F}^{i}_{\mu\nu}=\partial_{\mu}A^i_{\nu}-\partial_{\nu}A^i_{\mu}.
\end{equation}

The minimally coupled Lagrangian for the $q=4$ model is then:
\begin{widetext}
    \begin{multline}
    \mathcal{L} = \sum_{\alpha>0} \left\{(D^{\alpha}_{\mu}\phi^{\alpha})^*(D^{\alpha ;\mu}\phi^{\alpha})-m^2|\phi^{\alpha}|^2-\frac{u_1}{4}|\phi^{\alpha}|^4\right\} -u_{2;a}\sum_{\substack{\alpha,\beta\neq\alpha \\ (\alpha,\beta)\neq0}} |\phi^{\alpha}|^2|\phi^{\beta}|^2 - u_{2;b}\sum_{\substack{\alpha,\beta\neq\alpha \\ (\alpha,\beta)=0}} |\phi^{\alpha}|^2|\phi^{\beta}|^2 \\ -g\sum_{\text{3-cycles}}\left(\phi^{\alpha+\beta*}\phi^{\alpha}\phi^{\beta}+\text{h.c.}\right)-u_4\sum_{\text{4-cycles}}\left(\phi^{\alpha+\beta+\gamma*}\phi^{\alpha}\phi^{\beta}\phi^{\gamma} + \text{h.c.}\right)-\frac{1}{4}\mathcal{F}^{i;\mu\nu}\mathcal{F}^{i}_{\mu\nu}.
    \end{multline}
\end{widetext}
The set of relevant couplings is much more complex than for QED with a single complex scalar field. Notably, the $g$ and $u_4$ terms cannot be generated by the renormalization of pure electromagnetic interactions, and so represent novel direct interactions between the fields, including direct flavor-changing interactions via the exchange of a virtual rooton.

The RG flow equations for the pure GMFT action in the absence of a dynamical gauge field are modified by electromagnetic diagrams. At tree level, the $u_1$ and $u_{2;a}$ flow equations are modified, reflecting the fact that rootons whose roots $\alpha$ and $\beta$ are not orthogonal have direct coulomb interactions in the classical model. 

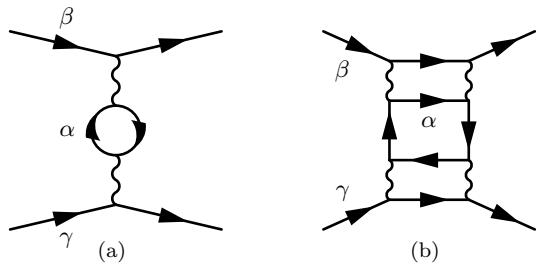
\begin{figure}
    \centering
    \subfigure[]{
    
    \begin{fmffile}{u2_b_one_loop_renormalisation}
        \begin{fmfgraph*}(100,75)
            \fmfleft{i,i2}
            \fmfright{o,o2}
            \fmf{fermion,label=$\gamma$}{i,v}
            \fmf{fermion}{v,o}
    
            \fmf{photon}{v,v3}
            \fmf{fermion,left=1,tension=0.5,label=$\alpha$}{v3,v4}
            \fmf{fermion,left=1,tension=0.5}{v4,v3}
            \fmf{photon}{v4,v2}

            \fmf{fermion,label=$\beta$}{i2,v2}
            \fmf{fermion}{v2,o2}
        
        \end{fmfgraph*}
    \end{fmffile}

    }
    \subfigure[]{
    
    \begin{fmffile}{u2_b_three_loop_renormalisation}
        \begin{fmfgraph*}(100,75)
            \fmfleft{i,i2}
            \fmfright{o,o2}
            \fmf{fermion,label=$\gamma$}{i,v}
            \fmf{fermion}{v,v2}
            \fmf{fermion}{v2,o}
            \fmfforce{(0.35w,0.15h)}{v}
            \fmfforce{(0.65w,0.15h)}{v2}
    
            \fmf{photon}{v3,v5}
            \fmf{photon}{v4,v6}
            
            \fmf{fermion,label=$\alpha$}{v5,v6}
            \fmf{fermion}{v6,v8}
            \fmf{fermion}{v8,v7}
            \fmf{fermion}{v7,v5}

            \fmfforce{(0.35w,0.65h)}{v5}
            \fmfforce{(0.65w,0.65h)}{v6}
            \fmfforce{(0.35w,0.35h)}{v7}
            \fmfforce{(0.65w,0.35h)}{v8}

            \fmf{photon}{v7,v}
            \fmf{photon}{v8,v2}

            \fmf{fermion,label=$\beta$}{i2,v3}
            \fmf{fermion}{v3,v4}
            \fmf{fermion}{v4,o2}
            \fmfforce{(0.35w,0.85h)}{v3}
            \fmfforce{(0.65w,0.85h)}{v4}
        
        \end{fmfgraph*}
    \end{fmffile}

    }
    \caption{Feynman diagrams contributing to the renormalization of the coupling constant $u_{2;b}$. The inner product between the two roots $\beta$ and $\gamma$ is zero, and so these rootons have no tree-level electromagnetic interaction. (a) a diagram contributing to the interaction between the rootons at one-loop level. Once the sum over intermediate roots $\alpha$ is performed, the total contribution from these diagrams vanishes. (b) a higher loop-order diagram with a non-vanishing contribution to $u_{2;b}$.}
    \label{fig:u2_b_feynman_diagrams}
\end{figure}

One must go to several loop orders to generate contributions to the RG flow of the interaction between orthogonal rootons, $u_{2;b}$, from pure electromagnetic diagrams; these excitations remain electromagnetically `transparent' to leading order in quantum fluctuations. Whilst there are several one-loop diagrams that contribute of the type shown in \ref{fig:u2_b_feynman_diagrams} (a), summing over the intermediate rooton $\alpha$ gives a total contribution proportional to $\sum_{\alpha>0}\beta^i\alpha^i\alpha^j\gamma^j\propto\beta^i\gamma^i$, which vanishes if $\beta$ and $\gamma$ are orthogonal.

The first non-vanishing contributions are of the form shown in Fig. \ref{fig:u2_b_feynman_diagrams} (b), and are proportional to $\sum_{\alpha>0}(\vec{\alpha}\cdot\vec{\beta})^2(\vec{\alpha}\cdot\vec{\gamma})^2$, which is non-zero even when $\gamma$ and $\beta$ are orthogonal. 

\subsection{Flux frustration}  \label{sec:flux}

\begin{figure}
    \centering
    \subfigure[]{
    \includegraphics[width=0.9\linewidth]{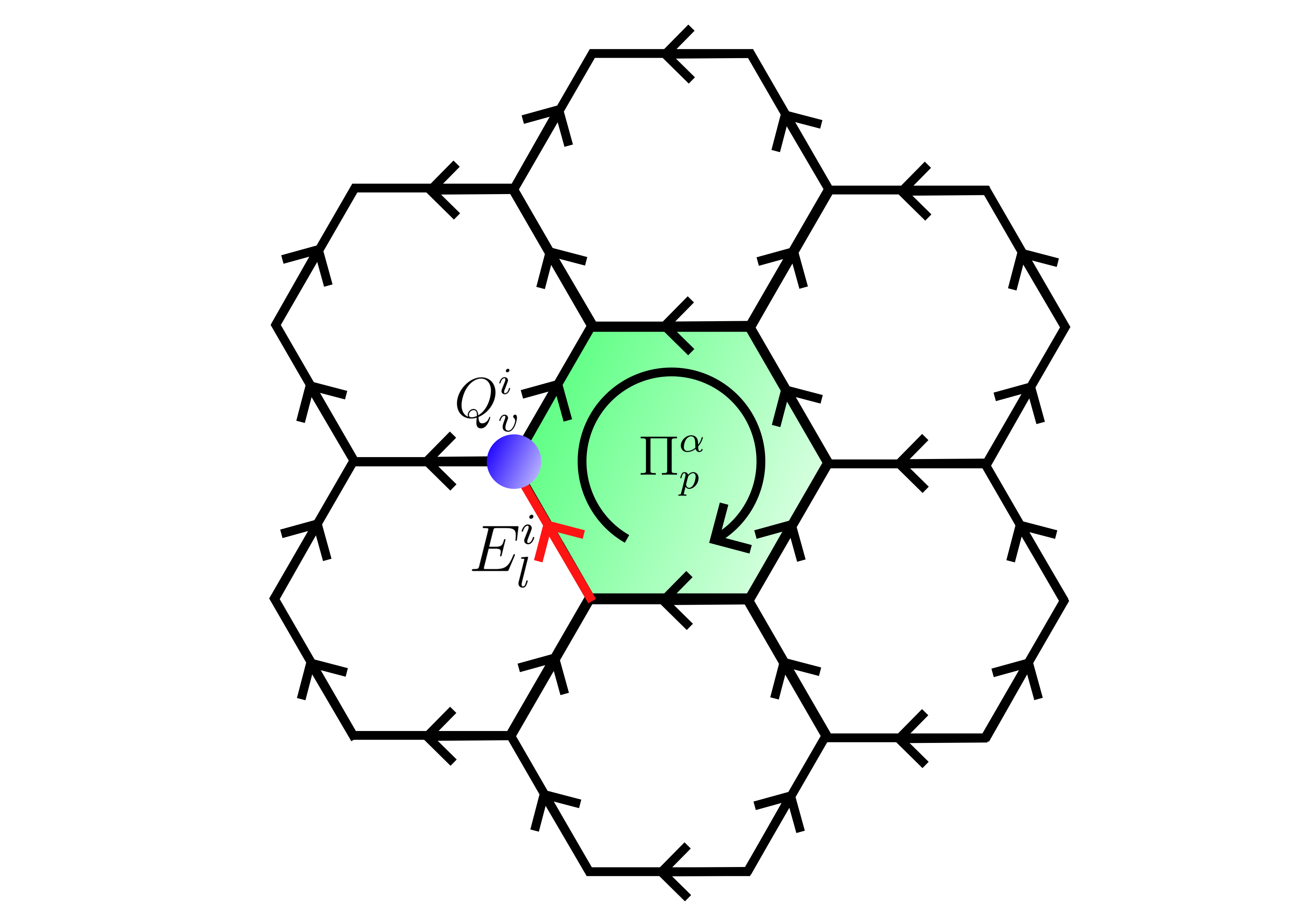}
    }
    \subfigure[]{
    \includegraphics[width=0.9\linewidth]{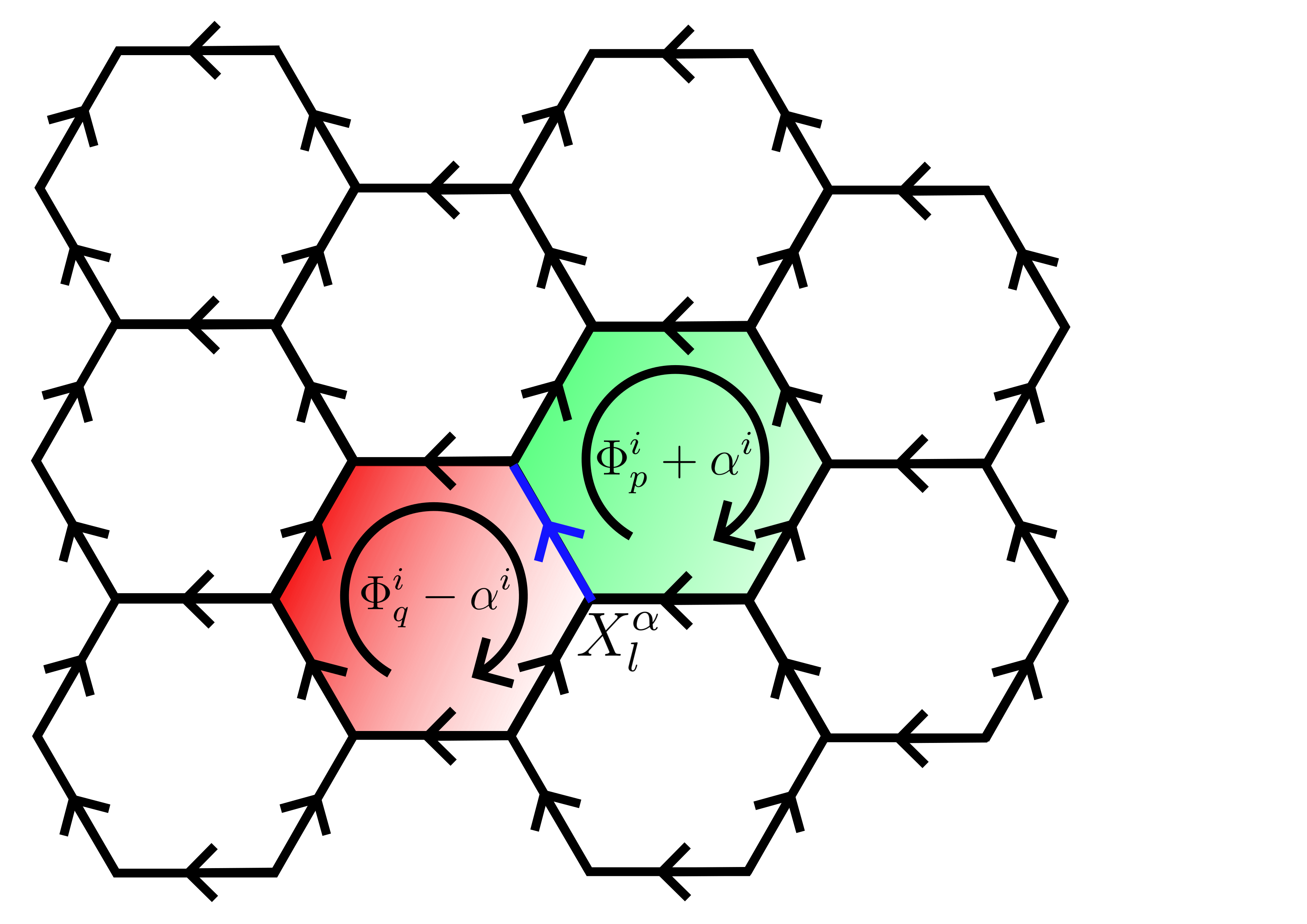}
    }
    \caption{((a)) Cell complex/ lattice in two dimensions with directed edges with charge operators $Q^i_v$ defined at vertices, field operators $E^i_l$ defined on links, and flux operators $\Pi^{\alpha}_p$ defined on plaquettes (with orientation indicated). ((b)) action of $X^{\alpha}_l=e^{i\alpha^iE^i_l}$ on the flux eigenstates of adjoining plaquettes. Flux is shifted by $\pm \alpha$ depending on the relative orientation of the plaquette and link.}
    \label{fig:Toric_code_figures}
\end{figure}

In this subsection, we return to the issue of flux frustration raised in section \ref{sec:GMFT}. In searching for energy minimizing fluxes $B^i_p$ in the effective Hamiltonian of Eq (\ref{eq:Gauge_eff_H}), it was highlighted that for $V<0$, there is generically not a unique solution for the plaquette flux that minimizes the magnetic part of the energy for $q>2$. We are then left without an obvious vacuum state to expand around for small fluctuations, unlike in the case of quantum spin ice ($q=2$) where the magnetic energy is uniquely minimized for $V<0$ by the insertion of $\pi$ units of flux through each plaquette.

Degeneracy of plaquette fluxes means that in the absence of an energetic restriction on the electric fields-- the first term in Eq. (\ref{eq:Gauge_eff_H})-- the ground state would be extensively degenerate. As the $E^i_l$ and $B^i_p$ operators do not commute, this extensive degeneracy is broken. We conjecture that the resulting ground state is a coherent superposition of the extensively degenerate minimizing flux states in the $V<0$ phase, as opposed to a symmetry broken product state or other flux ordered state.

We will support this conjecture by examining the flux frustration phenomenon from two more tractable directions. We will first apply the electric field term as a perturbation to an exactly solvable formulation of the model analogous to the Kitaev toric code \cite{10.1093/oso/9780198886723.001.0001,KITAEV20032}, and examine the states favored by the resulting effective Hamiltonians. We will then also solve for the ground state of the original system on a finite cluster, and show that this corroborates with the suggested ground state in the perturbative limit.

First, to concretely construct the exactly solvable models we examine in this section. We will generalize from $U(1)$ gauge groups to $\mathbb{Z}_N$, and thus desire quantum liquids with $\mathbb{Z}_{N}^{\otimes q-1}|_{\mathbb{Z}_2 \times S_{q}}$ emergent gauge fields, existing in some $d$ dimensions.

Degrees of freedom are electric fields $E^i_l$, living on the links of a bipartite lattice, as before, and canonically conjugate to a gauge field operator $A^i_l$ through $[A^i_l,E^j_m]=i\delta^{ij}_{lm}$. The electric field now takes eigenvalues from a $N-$periodic section of the root lattice. Addition of roots is thus to be understood as modulo $N$, i.e.:
\begin{equation}
    N \alpha \equiv 0 \ \forall \alpha \in \Delta,
\end{equation}
with the charge defect operators modified to be:
\begin{equation}
    Q^i_C= \left\{\sum_{l \in C_{\text{in}}} E^i_l - \sum_{l' \in C_{\text{out}}} E^i_{l'}\right\} \ \text{mod}(N).
\end{equation}
The form of the plaquette operators $\Pi^{\alpha}_p$ is unaltered:
\begin{equation}
    \Pi^{\alpha}_p = \exp\left(i\alpha^i\sum_{l\in \partial p}s_lA^i_l\right)=\exp\left(i\alpha^iB^i_p\right),
\end{equation}
however these now satisfy $\left(\Pi^{\alpha}_p\right)^N=\mathbb{1}$, and possess a finite set of eigenvalues. A schematic of the electric field, charge, and plaquette operators defined on a Hexagonal lattice is shown in Fig. \ref{fig:Toric_code_figures}. As usual for a toric code-like model, the exact solubility of the system follows from the commutativity of the $Q^i_C$ and $B^i_p$ operators, allowing one to write down a Hamiltonian promoting a color liquid ground state as:
\begin{equation}
    H = J_{Q} \sum_{v} \sum_i \left(Q^i_v\right)^2 -V \sum_p \sum_{\alpha} \cos\left(\sum_i\alpha^iB^i_p\right).
\end{equation}
For $J_{Q},V>0$, the ground states of this Hamiltonian are equal weighted superpositions of all charge neutral configurations of fields. With periodic boundary conditions, the ground state has a topological degeneracy $N^{d(q-1)}$ associated with the threading of flux through each periodic boundary.

Our interest lies in the case of $V<0$. Even for $q=2$, if $N$, the periodicity of the gauge group, is odd, there is no unique plaquette flux that minimizes the energy per plaquette, and the model possesses an extensive degeneracy. In $d>2$, topological constraints ensure that the total flux through any closed surface is equivalent to a reciprocal lattice vector of the root lattice, but these local constraints are still not sufficient to remove the extensive degeneracy.

To take a concrete example, consider a $q=3$ model on the Honeycomb (kagome) lattice with $N=2$. The emergent gauge theory is $\mathbb{Z}_2^{\otimes 2}|_{\mathbb{Z}_2\times S_3}$. The flux through each plaquette takes one of four values from the hexagonal Brillouin zone of the full root lattice, those being the origin $B=0$ or one of the three $M$ high symmetry points, $B_{M_i}$. If $V<0$, then the energy per plaquette is minimized by any of the fluxes $B_{M_i}$. Their energies must be equal if the Hamiltonian respects the root point group symmetry. Whilst there is a global constraint that the net flux through the surface of the torus must be a reciprocal lattice vector, there is still an extensive ground state degeneracy, and hence no topological order in the $V<0$ phase.

We can now apply a suitable perturbation to the Hamiltonian to break this degeneracy. Such an operator is of the form $e^{i \alpha^i E^i_l}$, which acts to shift the flux through neighboring plaquettes by an amount $\pm\alpha^i$ (the sign is dependent on the relative orientation of the link $l$ and the adjoining plaquettes, see Fig. \ref{fig:Toric_code_figures} (b)). This operator naturally generates flux excitations, and its negative real part can be understood as a compactification of the $\sum_i (E^i_l)^2$ operator once one sums over all roots, and so we might hope to gain some insight on the ground state of the original model. We also allow more general perturbations $e^{i\theta \alpha^iE^i_l}$, which rescale the flux shift parallel to root $\alpha$ by an amount $\theta$ to allow for all possible distinct couplings.

\begin{figure}
    \centering
    \includegraphics[width=0.8\linewidth]{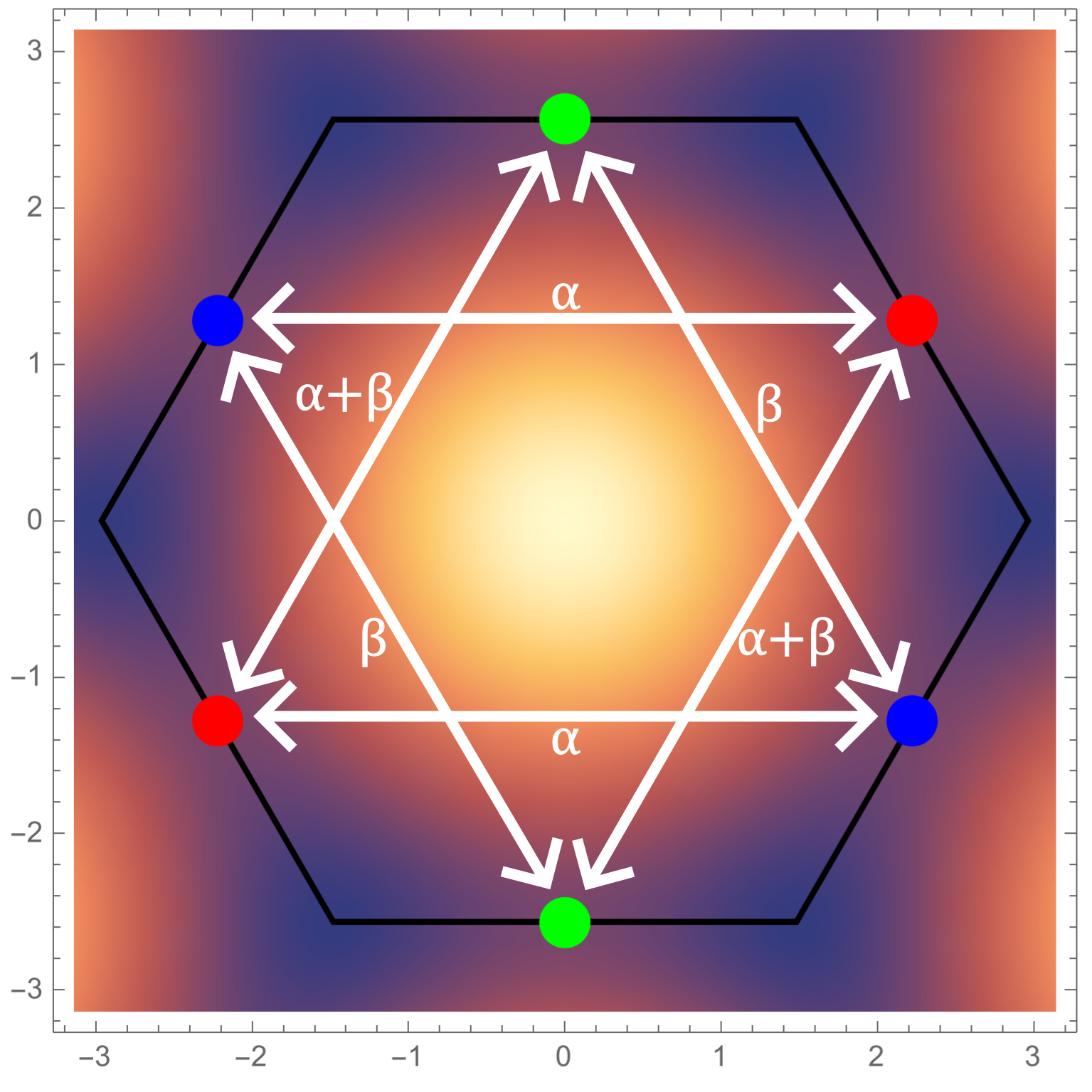}

    \caption{Action of $e^{i\alpha^iE^i_l}$ perturbation on degenerate flux states for $\mathbb{Z}_2$ $q=3$ theory. The $\mathbb{Z}_2$ model possess three minima, connected by each root flavored electric field operator as shown. The color plot upon which the diagram is superimposed shows the variation of the plaquette energy operator $\sum_{\alpha}\cos(\alpha^iB^i)$ across the full $\text{U}(1)$ theory Brillouin zone.}
    \label{fig:Nc3_flux_diagram_Z2}
\end{figure}

Returning to the $\mathbb{Z}_2^{\otimes 2}|_{\mathbb{Z}_2\times S_3}$ theory, and applying the perturbation:
\begin{equation}
    \delta H= - U \sum_{l}\sum_{\alpha} \cos\left(\sum_i\alpha^iE^i_l\right),
\end{equation}
the degeneracy originally found for $V<0$ is broken, with the different root contributions of $\delta H$ acting on the flux states as shown in Fig. \ref{fig:Nc3_flux_diagram_Z2}. 

We can treat the flux states $B_{M_i}$ within the degenerate subspace as 3-state Potts variables defined on each plaquette. The effective Hamiltonian can then be written in terms of operators $X^{\alpha}_p$ which perform the swaps indicated in Fig. \ref{fig:Nc3_flux_diagram_Z2} as:
\begin{equation}
    H_{\text{eff}} = P\left(-U\sum_{\langle p,q \rangle} \sum_{\alpha} X^{\alpha}_pX^{\alpha}_q \right)P, \label{eq:Z2_Nc3_eff_Hamiltonian}
\end{equation}
with $P$ a projector into the sector with $\sum_p B^i_p$ equal to a reciprocal lattice vector. 

The effective Hamiltonian \ref{eq:Z2_Nc3_eff_Hamiltonian} promotes the formation of $\text{SU}(3)$ singlet dimers across neighboring plaquettes, and suggests some correlated entangled flux ground state. Irrespective of the details of this state-- whether the ground state is some symmetry broken dimer covering or a long-ranged entangled liquid state-- the Hamiltonian generates entanglement between the flux through any given plaquette and its neighbors, such that the flux measured through any given plaquette $p$ is equally likely to be any of the states $B_{M_i}$.

This strongly fluctuating flux background has important implications for the dynamics of rootons. Moving a rooton around a plaquette no longer gives a definite Aharonov-Bohm phase, leading to changes in the internal phases of the wavefunction, and that the state is no longer an eigenstate of $\Pi^{\alpha}_p$:
\begin{align}
    \ket{\psi} = \frac{1}{\sqrt{3}}\{\ket{B_{M_1}}\ket{\Psi_1}+& \ket{B_{M_3}}\ket{\Psi_2}+\ket{B_{M_3}}\ket{\Psi_3}\} \\
    \Pi^{\alpha}_p\ket{\psi} = \frac{1}{\sqrt{3}}\{\ket{B_{M_1}}\ket{\Psi_1}-& \ket{B_{M_3}}\ket{\Psi_2}-\ket{B_{M_3}}\ket{\Psi_3}\} \\
    \bra{\psi} \Pi^{\alpha} \ket{\psi} =& -\frac{1}{3}.
\end{align}

With the assumption that the effective Hamiltonian is no longer extensively degenerate, such that local operators cannot lift any remaining ground state degeneracy, the motion of rootons around closed loops must then be associated with excitations of the flux background. This is reminiscent of the behavior of visons in quantum spin ice in the presence of a frustrated (or polarized) electric field vacuum \cite{PhysRevB.100.014417}, and suggests that the rootons may hybridize with flux fluctuations.

Further discussion of perturbations of exactly solvable models for different choices of $\mathbb{Z}_N$, number of colors $q$, and dimension $d$ can be found in supplementary material \cite{supp}, where a wide range of different effective Hamiltonians are shown to be generated by this mechanism of flux frustration, including spin-ice like physics in the flux sector of the $U(1)$ $q=3$ model on the cubic lattice.

\begin{figure}
    \centering
    \includegraphics[width=1.0\linewidth]{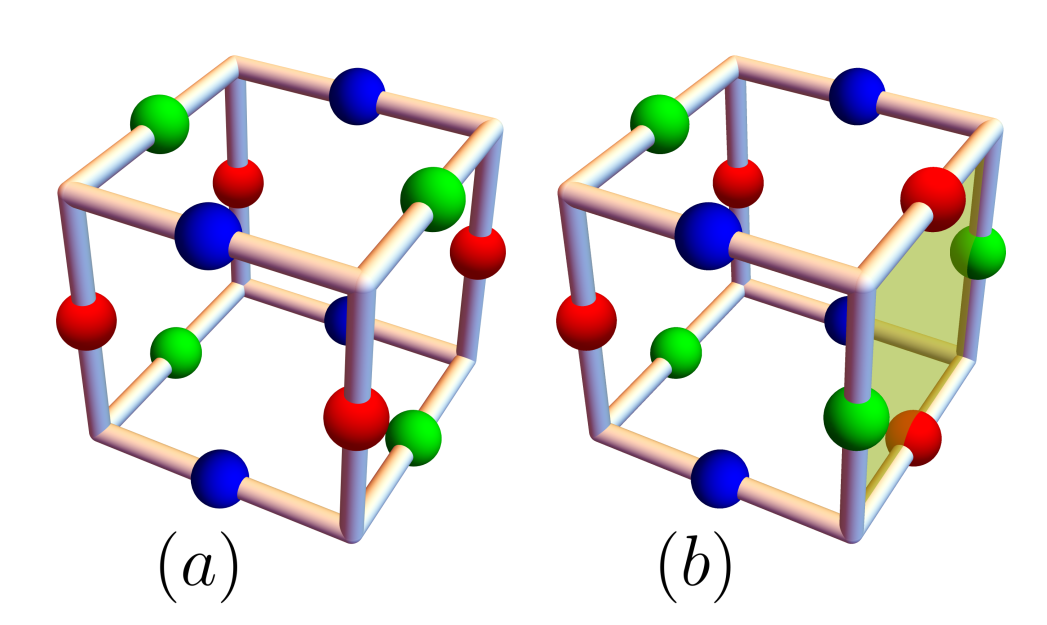}
    \caption{Primitive cluster of cubic lattice displaying color eigenstates contributing to the ground state of the perturbation Hamiltonian restricted to this finite cluster and to $q$ valued degrees of freedom. For $\pm V$, the ground state has equal weight $1/2\sqrt{3}$ over all maximally flippable configurations, such as those shown in (a), and is color symmetric. The ground state has also an equal weight $\pm 1/6$ over the `intermediate' configurations shown in $b$ that result from flipping a single plaquette of an (a) configuration.}
    \label{fig:Flipable_Cube}
\end{figure}

Further evidence in support of a correlated flux ground state for $V<0$ can be obtained by analyzing the effective perturbation Hamiltonian  for the pure gauge sector of the original model: $-2V\sum_{p,\alpha} \cos(\alpha^i B^i_p)$. This approach has been used to analyze similar flux-frustrated phases arising in QSI models in an external field \cite{sanders2024experimentallytunableqeddipolaroctupolar}. Degrees of freedom are restricted to the original $q$ colors, instead of the full root lattice. We examine the $q=3$ color model on the cubic lattice on a cluster of 8 sites, as shown in Fig. \ref{fig:Flipable_Cube}. The perturbation Hamiltonian acts to flip plaquettes whose edges host a worm of alternating colors. Accordingly, the ground state of the system is found in the Hilbert space fragment containing the maximally flippable color configurations, for both signs of $V$.

The resulting ground states for $\pm V$, respectively $\ket{\pm}$, have expectation values for the Plaquette operators $\Pi^{\alpha}_p$ of:
\begin{equation}
    \bra{\pm}\Pi^{\alpha}_p\ket{\pm} = \pm \frac{1}{6\sqrt{3}} \ \ \forall p \ \forall\alpha,
\end{equation}
indicating that neither state is a product state of flux eigenstates, though this would not be expected for so small a finite cluster of the lattice. What is notable however is that neither state breaks any symmetry of the lattice nor the $S_q$ color symmetry. If we seek a ground state of the full system that has such a property, then the only flux product state that achieves this is the $0-$flux state. If the symmetry of the ground state on this finite cluster is maintained to larger system sizes, and to the thermodynamic limit, then the ground state in the $V<0$ phase must be a correlated state of fluxes to avoid breaking the $S_q$ symmetry, and indeed must be a flux liquid state if lattice symmetries too are to be preserved, though the cluster we examine here may be too small to diagnose whether such symmetry breaking is preferred or not.

Taking both the behavior of exact models under small perturbations in flux frustrated phases, and the exact ground state of a small system into consideration, it seems suggestive that the true ground state for $V<0$ is a long-ranged entangled correlated flux phase, or flux liquid, coexisting with a color liquid, that breaks no lattice symmetry or symmetry of the degrees of freedom.

\section{Discussion}

The properties of the classical and quantum Potts ices are predicted to mirror those of more familiar spin ice in many regards, hosting Coulombic correlations in the classical model, and emergent photons in the quantum case. However, we find that the shadow of the non-abelian SU($q$) Lie group present in the Potts ices' emergent gauge fields imbues these models with many distinctive physical phenomena. Their novel interaction channels between excitations, strong coupling, and flux frustration result directly from the connection to the non-abelian group, and make the generalization from two-state `qubit' ice to $q$-state `qudit' ice non-trivial.

Further, that the Potts ices are such a natural continuation of spin ice means that they provide a relatively simple  lattice model of abelian projections of SU($q$) field theories-- a limit of interest to the high energy physics community in the study of quark confinement in the case of $q=3$ \cite{HOOFT1981455,PhysRevD.57.7467}.

Though spin$-1/2$ degrees of freedom are ubiquitous in the material world, higher $q$ qudits are most natural to ultracold atom and other synthetic quantum platforms, and it is in these settings we might hope first to realize quantum Potts ices. Previous work on the $q=4$ classical Potts ice presented a formulation of the model as an anisotropic Heisenberg spin model on the pyrochlore lattice \cite{PhysRevB.86.054411}. This realization of the Potts model relies on the point group symmetry of the lattice being the same as a root lattice, and so a similar construction can be used to realize the $q=3$ model in two dimensions with classical two-dimensional spins on the kagome lattice. Such models do not however lend themselves naturally to quantum extensions as they are motivated in the limit $S \rightarrow \infty$, though perhaps quantum fluctuations may be significant at intermediate values of $S$. In principle, $S=(q-1)/2$ models may be written down that realize the classical and quantum Potts ice Hamiltonians. However as spin degrees of freedom do not naturally possess the $S_{q}$ permutation symmetry between states of true Potts variables, all such models will require a degree of fine tuning if this symmetry is to be explicit in the Hamiltonian. More promising then would be to look to spin systems where the $S_{q}$ symmetry and classical ice Hamiltonian is are \textit{emergent} properties, as has been found for certain spin$-1/2$ models on tri-coordinated lattices \cite{PhysRevLett.120.117202,PhysRevB.99.104433,PhysRevB.103.144414}, and a spin$-1$ model on the pyrochlore lattice \cite{pohle2025abundancespinliquidss1}.

This work has presented a predominantly analytic treatment of the classical and quantum Potts models, alongside Monte Carlo numerics gathered to demonstrate the existence of Coulomb physics in the classical case. 
The strong coupling we predict for the quantum model on the one hand limits the utility of the analytical approach we have employed here, but leaves open the possibility of more novel strongly interacting physical effects that may be discovered in these models through a numerical or experimental study. 

Such future studies should also seek to clarify the nature of the ground state, and ascertain the existence and location of the likely first order phase transition from the liquid to the ordered phase for a given quantum Potts ice. In particular, the fate of the system for $V<0$, where we conjecture flux frustration promotes a liquid-like flux state instead of a simple flux product state, should be verified. The phenomenon may provide a new mechanism generating many nearly degenerate states above the ground state in a quantum model, providing a platform for interesting low-energy physics such as exotic solid or liquid states.

\section*{Acknowledgements}
We thank Josef Willsher and Han Yan for helpful discussions. This work was supported in part by
the Deutsche Forschungsgemeinschaft under Grant No. SFB
1143 (Project-ID No. 247310070) and by
the Deutsche Forschungsgemeinschaft  under cluster of excellence
ctd.qmat (EXC 2147, Project-ID No. 390858490). SAP acknowledges support from the  European Research Council under the European Union Horizon 2020 Research and Innovation Programme, Grant Agreement No. 804213-TMCS, and a Gutzwiller Fellowship at the Max Planck Institute for Complex Systems where this work was initiated. 

\bibliography{PottsBib.bib}

\begin{thebibliography}{56}%
\makeatletter
\providecommand \@ifxundefined [1]{%
 \@ifx{#1\undefined}
}%
\providecommand \@ifnum [1]{%
 \ifnum #1\expandafter \@firstoftwo
 \else \expandafter \@secondoftwo
 \fi
}%
\providecommand \@ifx [1]{%
 \ifx #1\expandafter \@firstoftwo
 \else \expandafter \@secondoftwo
 \fi
}%
\providecommand \natexlab [1]{#1}%
\providecommand \enquote  [1]{``#1''}%
\providecommand \bibnamefont  [1]{#1}%
\providecommand \bibfnamefont [1]{#1}%
\providecommand \citenamefont [1]{#1}%
\providecommand \href@noop [0]{\@secondoftwo}%
\providecommand \href [0]{\begingroup \@sanitize@url \@href}%
\providecommand \@href[1]{\@@startlink{#1}\@@href}%
\providecommand \@@href[1]{\endgroup#1\@@endlink}%
\providecommand \@sanitize@url [0]{\catcode `\\12\catcode `\$12\catcode `\&12\catcode `\#12\catcode `\^12\catcode `\_12\catcode `\%12\relax}%
\providecommand \@@startlink[1]{}%
\providecommand \@@endlink[0]{}%
\providecommand \url  [0]{\begingroup\@sanitize@url \@url }%
\providecommand \@url [1]{\endgroup\@href {#1}{\urlprefix }}%
\providecommand \urlprefix  [0]{URL }%
\providecommand \Eprint [0]{\href }%
\providecommand \doibase [0]{http://dx.doi.org/}%
\providecommand \selectlanguage [0]{\@gobble}%
\providecommand \bibinfo  [0]{\@secondoftwo}%
\providecommand \bibfield  [0]{\@secondoftwo}%
\providecommand \translation [1]{[#1]}%
\providecommand \BibitemOpen [0]{}%
\providecommand \bibitemStop [0]{}%
\providecommand \bibitemNoStop [0]{.\EOS\space}%
\providecommand \EOS [0]{\spacefactor3000\relax}%
\providecommand \BibitemShut  [1]{\csname bibitem#1\endcsname}%
\let\auto@bib@innerbib\@empty
\bibitem [{\citenamefont {Tsui}\ \emph {et~al.}(1982)\citenamefont {Tsui}, \citenamefont {Stormer},\ and\ \citenamefont {Gossard}}]{PhysRevLett.48.1559}%
  \BibitemOpen
  \bibfield  {author} {\bibinfo {author} {\bibfnamefont {D.~C.}\ \bibnamefont {Tsui}}, \bibinfo {author} {\bibfnamefont {H.~L.}\ \bibnamefont {Stormer}}, \ and\ \bibinfo {author} {\bibfnamefont {A.~C.}\ \bibnamefont {Gossard}},\ }\href {\doibase 10.1103/PhysRevLett.48.1559} {\bibfield  {journal} {\bibinfo  {journal} {Phys. Rev. Lett.}\ }\textbf {\bibinfo {volume} {48}},\ \bibinfo {pages} {1559} (\bibinfo {year} {1982})}\BibitemShut {NoStop}%
\bibitem [{\citenamefont {Zhang}\ \emph {et~al.}(1989)\citenamefont {Zhang}, \citenamefont {Hansson},\ and\ \citenamefont {Kivelson}}]{PhysRevLett.62.82}%
  \BibitemOpen
  \bibfield  {author} {\bibinfo {author} {\bibfnamefont {S.~C.}\ \bibnamefont {Zhang}}, \bibinfo {author} {\bibfnamefont {T.~H.}\ \bibnamefont {Hansson}}, \ and\ \bibinfo {author} {\bibfnamefont {S.}~\bibnamefont {Kivelson}},\ }\href {\doibase 10.1103/PhysRevLett.62.82} {\bibfield  {journal} {\bibinfo  {journal} {Phys. Rev. Lett.}\ }\textbf {\bibinfo {volume} {62}},\ \bibinfo {pages} {82} (\bibinfo {year} {1989})}\BibitemShut {NoStop}%
\bibitem [{\citenamefont {Lopez}\ and\ \citenamefont {Fradkin}(1991)}]{PhysRevB.44.5246}%
  \BibitemOpen
  \bibfield  {author} {\bibinfo {author} {\bibfnamefont {A.}~\bibnamefont {Lopez}}\ and\ \bibinfo {author} {\bibfnamefont {E.}~\bibnamefont {Fradkin}},\ }\href {\doibase 10.1103/PhysRevB.44.5246} {\bibfield  {journal} {\bibinfo  {journal} {Phys. Rev. B}\ }\textbf {\bibinfo {volume} {44}},\ \bibinfo {pages} {5246} (\bibinfo {year} {1991})}\BibitemShut {NoStop}%
\bibitem [{\citenamefont {ZHANG}(1992)}]{doi:10.1142/S0217979292000037}%
  \BibitemOpen
  \bibfield  {author} {\bibinfo {author} {\bibfnamefont {S.~C.}\ \bibnamefont {ZHANG}},\ }\href {\doibase 10.1142/S0217979292000037} {\bibfield  {journal} {\bibinfo  {journal} {International Journal of Modern Physics B}\ }\textbf {\bibinfo {volume} {06}},\ \bibinfo {pages} {25} (\bibinfo {year} {1992})},\ \Eprint {http://arxiv.org/abs/https://doi.org/10.1142/S0217979292000037} {https://doi.org/10.1142/S0217979292000037} \BibitemShut {NoStop}%
\bibitem [{\citenamefont {Bramwell}\ and\ \citenamefont {Gingras}(2001)}]{doi:10.1126/science.1064761}%
  \BibitemOpen
  \bibfield  {author} {\bibinfo {author} {\bibfnamefont {S.~T.}\ \bibnamefont {Bramwell}}\ and\ \bibinfo {author} {\bibfnamefont {M.~J.~P.}\ \bibnamefont {Gingras}},\ }\href {\doibase 10.1126/science.1064761} {\bibfield  {journal} {\bibinfo  {journal} {Science}\ }\textbf {\bibinfo {volume} {294}},\ \bibinfo {pages} {1495} (\bibinfo {year} {2001})},\ \Eprint {http://arxiv.org/abs/https://www.science.org/doi/pdf/10.1126/science.1064761} {https://www.science.org/doi/pdf/10.1126/science.1064761} \BibitemShut {NoStop}%
\bibitem [{\citenamefont {Harris}\ \emph {et~al.}(1997)\citenamefont {Harris}, \citenamefont {Bramwell}, \citenamefont {McMorrow}, \citenamefont {Zeiske},\ and\ \citenamefont {Godfrey}}]{PhysRevLett.79.2554}%
  \BibitemOpen
  \bibfield  {author} {\bibinfo {author} {\bibfnamefont {M.~J.}\ \bibnamefont {Harris}}, \bibinfo {author} {\bibfnamefont {S.~T.}\ \bibnamefont {Bramwell}}, \bibinfo {author} {\bibfnamefont {D.~F.}\ \bibnamefont {McMorrow}}, \bibinfo {author} {\bibfnamefont {T.}~\bibnamefont {Zeiske}}, \ and\ \bibinfo {author} {\bibfnamefont {K.~W.}\ \bibnamefont {Godfrey}},\ }\href {\doibase 10.1103/PhysRevLett.79.2554} {\bibfield  {journal} {\bibinfo  {journal} {Phys. Rev. Lett.}\ }\textbf {\bibinfo {volume} {79}},\ \bibinfo {pages} {2554} (\bibinfo {year} {1997})}\BibitemShut {NoStop}%
\bibitem [{\citenamefont {Morris}\ \emph {et~al.}(2009)\citenamefont {Morris}, \citenamefont {Tennant}, \citenamefont {Grigera}, \citenamefont {Klemke}, \citenamefont {Castelnovo}, \citenamefont {Moessner}, \citenamefont {Czternasty}, \citenamefont {Meissner}, \citenamefont {Rule}, \citenamefont {Hoffmann}, \citenamefont {Kiefer}, \citenamefont {Gerischer}, \citenamefont {Slobinsky},\ and\ \citenamefont {Perry}}]{doi:10.1126/science.1178868}%
  \BibitemOpen
  \bibfield  {author} {\bibinfo {author} {\bibfnamefont {D.~J.~P.}\ \bibnamefont {Morris}}, \bibinfo {author} {\bibfnamefont {D.~A.}\ \bibnamefont {Tennant}}, \bibinfo {author} {\bibfnamefont {S.~A.}\ \bibnamefont {Grigera}}, \bibinfo {author} {\bibfnamefont {B.}~\bibnamefont {Klemke}}, \bibinfo {author} {\bibfnamefont {C.}~\bibnamefont {Castelnovo}}, \bibinfo {author} {\bibfnamefont {R.}~\bibnamefont {Moessner}}, \bibinfo {author} {\bibfnamefont {C.}~\bibnamefont {Czternasty}}, \bibinfo {author} {\bibfnamefont {M.}~\bibnamefont {Meissner}}, \bibinfo {author} {\bibfnamefont {K.~C.}\ \bibnamefont {Rule}}, \bibinfo {author} {\bibfnamefont {J.-U.}\ \bibnamefont {Hoffmann}}, \bibinfo {author} {\bibfnamefont {K.}~\bibnamefont {Kiefer}}, \bibinfo {author} {\bibfnamefont {S.}~\bibnamefont {Gerischer}}, \bibinfo {author} {\bibfnamefont {D.}~\bibnamefont {Slobinsky}}, \ and\ \bibinfo {author} {\bibfnamefont {R.~S.}\ \bibnamefont {Perry}},\ }\href {\doibase 10.1126/science.1178868} {\bibfield  {journal} {\bibinfo
  {journal} {Science}\ }\textbf {\bibinfo {volume} {326}},\ \bibinfo {pages} {411} (\bibinfo {year} {2009})},\ \Eprint {http://arxiv.org/abs/https://www.science.org/doi/pdf/10.1126/science.1178868} {https://www.science.org/doi/pdf/10.1126/science.1178868} \BibitemShut {NoStop}%
\bibitem [{\citenamefont {Smith}\ \emph {et~al.}(2022)\citenamefont {Smith}, \citenamefont {Benton}, \citenamefont {Yahne}, \citenamefont {Placke}, \citenamefont {Sch\"afer}, \citenamefont {Gaudet}, \citenamefont {Dudemaine}, \citenamefont {Fitterman}, \citenamefont {Beare}, \citenamefont {Wildes}, \citenamefont {Bhattacharya}, \citenamefont {DeLazzer}, \citenamefont {Buhariwalla}, \citenamefont {Butch}, \citenamefont {Movshovich}, \citenamefont {Garrett}, \citenamefont {Marjerrison}, \citenamefont {Clancy}, \citenamefont {Kermarrec}, \citenamefont {Luke}, \citenamefont {Bianchi}, \citenamefont {Ross},\ and\ \citenamefont {Gaulin}}]{PhysRevX.12.021015}%
  \BibitemOpen
  \bibfield  {author} {\bibinfo {author} {\bibfnamefont {E.~M.}\ \bibnamefont {Smith}}, \bibinfo {author} {\bibfnamefont {O.}~\bibnamefont {Benton}}, \bibinfo {author} {\bibfnamefont {D.~R.}\ \bibnamefont {Yahne}}, \bibinfo {author} {\bibfnamefont {B.}~\bibnamefont {Placke}}, \bibinfo {author} {\bibfnamefont {R.}~\bibnamefont {Sch\"afer}}, \bibinfo {author} {\bibfnamefont {J.}~\bibnamefont {Gaudet}}, \bibinfo {author} {\bibfnamefont {J.}~\bibnamefont {Dudemaine}}, \bibinfo {author} {\bibfnamefont {A.}~\bibnamefont {Fitterman}}, \bibinfo {author} {\bibfnamefont {J.}~\bibnamefont {Beare}}, \bibinfo {author} {\bibfnamefont {A.~R.}\ \bibnamefont {Wildes}}, \bibinfo {author} {\bibfnamefont {S.}~\bibnamefont {Bhattacharya}}, \bibinfo {author} {\bibfnamefont {T.}~\bibnamefont {DeLazzer}}, \bibinfo {author} {\bibfnamefont {C.~R.~C.}\ \bibnamefont {Buhariwalla}}, \bibinfo {author} {\bibfnamefont {N.~P.}\ \bibnamefont {Butch}}, \bibinfo {author} {\bibfnamefont {R.}~\bibnamefont {Movshovich}}, \bibinfo {author}
  {\bibfnamefont {J.~D.}\ \bibnamefont {Garrett}}, \bibinfo {author} {\bibfnamefont {C.~A.}\ \bibnamefont {Marjerrison}}, \bibinfo {author} {\bibfnamefont {J.~P.}\ \bibnamefont {Clancy}}, \bibinfo {author} {\bibfnamefont {E.}~\bibnamefont {Kermarrec}}, \bibinfo {author} {\bibfnamefont {G.~M.}\ \bibnamefont {Luke}}, \bibinfo {author} {\bibfnamefont {A.~D.}\ \bibnamefont {Bianchi}}, \bibinfo {author} {\bibfnamefont {K.~A.}\ \bibnamefont {Ross}}, \ and\ \bibinfo {author} {\bibfnamefont {B.~D.}\ \bibnamefont {Gaulin}},\ }\href {\doibase 10.1103/PhysRevX.12.021015} {\bibfield  {journal} {\bibinfo  {journal} {Phys. Rev. X}\ }\textbf {\bibinfo {volume} {12}},\ \bibinfo {pages} {021015} (\bibinfo {year} {2022})}\BibitemShut {NoStop}%
\bibitem [{\citenamefont {Por\'ee}\ \emph {et~al.}(2025)\citenamefont {Por\'ee}, \citenamefont {Bhardwaj}, \citenamefont {Lhotel}, \citenamefont {Petit}, \citenamefont {Gauthier}, \citenamefont {Yan}, \citenamefont {Pomjakushin}, \citenamefont {Ollivier}, \citenamefont {Quilliam}, \citenamefont {Nevidomskyy}, \citenamefont {Changlani},\ and\ \citenamefont {Sibille}}]{j451-ztvr}%
  \BibitemOpen
  \bibfield  {author} {\bibinfo {author} {\bibfnamefont {V.}~\bibnamefont {Por\'ee}}, \bibinfo {author} {\bibfnamefont {A.}~\bibnamefont {Bhardwaj}}, \bibinfo {author} {\bibfnamefont {E.}~\bibnamefont {Lhotel}}, \bibinfo {author} {\bibfnamefont {S.}~\bibnamefont {Petit}}, \bibinfo {author} {\bibfnamefont {N.}~\bibnamefont {Gauthier}}, \bibinfo {author} {\bibfnamefont {H.}~\bibnamefont {Yan}}, \bibinfo {author} {\bibfnamefont {V.}~\bibnamefont {Pomjakushin}}, \bibinfo {author} {\bibfnamefont {J.}~\bibnamefont {Ollivier}}, \bibinfo {author} {\bibfnamefont {J.~A.}\ \bibnamefont {Quilliam}}, \bibinfo {author} {\bibfnamefont {A.~H.}\ \bibnamefont {Nevidomskyy}}, \bibinfo {author} {\bibfnamefont {H.~J.}\ \bibnamefont {Changlani}}, \ and\ \bibinfo {author} {\bibfnamefont {R.}~\bibnamefont {Sibille}},\ }\href {\doibase 10.1103/j451-ztvr} {\bibfield  {journal} {\bibinfo  {journal} {Phys. Rev. B}\ }\textbf {\bibinfo {volume} {112}},\ \bibinfo {pages} {L180404} (\bibinfo {year} {2025})}\BibitemShut {NoStop}%
\bibitem [{\citenamefont {Sibille}\ \emph {et~al.}(2020)\citenamefont {Sibille}, \citenamefont {Gauthier}, \citenamefont {Lhotel}, \citenamefont {Por{\'e}e}, \citenamefont {Pomjakushin}, \citenamefont {Ewings}, \citenamefont {Perring}, \citenamefont {Ollivier}, \citenamefont {Wildes}, \citenamefont {Ritter}, \citenamefont {Hansen}, \citenamefont {Keen}, \citenamefont {Nilsen}, \citenamefont {Keller}, \citenamefont {Petit},\ and\ \citenamefont {Fennell}}]{Sibille2020}%
  \BibitemOpen
  \bibfield  {author} {\bibinfo {author} {\bibfnamefont {R.}~\bibnamefont {Sibille}}, \bibinfo {author} {\bibfnamefont {N.}~\bibnamefont {Gauthier}}, \bibinfo {author} {\bibfnamefont {E.}~\bibnamefont {Lhotel}}, \bibinfo {author} {\bibfnamefont {V.}~\bibnamefont {Por{\'e}e}}, \bibinfo {author} {\bibfnamefont {V.}~\bibnamefont {Pomjakushin}}, \bibinfo {author} {\bibfnamefont {R.~A.}\ \bibnamefont {Ewings}}, \bibinfo {author} {\bibfnamefont {T.~G.}\ \bibnamefont {Perring}}, \bibinfo {author} {\bibfnamefont {J.}~\bibnamefont {Ollivier}}, \bibinfo {author} {\bibfnamefont {A.}~\bibnamefont {Wildes}}, \bibinfo {author} {\bibfnamefont {C.}~\bibnamefont {Ritter}}, \bibinfo {author} {\bibfnamefont {T.~C.}\ \bibnamefont {Hansen}}, \bibinfo {author} {\bibfnamefont {D.~A.}\ \bibnamefont {Keen}}, \bibinfo {author} {\bibfnamefont {G.~J.}\ \bibnamefont {Nilsen}}, \bibinfo {author} {\bibfnamefont {L.}~\bibnamefont {Keller}}, \bibinfo {author} {\bibfnamefont {S.}~\bibnamefont {Petit}}, \ and\ \bibinfo {author} {\bibfnamefont
  {T.}~\bibnamefont {Fennell}},\ }\href {\doibase 10.1038/s41567-020-0827-7} {\bibfield  {journal} {\bibinfo  {journal} {Nature Physics}\ }\textbf {\bibinfo {volume} {16}},\ \bibinfo {pages} {546} (\bibinfo {year} {2020})}\BibitemShut {NoStop}%
\bibitem [{\citenamefont {Balents}(2010)}]{Balents2010}%
  \BibitemOpen
  \bibfield  {author} {\bibinfo {author} {\bibfnamefont {L.}~\bibnamefont {Balents}},\ }\href {\doibase 10.1038/nature08917} {\bibfield  {journal} {\bibinfo  {journal} {Nature}\ }\textbf {\bibinfo {volume} {464}},\ \bibinfo {pages} {199} (\bibinfo {year} {2010})}\BibitemShut {NoStop}%
\bibitem [{\citenamefont {Savary}\ and\ \citenamefont {Balents}(2016)}]{Savary_2017}%
  \BibitemOpen
  \bibfield  {author} {\bibinfo {author} {\bibfnamefont {L.}~\bibnamefont {Savary}}\ and\ \bibinfo {author} {\bibfnamefont {L.}~\bibnamefont {Balents}},\ }\href {\doibase 10.1088/0034-4885/80/1/016502} {\bibfield  {journal} {\bibinfo  {journal} {Reports on Progress in Physics}\ }\textbf {\bibinfo {volume} {80}},\ \bibinfo {pages} {016502} (\bibinfo {year} {2016})}\BibitemShut {NoStop}%
\bibitem [{\citenamefont {Knolle}\ and\ \citenamefont {Moessner}(2019)}]{Knolle19}%
  \BibitemOpen
  \bibfield  {author} {\bibinfo {author} {\bibfnamefont {J.}~\bibnamefont {Knolle}}\ and\ \bibinfo {author} {\bibfnamefont {R.}~\bibnamefont {Moessner}},\ }\href {\doibase 10.1146/annurev-conmatphys-031218-013401} {\bibfield  {journal} {\bibinfo  {journal} {Annual Review of Condensed Matter Physics}\ }\textbf {\bibinfo {volume} {10}},\ \bibinfo {pages} {451} (\bibinfo {year} {2019})},\ \Eprint {http://arxiv.org/abs/https://doi.org/10.1146/annurev-conmatphys-031218-013401} {https://doi.org/10.1146/annurev-conmatphys-031218-013401} \BibitemShut {NoStop}%
\bibitem [{\citenamefont {Zhou}\ \emph {et~al.}(2017)\citenamefont {Zhou}, \citenamefont {Kanoda},\ and\ \citenamefont {Ng}}]{RevModPhys.89.025003}%
  \BibitemOpen
  \bibfield  {author} {\bibinfo {author} {\bibfnamefont {Y.}~\bibnamefont {Zhou}}, \bibinfo {author} {\bibfnamefont {K.}~\bibnamefont {Kanoda}}, \ and\ \bibinfo {author} {\bibfnamefont {T.-K.}\ \bibnamefont {Ng}},\ }\href {\doibase 10.1103/RevModPhys.89.025003} {\bibfield  {journal} {\bibinfo  {journal} {Rev. Mod. Phys.}\ }\textbf {\bibinfo {volume} {89}},\ \bibinfo {pages} {025003} (\bibinfo {year} {2017})}\BibitemShut {NoStop}%
\bibitem [{\citenamefont {Savary}\ and\ \citenamefont {Balents}(2012)}]{PhysRevLett.108.037202}%
  \BibitemOpen
  \bibfield  {author} {\bibinfo {author} {\bibfnamefont {L.}~\bibnamefont {Savary}}\ and\ \bibinfo {author} {\bibfnamefont {L.}~\bibnamefont {Balents}},\ }\href {\doibase 10.1103/PhysRevLett.108.037202} {\bibfield  {journal} {\bibinfo  {journal} {Phys. Rev. Lett.}\ }\textbf {\bibinfo {volume} {108}},\ \bibinfo {pages} {037202} (\bibinfo {year} {2012})}\BibitemShut {NoStop}%
\bibitem [{\citenamefont {Hermele}\ \emph {et~al.}(2004)\citenamefont {Hermele}, \citenamefont {Fisher},\ and\ \citenamefont {Balents}}]{PhysRevB.69.064404}%
  \BibitemOpen
  \bibfield  {author} {\bibinfo {author} {\bibfnamefont {M.}~\bibnamefont {Hermele}}, \bibinfo {author} {\bibfnamefont {M.~P.~A.}\ \bibnamefont {Fisher}}, \ and\ \bibinfo {author} {\bibfnamefont {L.}~\bibnamefont {Balents}},\ }\href {\doibase 10.1103/PhysRevB.69.064404} {\bibfield  {journal} {\bibinfo  {journal} {Phys. Rev. B}\ }\textbf {\bibinfo {volume} {69}},\ \bibinfo {pages} {064404} (\bibinfo {year} {2004})}\BibitemShut {NoStop}%
\bibitem [{\citenamefont {Gingras}\ and\ \citenamefont {McClarty}(2014)}]{Gingras_2014}%
  \BibitemOpen
  \bibfield  {author} {\bibinfo {author} {\bibfnamefont {M.~J.~P.}\ \bibnamefont {Gingras}}\ and\ \bibinfo {author} {\bibfnamefont {P.~A.}\ \bibnamefont {McClarty}},\ }\href {\doibase 10.1088/0034-4885/77/5/056501} {\bibfield  {journal} {\bibinfo  {journal} {Reports on Progress in Physics}\ }\textbf {\bibinfo {volume} {77}},\ \bibinfo {pages} {056501} (\bibinfo {year} {2014})}\BibitemShut {NoStop}%
\bibitem [{\citenamefont {Castelnovo}\ \emph {et~al.}(2008)\citenamefont {Castelnovo}, \citenamefont {Moessner},\ and\ \citenamefont {Sondhi}}]{Castelnovo2008}%
  \BibitemOpen
  \bibfield  {author} {\bibinfo {author} {\bibfnamefont {C.}~\bibnamefont {Castelnovo}}, \bibinfo {author} {\bibfnamefont {R.}~\bibnamefont {Moessner}}, \ and\ \bibinfo {author} {\bibfnamefont {S.~L.}\ \bibnamefont {Sondhi}},\ }\href {\doibase 10.1038/nature06433} {\bibfield  {journal} {\bibinfo  {journal} {Nature}\ }\textbf {\bibinfo {volume} {451}},\ \bibinfo {pages} {42} (\bibinfo {year} {2008})}\BibitemShut {NoStop}%
\bibitem [{\citenamefont {Moessner}\ and\ \citenamefont {Chalker}(1998)}]{PhysRevLett.80.2929}%
  \BibitemOpen
  \bibfield  {author} {\bibinfo {author} {\bibfnamefont {R.}~\bibnamefont {Moessner}}\ and\ \bibinfo {author} {\bibfnamefont {J.~T.}\ \bibnamefont {Chalker}},\ }\href {\doibase 10.1103/PhysRevLett.80.2929} {\bibfield  {journal} {\bibinfo  {journal} {Phys. Rev. Lett.}\ }\textbf {\bibinfo {volume} {80}},\ \bibinfo {pages} {2929} (\bibinfo {year} {1998})}\BibitemShut {NoStop}%
\bibitem [{\citenamefont {Kondev}\ and\ \citenamefont {Henley}(1995)}]{PhysRevB.52.6628}%
  \BibitemOpen
  \bibfield  {author} {\bibinfo {author} {\bibfnamefont {J.}~\bibnamefont {Kondev}}\ and\ \bibinfo {author} {\bibfnamefont {C.~L.}\ \bibnamefont {Henley}},\ }\href {\doibase 10.1103/PhysRevB.52.6628} {\bibfield  {journal} {\bibinfo  {journal} {Phys. Rev. B}\ }\textbf {\bibinfo {volume} {52}},\ \bibinfo {pages} {6628} (\bibinfo {year} {1995})}\BibitemShut {NoStop}%
\bibitem [{\citenamefont {Khemani}\ \emph {et~al.}(2012)\citenamefont {Khemani}, \citenamefont {Moessner}, \citenamefont {Parameswaran},\ and\ \citenamefont {Sondhi}}]{PhysRevB.86.054411}%
  \BibitemOpen
  \bibfield  {author} {\bibinfo {author} {\bibfnamefont {V.}~\bibnamefont {Khemani}}, \bibinfo {author} {\bibfnamefont {R.}~\bibnamefont {Moessner}}, \bibinfo {author} {\bibfnamefont {S.~A.}\ \bibnamefont {Parameswaran}}, \ and\ \bibinfo {author} {\bibfnamefont {S.~L.}\ \bibnamefont {Sondhi}},\ }\href {\doibase 10.1103/PhysRevB.86.054411} {\bibfield  {journal} {\bibinfo  {journal} {Phys. Rev. B}\ }\textbf {\bibinfo {volume} {86}},\ \bibinfo {pages} {054411} (\bibinfo {year} {2012})}\BibitemShut {NoStop}%
\bibitem [{\citenamefont {Chern}\ and\ \citenamefont {Wu}(2014)}]{PhysRevLett.112.020601}%
  \BibitemOpen
  \bibfield  {author} {\bibinfo {author} {\bibfnamefont {G.-W.}\ \bibnamefont {Chern}}\ and\ \bibinfo {author} {\bibfnamefont {C.}~\bibnamefont {Wu}},\ }\href {\doibase 10.1103/PhysRevLett.112.020601} {\bibfield  {journal} {\bibinfo  {journal} {Phys. Rev. Lett.}\ }\textbf {\bibinfo {volume} {112}},\ \bibinfo {pages} {020601} (\bibinfo {year} {2014})}\BibitemShut {NoStop}%
\bibitem [{\citenamefont {Wan}\ and\ \citenamefont {Gingras}(2016)}]{PhysRevB.94.174417}%
  \BibitemOpen
  \bibfield  {author} {\bibinfo {author} {\bibfnamefont {Y.}~\bibnamefont {Wan}}\ and\ \bibinfo {author} {\bibfnamefont {M.~J.~P.}\ \bibnamefont {Gingras}},\ }\href {\doibase 10.1103/PhysRevB.94.174417} {\bibfield  {journal} {\bibinfo  {journal} {Phys. Rev. B}\ }\textbf {\bibinfo {volume} {94}},\ \bibinfo {pages} {174417} (\bibinfo {year} {2016})}\BibitemShut {NoStop}%
\bibitem [{\citenamefont {Lozano-G{\'o}mez}\ \emph {et~al.}(2024)\citenamefont {Lozano-G{\'o}mez}, \citenamefont {Iqbal},\ and\ \citenamefont {Vojta}}]{Lozano-Gómez2024}%
  \BibitemOpen
  \bibfield  {author} {\bibinfo {author} {\bibfnamefont {D.}~\bibnamefont {Lozano-G{\'o}mez}}, \bibinfo {author} {\bibfnamefont {Y.}~\bibnamefont {Iqbal}}, \ and\ \bibinfo {author} {\bibfnamefont {M.}~\bibnamefont {Vojta}},\ }\href {\doibase 10.1038/s41467-024-54558-7} {\bibfield  {journal} {\bibinfo  {journal} {Nature Communications}\ }\textbf {\bibinfo {volume} {15}},\ \bibinfo {pages} {10162} (\bibinfo {year} {2024})}\BibitemShut {NoStop}%
\bibitem [{\citenamefont {Hu}(2013)}]{https://doi.org/10.1155/2013/836168}%
  \BibitemOpen
  \bibfield  {author} {\bibinfo {author} {\bibfnamefont {C.-R.}\ \bibnamefont {Hu}},\ }\href {\doibase https://doi.org/10.1155/2013/836168} {\bibfield  {journal} {\bibinfo  {journal} {Journal of Materials}\ }\textbf {\bibinfo {volume} {2013}},\ \bibinfo {pages} {836168} (\bibinfo {year} {2013})},\ \Eprint {http://arxiv.org/abs/https://onlinelibrary.wiley.com/doi/pdf/10.1155/2013/836168} {https://onlinelibrary.wiley.com/doi/pdf/10.1155/2013/836168} \BibitemShut {NoStop}%
\bibitem [{\citenamefont {Changlani}\ \emph {et~al.}(2018)\citenamefont {Changlani}, \citenamefont {Kochkov}, \citenamefont {Kumar}, \citenamefont {Clark},\ and\ \citenamefont {Fradkin}}]{PhysRevLett.120.117202}%
  \BibitemOpen
  \bibfield  {author} {\bibinfo {author} {\bibfnamefont {H.~J.}\ \bibnamefont {Changlani}}, \bibinfo {author} {\bibfnamefont {D.}~\bibnamefont {Kochkov}}, \bibinfo {author} {\bibfnamefont {K.}~\bibnamefont {Kumar}}, \bibinfo {author} {\bibfnamefont {B.~K.}\ \bibnamefont {Clark}}, \ and\ \bibinfo {author} {\bibfnamefont {E.}~\bibnamefont {Fradkin}},\ }\href {\doibase 10.1103/PhysRevLett.120.117202} {\bibfield  {journal} {\bibinfo  {journal} {Phys. Rev. Lett.}\ }\textbf {\bibinfo {volume} {120}},\ \bibinfo {pages} {117202} (\bibinfo {year} {2018})}\BibitemShut {NoStop}%
\bibitem [{\citenamefont {Changlani}\ \emph {et~al.}(2019)\citenamefont {Changlani}, \citenamefont {Pujari}, \citenamefont {Chung},\ and\ \citenamefont {Clark}}]{PhysRevB.99.104433}%
  \BibitemOpen
  \bibfield  {author} {\bibinfo {author} {\bibfnamefont {H.~J.}\ \bibnamefont {Changlani}}, \bibinfo {author} {\bibfnamefont {S.}~\bibnamefont {Pujari}}, \bibinfo {author} {\bibfnamefont {C.-M.}\ \bibnamefont {Chung}}, \ and\ \bibinfo {author} {\bibfnamefont {B.~K.}\ \bibnamefont {Clark}},\ }\href {\doibase 10.1103/PhysRevB.99.104433} {\bibfield  {journal} {\bibinfo  {journal} {Phys. Rev. B}\ }\textbf {\bibinfo {volume} {99}},\ \bibinfo {pages} {104433} (\bibinfo {year} {2019})}\BibitemShut {NoStop}%
\bibitem [{\citenamefont {Pal}\ \emph {et~al.}(2021)\citenamefont {Pal}, \citenamefont {Sharma}, \citenamefont {Changlani},\ and\ \citenamefont {Pujari}}]{PhysRevB.103.144414}%
  \BibitemOpen
  \bibfield  {author} {\bibinfo {author} {\bibfnamefont {S.}~\bibnamefont {Pal}}, \bibinfo {author} {\bibfnamefont {P.}~\bibnamefont {Sharma}}, \bibinfo {author} {\bibfnamefont {H.~J.}\ \bibnamefont {Changlani}}, \ and\ \bibinfo {author} {\bibfnamefont {S.}~\bibnamefont {Pujari}},\ }\href {\doibase 10.1103/PhysRevB.103.144414} {\bibfield  {journal} {\bibinfo  {journal} {Phys. Rev. B}\ }\textbf {\bibinfo {volume} {103}},\ \bibinfo {pages} {144414} (\bibinfo {year} {2021})}\BibitemShut {NoStop}%
\bibitem [{\citenamefont {Zhang}\ \emph {et~al.}(2025)\citenamefont {Zhang}, \citenamefont {Mao}, \citenamefont {Kim},\ and\ \citenamefont {Moessner}}]{Zhang2025}%
  \BibitemOpen
  \bibfield  {author} {\bibinfo {author} {\bibfnamefont {K.}~\bibnamefont {Zhang}}, \bibinfo {author} {\bibfnamefont {D.}~\bibnamefont {Mao}}, \bibinfo {author} {\bibfnamefont {E.-A.}\ \bibnamefont {Kim}}, \ and\ \bibinfo {author} {\bibfnamefont {R.}~\bibnamefont {Moessner}},\ }\href {\doibase 10.1038/s43246-025-00849-5} {\bibfield  {journal} {\bibinfo  {journal} {Communications Materials}\ }\textbf {\bibinfo {volume} {6}},\ \bibinfo {pages} {124} (\bibinfo {year} {2025})}\BibitemShut {NoStop}%
\bibitem [{\citenamefont {Ryzhkin}(2005)}]{Ryzhkin2005}%
  \BibitemOpen
  \bibfield  {author} {\bibinfo {author} {\bibfnamefont {I.~A.}\ \bibnamefont {Ryzhkin}},\ }\href {\doibase 10.1134/1.2103216} {\bibfield  {journal} {\bibinfo  {journal} {Journal of Experimental and Theoretical Physics}\ }\textbf {\bibinfo {volume} {101}},\ \bibinfo {pages} {481} (\bibinfo {year} {2005})}\BibitemShut {NoStop}%
\bibitem [{\citenamefont {Jaubert}\ and\ \citenamefont {Holdsworth}(2009)}]{Jaubert-dynamics}%
  \BibitemOpen
  \bibfield  {author} {\bibinfo {author} {\bibfnamefont {L.~D.~C.}\ \bibnamefont {Jaubert}}\ and\ \bibinfo {author} {\bibfnamefont {P.~C.~W.}\ \bibnamefont {Holdsworth}},\ }\href {https://doi.org/10.1038/nphys1227} {\bibfield  {journal} {\bibinfo  {journal} {Nature Physics}\ }\textbf {\bibinfo {volume} {5}},\ \bibinfo {pages} {258–261} (\bibinfo {year} {2009})}\BibitemShut {NoStop}%
\bibitem [{\citenamefont {Hallén}\ \emph {et~al.}(2022)\citenamefont {Hallén}, \citenamefont {Grigera}, \citenamefont {Tennant}, \citenamefont {Castelnovo},\ and\ \citenamefont {Moessner}}]{doi:10.1126/science.add1644}%
  \BibitemOpen
  \bibfield  {author} {\bibinfo {author} {\bibfnamefont {J.~N.}\ \bibnamefont {Hallén}}, \bibinfo {author} {\bibfnamefont {S.~A.}\ \bibnamefont {Grigera}}, \bibinfo {author} {\bibfnamefont {D.~A.}\ \bibnamefont {Tennant}}, \bibinfo {author} {\bibfnamefont {C.}~\bibnamefont {Castelnovo}}, \ and\ \bibinfo {author} {\bibfnamefont {R.}~\bibnamefont {Moessner}},\ }\href {\doibase 10.1126/science.add1644} {\bibfield  {journal} {\bibinfo  {journal} {Science}\ }\textbf {\bibinfo {volume} {378}},\ \bibinfo {pages} {1218} (\bibinfo {year} {2022})},\ \Eprint {http://arxiv.org/abs/https://www.science.org/doi/pdf/10.1126/science.add1644} {https://www.science.org/doi/pdf/10.1126/science.add1644} \BibitemShut {NoStop}%
\bibitem [{\citenamefont {Fulton}\ and\ \citenamefont {Harris}(1991)}]{Groups_Fulton-Harris}%
  \BibitemOpen
  \bibfield  {author} {\bibinfo {author} {\bibfnamefont {W.}~\bibnamefont {Fulton}}\ and\ \bibinfo {author} {\bibfnamefont {J.}~\bibnamefont {Harris}},\ }\href@noop {} {\emph {\bibinfo {title} {Representation Theory a First Course}}}\ (\bibinfo  {publisher} {Springer New York, NY},\ \bibinfo {year} {1991})\ Chap.~\bibinfo {chapter} {21}\BibitemShut {NoStop}%
\bibitem [{\citenamefont {Jaubert}\ \emph {et~al.}(2011)\citenamefont {Jaubert}, \citenamefont {Haque},\ and\ \citenamefont {Moessner}}]{PhysRevLett.107.177202}%
  \BibitemOpen
  \bibfield  {author} {\bibinfo {author} {\bibfnamefont {L.~D.~C.}\ \bibnamefont {Jaubert}}, \bibinfo {author} {\bibfnamefont {M.}~\bibnamefont {Haque}}, \ and\ \bibinfo {author} {\bibfnamefont {R.}~\bibnamefont {Moessner}},\ }\href {\doibase 10.1103/PhysRevLett.107.177202} {\bibfield  {journal} {\bibinfo  {journal} {Phys. Rev. Lett.}\ }\textbf {\bibinfo {volume} {107}},\ \bibinfo {pages} {177202} (\bibinfo {year} {2011})}\BibitemShut {NoStop}%
\bibitem [{\citenamefont {Cépas}\ and\ \citenamefont {Akhmetiev}(2019)}]{10.21468/SciPostPhys.7.3.032}%
  \BibitemOpen
  \bibfield  {author} {\bibinfo {author} {\bibfnamefont {O.}~\bibnamefont {Cépas}}\ and\ \bibinfo {author} {\bibfnamefont {P.~M.}\ \bibnamefont {Akhmetiev}},\ }\href {\doibase 10.21468/SciPostPhys.7.3.032} {\bibfield  {journal} {\bibinfo  {journal} {SciPost Phys.}\ }\textbf {\bibinfo {volume} {7}},\ \bibinfo {pages} {032} (\bibinfo {year} {2019})}\BibitemShut {NoStop}%
\bibitem [{\citenamefont {Cépas}\ and\ \citenamefont {Akhmetiev}(2021)}]{10.21468/SciPostPhys.10.2.042}%
  \BibitemOpen
  \bibfield  {author} {\bibinfo {author} {\bibfnamefont {O.}~\bibnamefont {Cépas}}\ and\ \bibinfo {author} {\bibfnamefont {P.~M.}\ \bibnamefont {Akhmetiev}},\ }\href {\doibase 10.21468/SciPostPhys.10.2.042} {\bibfield  {journal} {\bibinfo  {journal} {SciPost Phys.}\ }\textbf {\bibinfo {volume} {10}},\ \bibinfo {pages} {042} (\bibinfo {year} {2021})}\BibitemShut {NoStop}%
\bibitem [{\citenamefont {Jr}\ and\ \citenamefont {Henley}(1997)}]{J_K_Burton_Jr_1997}%
  \BibitemOpen
  \bibfield  {author} {\bibinfo {author} {\bibfnamefont {J.~K.~B.}\ \bibnamefont {Jr}}\ and\ \bibinfo {author} {\bibfnamefont {C.~L.}\ \bibnamefont {Henley}},\ }\href {\doibase 10.1088/0305-4470/30/24/007} {\bibfield  {journal} {\bibinfo  {journal} {Journal of Physics A: Mathematical and General}\ }\textbf {\bibinfo {volume} {30}},\ \bibinfo {pages} {8385} (\bibinfo {year} {1997})}\BibitemShut {NoStop}%
\bibitem [{\citenamefont {Ferreira}\ and\ \citenamefont {Sokal}(1999)}]{Ferreira1999}%
  \BibitemOpen
  \bibfield  {author} {\bibinfo {author} {\bibfnamefont {S.~J.}\ \bibnamefont {Ferreira}}\ and\ \bibinfo {author} {\bibfnamefont {A.~D.}\ \bibnamefont {Sokal}},\ }\href {\doibase 10.1023/A:1004599121565} {\bibfield  {journal} {\bibinfo  {journal} {Journal of Statistical Physics}\ }\textbf {\bibinfo {volume} {96}},\ \bibinfo {pages} {461} (\bibinfo {year} {1999})}\BibitemShut {NoStop}%
\bibitem [{\citenamefont {Rokhsar}\ and\ \citenamefont {Kivelson}(1988)}]{PhysRevLett.61.2376}%
  \BibitemOpen
  \bibfield  {author} {\bibinfo {author} {\bibfnamefont {D.~S.}\ \bibnamefont {Rokhsar}}\ and\ \bibinfo {author} {\bibfnamefont {S.~A.}\ \bibnamefont {Kivelson}},\ }\href {\doibase 10.1103/PhysRevLett.61.2376} {\bibfield  {journal} {\bibinfo  {journal} {Phys. Rev. Lett.}\ }\textbf {\bibinfo {volume} {61}},\ \bibinfo {pages} {2376} (\bibinfo {year} {1988})}\BibitemShut {NoStop}%
\bibitem [{\citenamefont {Shannon}\ \emph {et~al.}(2012)\citenamefont {Shannon}, \citenamefont {Sikora}, \citenamefont {Pollmann}, \citenamefont {Penc},\ and\ \citenamefont {Fulde}}]{PhysRevLett.108.067204}%
  \BibitemOpen
  \bibfield  {author} {\bibinfo {author} {\bibfnamefont {N.}~\bibnamefont {Shannon}}, \bibinfo {author} {\bibfnamefont {O.}~\bibnamefont {Sikora}}, \bibinfo {author} {\bibfnamefont {F.}~\bibnamefont {Pollmann}}, \bibinfo {author} {\bibfnamefont {K.}~\bibnamefont {Penc}}, \ and\ \bibinfo {author} {\bibfnamefont {P.}~\bibnamefont {Fulde}},\ }\href {\doibase 10.1103/PhysRevLett.108.067204} {\bibfield  {journal} {\bibinfo  {journal} {Phys. Rev. Lett.}\ }\textbf {\bibinfo {volume} {108}},\ \bibinfo {pages} {067204} (\bibinfo {year} {2012})}\BibitemShut {NoStop}%
\bibitem [{\citenamefont {Kato}\ and\ \citenamefont {Onoda}(2015)}]{PhysRevLett.115.077202}%
  \BibitemOpen
  \bibfield  {author} {\bibinfo {author} {\bibfnamefont {Y.}~\bibnamefont {Kato}}\ and\ \bibinfo {author} {\bibfnamefont {S.}~\bibnamefont {Onoda}},\ }\href {\doibase 10.1103/PhysRevLett.115.077202} {\bibfield  {journal} {\bibinfo  {journal} {Phys. Rev. Lett.}\ }\textbf {\bibinfo {volume} {115}},\ \bibinfo {pages} {077202} (\bibinfo {year} {2015})}\BibitemShut {NoStop}%
\bibitem [{\citenamefont {Huang}\ \emph {et~al.}(2018)\citenamefont {Huang}, \citenamefont {Deng}, \citenamefont {Wan},\ and\ \citenamefont {Meng}}]{PhysRevLett.120.167202}%
  \BibitemOpen
  \bibfield  {author} {\bibinfo {author} {\bibfnamefont {C.-J.}\ \bibnamefont {Huang}}, \bibinfo {author} {\bibfnamefont {Y.}~\bibnamefont {Deng}}, \bibinfo {author} {\bibfnamefont {Y.}~\bibnamefont {Wan}}, \ and\ \bibinfo {author} {\bibfnamefont {Z.~Y.}\ \bibnamefont {Meng}},\ }\href {\doibase 10.1103/PhysRevLett.120.167202} {\bibfield  {journal} {\bibinfo  {journal} {Phys. Rev. Lett.}\ }\textbf {\bibinfo {volume} {120}},\ \bibinfo {pages} {167202} (\bibinfo {year} {2018})}\BibitemShut {NoStop}%
\bibitem [{\citenamefont {Lee}\ \emph {et~al.}(2012)\citenamefont {Lee}, \citenamefont {Onoda},\ and\ \citenamefont {Balents}}]{PhysRevB.86.104412}%
  \BibitemOpen
  \bibfield  {author} {\bibinfo {author} {\bibfnamefont {S.}~\bibnamefont {Lee}}, \bibinfo {author} {\bibfnamefont {S.}~\bibnamefont {Onoda}}, \ and\ \bibinfo {author} {\bibfnamefont {L.}~\bibnamefont {Balents}},\ }\href {\doibase 10.1103/PhysRevB.86.104412} {\bibfield  {journal} {\bibinfo  {journal} {Phys. Rev. B}\ }\textbf {\bibinfo {volume} {86}},\ \bibinfo {pages} {104412} (\bibinfo {year} {2012})}\BibitemShut {NoStop}%
\bibitem [{\citenamefont {Savary}\ and\ \citenamefont {Balents}(2013)}]{PhysRevB.87.205130}%
  \BibitemOpen
  \bibfield  {author} {\bibinfo {author} {\bibfnamefont {L.}~\bibnamefont {Savary}}\ and\ \bibinfo {author} {\bibfnamefont {L.}~\bibnamefont {Balents}},\ }\href {\doibase 10.1103/PhysRevB.87.205130} {\bibfield  {journal} {\bibinfo  {journal} {Phys. Rev. B}\ }\textbf {\bibinfo {volume} {87}},\ \bibinfo {pages} {205130} (\bibinfo {year} {2013})}\BibitemShut {NoStop}%
\bibitem [{\citenamefont {Desrochers}\ and\ \citenamefont {Kim}(2024)}]{PhysRevLett.132.066502}%
  \BibitemOpen
  \bibfield  {author} {\bibinfo {author} {\bibfnamefont {F.}~\bibnamefont {Desrochers}}\ and\ \bibinfo {author} {\bibfnamefont {Y.~B.}\ \bibnamefont {Kim}},\ }\href {\doibase 10.1103/PhysRevLett.132.066502} {\bibfield  {journal} {\bibinfo  {journal} {Phys. Rev. Lett.}\ }\textbf {\bibinfo {volume} {132}},\ \bibinfo {pages} {066502} (\bibinfo {year} {2024})}\BibitemShut {NoStop}%
\bibitem [{\citenamefont {Hao}\ \emph {et~al.}(2014)\citenamefont {Hao}, \citenamefont {Day},\ and\ \citenamefont {Gingras}}]{PhysRevB.90.214430}%
  \BibitemOpen
  \bibfield  {author} {\bibinfo {author} {\bibfnamefont {Z.}~\bibnamefont {Hao}}, \bibinfo {author} {\bibfnamefont {A.~G.~R.}\ \bibnamefont {Day}}, \ and\ \bibinfo {author} {\bibfnamefont {M.~J.~P.}\ \bibnamefont {Gingras}},\ }\href {\doibase 10.1103/PhysRevB.90.214430} {\bibfield  {journal} {\bibinfo  {journal} {Phys. Rev. B}\ }\textbf {\bibinfo {volume} {90}},\ \bibinfo {pages} {214430} (\bibinfo {year} {2014})}\BibitemShut {NoStop}%
\bibitem [{\citenamefont {Hooft}(1981)}]{HOOFT1981455}%
  \BibitemOpen
  \bibfield  {author} {\bibinfo {author} {\bibfnamefont {G.}~\bibnamefont {Hooft}},\ }\href {\doibase https://doi.org/10.1016/0550-3213(81)90442-9} {\bibfield  {journal} {\bibinfo  {journal} {Nuclear Physics B}\ }\textbf {\bibinfo {volume} {190}},\ \bibinfo {pages} {455} (\bibinfo {year} {1981})}\BibitemShut {NoStop}%
\bibitem [{\citenamefont {Kondo}(1998)}]{PhysRevD.57.7467}%
  \BibitemOpen
  \bibfield  {author} {\bibinfo {author} {\bibfnamefont {K.-I.}\ \bibnamefont {Kondo}},\ }\href {\doibase 10.1103/PhysRevD.57.7467} {\bibfield  {journal} {\bibinfo  {journal} {Phys. Rev. D}\ }\textbf {\bibinfo {volume} {57}},\ \bibinfo {pages} {7467} (\bibinfo {year} {1998})}\BibitemShut {NoStop}%
\bibitem [{\citenamefont {Polyakov}(1975)}]{POLYAKOV197582}%
  \BibitemOpen
  \bibfield  {author} {\bibinfo {author} {\bibfnamefont {A.}~\bibnamefont {Polyakov}},\ }\href {\doibase https://doi.org/10.1016/0370-2693(75)90162-8} {\bibfield  {journal} {\bibinfo  {journal} {Physics Letters B}\ }\textbf {\bibinfo {volume} {59}},\ \bibinfo {pages} {82} (\bibinfo {year} {1975})}\BibitemShut {NoStop}%
\bibitem [{\citenamefont {Polyakov}(1987)}]{Polyakov_book}%
  \BibitemOpen
  \bibfield  {author} {\bibinfo {author} {\bibfnamefont {A.}~\bibnamefont {Polyakov}},\ }\href {\doibase https://doi.org/10.1201/9780203755082} {\emph {\bibinfo {title} {Gauge Fields and Strings (1st ed.)}}}\ (\bibinfo  {publisher} {Routledge},\ \bibinfo {year} {1987})\BibitemShut {NoStop}%
\bibitem [{sup()}]{supp}%
  \BibitemOpen
  \href@noop {} {}\bibinfo {note} {See Supplemental Material at URL-will-be-inserted-by-publisher for details on one-loop renormalization of the matter-only GMFT and further details on flux frustration}\BibitemShut {NoStop}%
\bibitem [{\citenamefont {Simon}(2023)}]{10.1093/oso/9780198886723.001.0001}%
  \BibitemOpen
  \bibfield  {author} {\bibinfo {author} {\bibfnamefont {S.~H.}\ \bibnamefont {Simon}},\ }\href {\doibase 10.1093/oso/9780198886723.001.0001} {\emph {\bibinfo {title} {Topological Quantum}}}\ (\bibinfo  {publisher} {Oxford University Press},\ \bibinfo {year} {2023})\BibitemShut {NoStop}%
\bibitem [{\citenamefont {Kitaev}(2003)}]{KITAEV20032}%
  \BibitemOpen
  \bibfield  {author} {\bibinfo {author} {\bibfnamefont {A.}~\bibnamefont {Kitaev}},\ }\href {\doibase https://doi.org/10.1016/S0003-4916(02)00018-0} {\bibfield  {journal} {\bibinfo  {journal} {Annals of Physics}\ }\textbf {\bibinfo {volume} {303}},\ \bibinfo {pages} {2} (\bibinfo {year} {2003})}\BibitemShut {NoStop}%
\bibitem [{\citenamefont {Szab\'o}\ and\ \citenamefont {Castelnovo}(2019)}]{PhysRevB.100.014417}%
  \BibitemOpen
  \bibfield  {author} {\bibinfo {author} {\bibfnamefont {A.}~\bibnamefont {Szab\'o}}\ and\ \bibinfo {author} {\bibfnamefont {C.}~\bibnamefont {Castelnovo}},\ }\href {\doibase 10.1103/PhysRevB.100.014417} {\bibfield  {journal} {\bibinfo  {journal} {Phys. Rev. B}\ }\textbf {\bibinfo {volume} {100}},\ \bibinfo {pages} {014417} (\bibinfo {year} {2019})}\BibitemShut {NoStop}%
\bibitem [{\citenamefont {Sanders}\ \emph {et~al.}(2024)\citenamefont {Sanders}, \citenamefont {Yan}, \citenamefont {Castelnovo},\ and\ \citenamefont {Nevidomskyy}}]{sanders2024experimentallytunableqeddipolaroctupolar}%
  \BibitemOpen
  \bibfield  {author} {\bibinfo {author} {\bibfnamefont {A.}~\bibnamefont {Sanders}}, \bibinfo {author} {\bibfnamefont {H.}~\bibnamefont {Yan}}, \bibinfo {author} {\bibfnamefont {C.}~\bibnamefont {Castelnovo}}, \ and\ \bibinfo {author} {\bibfnamefont {A.~H.}\ \bibnamefont {Nevidomskyy}},\ }\href {https://arxiv.org/abs/2312.11641} {\enquote {\bibinfo {title} {Experimentally tunable qed in dipolar-octupolar quantum spin ice},}\ } (\bibinfo {year} {2024}),\ \Eprint {http://arxiv.org/abs/2312.11641} {arXiv:2312.11641 [cond-mat.str-el]} \BibitemShut {NoStop}%
\bibitem [{\citenamefont {Pohle}\ and\ \citenamefont {Shannon}(2025)}]{pohle2025abundancespinliquidss1}%
  \BibitemOpen
  \bibfield  {author} {\bibinfo {author} {\bibfnamefont {R.}~\bibnamefont {Pohle}}\ and\ \bibinfo {author} {\bibfnamefont {N.}~\bibnamefont {Shannon}},\ }\href {https://arxiv.org/abs/2503.12776} {\enquote {\bibinfo {title} {Abundance of spin liquids in the $s=1$ bilinear-biquadratic model on the pyrochlore lattice, and its application to $\mathrm{NaCaNi}_2\mathrm{F}_7$},}\ } (\bibinfo {year} {2025}),\ \Eprint {http://arxiv.org/abs/2503.12776} {arXiv:2503.12776 [cond-mat.str-el]} \BibitemShut {NoStop}%
\end{thebibliography}%


\begin{thebibliography}{0}%
\makeatletter
\providecommand \@ifxundefined [1]{%
 \@ifx{#1\undefined}
}%
\providecommand \@ifnum [1]{%
 \ifnum #1\expandafter \@firstoftwo
 \else \expandafter \@secondoftwo
 \fi
}%
\providecommand \@ifx [1]{%
 \ifx #1\expandafter \@firstoftwo
 \else \expandafter \@secondoftwo
 \fi
}%
\providecommand \natexlab [1]{#1}%
\providecommand \enquote  [1]{``#1''}%
\providecommand \bibnamefont  [1]{#1}%
\providecommand \bibfnamefont [1]{#1}%
\providecommand \citenamefont [1]{#1}%
\providecommand \href@noop [0]{\@secondoftwo}%
\providecommand \href [0]{\begingroup \@sanitize@url \@href}%
\providecommand \@href[1]{\@@startlink{#1}\@@href}%
\providecommand \@@href[1]{\endgroup#1\@@endlink}%
\providecommand \@sanitize@url [0]{\catcode `\\12\catcode `\$12\catcode `\&12\catcode `\#12\catcode `\^12\catcode `\_12\catcode `\%12\relax}%
\providecommand \@@startlink[1]{}%
\providecommand \@@endlink[0]{}%
\providecommand \url  [0]{\begingroup\@sanitize@url \@url }%
\providecommand \@url [1]{\endgroup\@href {#1}{\urlprefix }}%
\providecommand \urlprefix  [0]{URL }%
\providecommand \Eprint [0]{\href }%
\providecommand \doibase [0]{https://doi.org/}%
\providecommand \selectlanguage [0]{\@gobble}%
\providecommand \bibinfo  [0]{\@secondoftwo}%
\providecommand \bibfield  [0]{\@secondoftwo}%
\providecommand \translation [1]{[#1]}%
\providecommand \BibitemOpen [0]{}%
\providecommand \bibitemStop [0]{}%
\providecommand \bibitemNoStop [0]{.\EOS\space}%
\providecommand \EOS [0]{\spacefactor3000\relax}%
\providecommand \BibitemShut  [1]{\csname bibitem#1\endcsname}%
\let\auto@bib@innerbib\@empty
\end{thebibliography}%

\end{document}


\title{Supplementary material for \\ `N-state Potts ices as a route to generalized spin liquids'}
\author{Mark Potts}
\affiliation{Max Planck Institute for the Physics of Complex Systems, N\"{o}thnitzer Str. 38, Dresden 01187, Germany}
\author{S.A. Parameswaran}
\affiliation{Rudolf Peierls Centre for Theoretical Physics, Parks Road, Oxford OX1 3PU, United Kingdom}
\affiliation{Hertford College, Catte Street, Oxford OX1 3BW, United Kingdom}
\author{Roderich Moessner}
\affiliation{Max Planck Institute for the Physics of Complex Systems, N\"{o}thnitzer Str. 38, Dresden 01187, Germany}
\affiliation{Hertford College, Catte Street, Oxford OX1 3BW, United Kingdom}

\maketitle

\section{Renormalisation group equations for $N_c=3,4$ models in (3+1) dimensions}

In this section we give a thorough exhibition of the momentum-shell renormalisation of complex scalar boson field theories possessing $U(1)^{\otimes (N_c-1)}\Bigr|_{\mathbb{Z}_2\times S_{N_c}}$ gauge symmetries for $N_C=3$ and $N_c=4$ in $(3+1)$ dimensions, up to single loop order. We shall work in imaginary time throughout.

\subsection{$N_c=3$ RG flow}
We write the action for the three color model as $S=S_0+\delta S$, and define:
\begin{equation}
    S_0=\int \frac{\text{d}^4k}{(2\pi)^4}\sum_{\alpha>0} \phi^{\alpha*}(k)\phi^{\alpha}(k)\left\{k^2r_2+r_0\right\},
\end{equation}
and:
\begin{multline}
    \delta S =g\prod_{i=1}^3\left(\int\frac{\text{d}^4k_i}{(2\pi)^4}\right) (2\pi)^4\delta(k_1-k_2-k_3)\left\{\phi^{\alpha+\beta*}(k_1)\phi^{\alpha}(k_2)\phi^{\beta}(k_3) + \text{h.c.}\right\} \\ \prod_{i=1}^4\left(\int\frac{\text{d}^4k_i}{(2\pi)^4}\right) (2\pi)^4\delta(k_1+k_2-k_3-k_4)\left\{\frac{u_1}{4}\sum_{\alpha>0}\phi^{\alpha*}(k_1)\phi^{\alpha*}(k_2)\phi^{\alpha}(k_3)\phi^{\alpha}(k_4) \right. \\ \left. +u_{2;a}\sum_{\langle \alpha, \beta \rangle}\phi^{\alpha*}(k_1)\phi^{\beta*}(k_2)\phi^{\alpha}(k_3)\phi^{\beta}(k_4)\right\}
\end{multline}

Rescaling momenta as $k'=k/s$, and fields as $\phi'^{\alpha}=\zeta\phi^{\alpha}$, the tree level scaling relations are:
\begin{align}
    r'_2&=\frac{\zeta^2}{s^6}r_2 \ \ ; \ \ r'_0=\frac{\zeta^2}{s^4}r_0 \ \ ; \\ g'&=\frac{\zeta^3}{s^{8}}g \ \ ; \\
    u'_{1}&=\frac{\zeta^4}{s^{12}}u_1 \ \ ; \ \ u'_{2;a}=\frac{\zeta^4}{s^{12}}u_{2;a} \ .
\end{align}
Holding $r_2$ constant, and writing $s=1+\text{d}l$, the flow equations before quantum corrections are thus:
\begin{align}
    \frac{\text{d}r_0}{dl}&=2r_0, \\
    \frac{\text{d}g}{dl}&=g, \\
    \frac{\text{d}u_1}{dl}&=\frac{\text{d}u_{2;a}}{dl}=0.
\end{align}
The coupling $g$ is a relevant parameter, and the $u$'s are marginal. At the mean field level, the combination $g^2/r_0$ is a constant under the RG flow, hence the phase boundaries presented in Figs. 5 and 6 are all parabolas. Further, the model flows to large coupling and so, without some conspiracy of fine tuning in the original color ice model, as $J_{\pm}/J_{z}$ is varied any rooton condensation transition/ ordering transition out of the liquid phase is expected to be first order, and not pass through the critical point. 

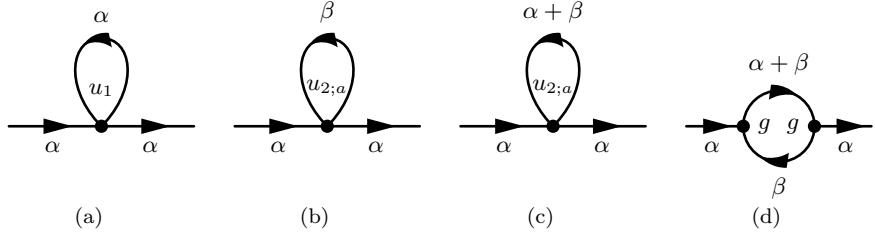
\begin{figure}
    \centering
    \subfigure[]{
    \begin{fmffile}{Tadpole_1}
        \begin{fmfgraph*}(70,52.5)
        \fmfleft{i}
        \fmfright{o}
        \fmf{fermion,label=$\alpha$}{i,v}
        \fmf{fermion,label=$\alpha$}{v,o}
        \fmf{fermion,right=1,tension=0.7,label=$\alpha$}{v,v}
        \fmfdot{v}

        \fmfv{label=$u_1$,label.angle=90, label.dist=4mm}{v}
        \end{fmfgraph*}
    \end{fmffile}}
    \subfigure[]{
    \begin{fmffile}{Tadpole_2}
        \begin{fmfgraph*}(70,52.5)
        \fmfleft{i}
        \fmfright{o}
        \fmf{fermion,label=$\alpha$}{i,v}
        \fmf{fermion,label=$\alpha$}{v,o}
        \fmf{fermion,right=1,tension=0.7,label=$\beta$}{v,v}
        \fmfdot{v}

        \fmfv{label=$u_{2;a}$,label.angle=90, label.dist=4mm}{v}
        \end{fmfgraph*}
    \end{fmffile}}
    \subfigure[]{
    \begin{fmffile}{Tadpole_3}
        \begin{fmfgraph*}(70,52.5)
        \fmfleft{i}
        \fmfright{o}
        \fmf{fermion,label=$\alpha$}{i,v}
        \fmf{fermion,label=$\alpha$}{v,o}
        \fmf{fermion,right=1,tension=0.7,label=$\alpha+\beta$}{v,v}
        \fmfdot{v}

        \fmfv{label=$u_{2;a}$,label.angle=90, label.dist=4mm}{v}
        \end{fmfgraph*}
    \end{fmffile}}
    \subfigure[]{
    \begin{fmffile}{fermion_circle_loop}
        \begin{fmfgraph*}(70,52.5)
        \fmfleft{i}
        \fmfright{o}
        \fmf{fermion,label=$\alpha$}{i,v}
        \fmf{fermion,label=$\alpha$}{v2,o}
        \fmf{fermion,left=1,tension=0.4,label=$\alpha+\beta$}{v,v2}
        \fmf{fermion,left=1,tension=0.4,label=$\beta$}{v2,v}
        \fmfdot{v}
        \fmfdot{v2}

        \fmfv{label=$g$,label.angle=0, label.dist=2mm}{v}
        \fmfv{label=$g$,label.angle=180, label.dist=2mm}{v2}
        \end{fmfgraph*}
    \end{fmffile}}
    
    \caption{Diagrams contributing to the renormalisation of the two-point function for $N_c=3$. Diagrams (a-c) are momentum independent, whilst diagram (d) is momentum dependent, and renormalises the field scale factor $\zeta$.}
    \label{fig:Nc_3_RG_diagrams}
\end{figure}

Quantum corrections to the above RG flow equations can be calculated perturbatively. Feynman diagrams for terms in the the single loop renormalisation of the two point function are shown in Fig. \ref{fig:Nc_3_RG_diagrams}. Diagrams (a-c) contribute only to the renormalisation of $r_0$, with total contribution:
\begin{equation}
    \delta r_0 = \left\{u_1+2u_{2;a}\right\} \frac{\zeta^2}{s^4}\int^{\Lambda}_{\Lambda/s}\frac{\text{d}^4k}{(2\pi)^4}G_0(k),
\end{equation}
where $\Lambda$ is the lattice scale momentum cut-off, and $G_0(k)$ is the non-interacting two-point function. This evaluates to:
\begin{equation}
    \delta r_0 = \left\{u_1+2u_{2;a}\right\} \frac{\zeta^2}{s^4} \frac{1}{16\pi^2} \left[\frac{\Lambda^2}{r_2}\left(1-\frac{1}{s^2}\right)-\frac{r_0}{r_2^2}\ln\left(\frac{\Lambda^2+\frac{r_0}{r_2}}{\frac{\Lambda^2}{s^2}+\frac{r_0}{r_2}}\right)\right] \approx \left\{u_1+2u_{2;a}\right\}\frac{\zeta^2}{s^4}\frac{1}{8\pi^2} \frac{\Lambda^4}{r_2\Lambda^2+r_0}\delta   l 
\end{equation}

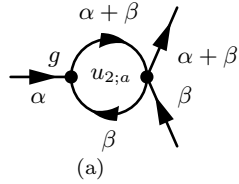
\begin{figure}[h!]
    \centering
    \subfigure[]{
    \begin{fmffile}{Nc_3_g_RG}
        \begin{fmfgraph*}(70,52.5)
            \fmfleft{i1}
            \fmfright{o1,o2}
           
            \fmf{fermion,left=1,tension=0.4,label=$\alpha+\beta$}{v1,v2}
            \fmf{fermion,left=1,tension=0.4,label=$\beta$}{v2,v1}

            \fmf{fermion,label=$\alpha$}{i1,v1}
            \fmf{fermion,label=$\beta$}{o1,v2}
            \fmf{fermion,label=$\alpha +\beta$,label.side=right}{v2,o2}

            \fmfdot{v1}
            \fmfdot{v2}

            \fmfv{label=$g$,label.angle=135, label.dist=2mm}{v1}
            \fmfv{label=$u_{2;a}$,label.angle=180, label.dist=2mm}{v2}
        
        \end{fmfgraph*}
    \end{fmffile}}
    \caption{Single loop diagram that renormalises the three-point function for $N_c=3$ (in addition to permutations of this diagram). The diagram is proportional to $u_{2;a}g$, involving one three-point vertex, and one four-point vertex between different rootons.}
    \label{fig:Nc_3_g_renormalisation}
\end{figure}

Diagram (d) contributes both to the renormalisation of $r_0$ and $r_2$. The $k-$dependent correction from the diagram can be approximated by the following integral for small slow-mode momentum $k'$ as:
\begin{equation}
    \delta r(k') = -g^2 \frac{\zeta^2}{s^4}\int_{\Lambda/s}^{\Lambda}\frac{\text{d}^4k}{(2\pi)^4}\frac{1}{r_2k^2+r_0}\frac{1}{r_2|k'/s+k|^2+r_0}.
\end{equation}
From this, the corrections to $r_0$ and $r_2$ are:
\begin{align}
    \delta r_0 &= -g^2\frac{\zeta^2}{s^4}\frac{1}{8\pi^2}\frac{\Lambda^4}{(r_2\Lambda^2+r_0)^2} \text{d}l,\\ \delta r_2 &= g^2 \frac{\zeta^2}{s^6}\frac{1}{8\pi^2}\frac{\Lambda^4r_0}{(r_2\Lambda^2+r_0)^4}\text{d}l.
\end{align}
Keeping $r_2=1$ constant, $\zeta$ thus picks up a contribution proportional to $\text{d}l$:
\begin{equation}
    \zeta=s^3\left\{1-\frac{g^2}{16\pi^2}\frac{\Lambda^4 r_0}{(r_2\Lambda^2+r_0)^4} \text{d}l\right\}.
\end{equation}
The diagrams contributing to the renormalisation of the parameters $g$, $u_1$, and $u_{2;a}$ are presented in Figs. \ref{fig:Nc_3_g_renormalisation},\ref{fig:Nc_3_u1_renormalisation}, and \ref{fig:Nc_3_u2a_renormalisation}.

Evaluating each of these, the full RG equations correct to first order in loops are given by (with $r_0$ in units of $\Lambda^2$, and $g$ in units of $\Lambda$):
\begin{align}
    \frac{\text{d}r_0}{dl}&=2r_0 -\frac{g^2}{8\pi^2}\frac{r_0^2}{(1+r_0)^4} +\frac{u_1+2u_{2;a}}{8\pi^2}\frac{1}{1+r_0}- \frac{g^2}{8\pi^2}\frac{1}{(1+r_0)^2}, \\
    \frac{\text{d}g}{dl}&=g -\frac{3g^3}{16\pi^2}\frac{r_0}{(1+r_0)^4} - \frac{3}{8\pi^2} \frac{u_{2;a}g}{(1+r_0)^3}, \\
    \frac{\text{d}u_1}{dl}&= -\frac{g^2u_1}{4 \pi^2}\frac{r_0}{(1+r_0)^4}-\frac{5u_1^2+8u_{2;a}^2}{16\pi^2}\frac{1}{(1+r_0)^2}+\frac{g^2u_{2;a}}{\pi^2}\frac{1}{(1+r_0)^3} \\ \frac{\text{d}u_{2;a}}{dl}&= -\frac{g^2u_{2;a}}{4\pi^2}\frac{r_0}{(1+r_0)^4}-\frac{2u_1u_{2;a}+3u_{2;a}^2}{8\pi^2}\frac{1}{(1+r_0)^2}+\frac{2u_1g^2+4u_{2;a}g^2}{8\pi^2}\frac{1}{(1+r_0)^3}.
\end{align}

These flow equations are valid within the rooton vacuum ($P_0$) phase. Whilst they are perturbative in $g$, a relevant parameter, all corrections to the RG flow beyond tree-level vanish at late stages of the RG flow within this phase, and thus the expansion remains controlled. The leading order effects of quantum fluctuations is to shift the location of the critical point parallel to the $r_0$ axis by $-\frac{u_1+2u_{2;a}}{16\pi^2}$.

\begin{figure}[h!]
    \centering
    \subfigure[]{
    \begin{fmffile}{Nc_3_u1_RG_1}
        \begin{fmfgraph*}(70,52.5)
            \fmfleft{i1,i2}
            \fmfright{o1,o2}
           
            \fmf{fermion,left=1,tension=0.4,label=$\alpha$}{v1,v2}
            \fmf{fermion,right=1,tension=0.4,label=$\alpha$}{v1,v2}
            
            \fmf{fermion,label=$\alpha$,label.side=left}{i1,v1}
            \fmf{fermion,label=$\alpha$}{v2,i2}

            \fmf{fermion,label=$\alpha$}{o1,v1}
            \fmf{fermion,label=$\alpha$,label.side=left}{v2,o2}

            \fmfdot{v1}
            \fmfdot{v2}

            \fmfv{label=$u_1$,label.angle=90, label.dist=2mm}{v1}
            \fmfv{label=$u_1$,label.angle=-90, label.dist=2mm}{v2}
        
        \end{fmfgraph*}
    \end{fmffile}}
    \subfigure[]{
    \begin{fmffile}{Nc_3_u1_RG_2}
        \begin{fmfgraph*}(70,52.5)
            \fmfleft{i1,i2}
            \fmfright{o1,o2}
           
            \fmf{fermion,left=1,tension=0.4,label=$\alpha_{\mu}$}{v1,v2}
            \fmf{fermion,left=1,tension=0.4,label=$\alpha_{\mu}$}{v2,v1}
            
            \fmf{fermion,label=$\alpha$}{i1,v1}
            \fmf{fermion,label=$\alpha$}{v1,i2}

            \fmf{fermion,label=$\alpha$,label.side=left}{o1,v2}
            \fmf{fermion,label=$\alpha$}{v2,o2}

            \fmfdot{v1}
            \fmfdot{v2}

            \fmfv{label=$u_{\mu}$,label.angle=180, label.dist=2mm}{v1}
            \fmfv{label=$u_{\mu}$,label.angle=0, label.dist=2mm}{v2}
        
        \end{fmfgraph*}
    \end{fmffile}}
    \subfigure[]{
    \begin{fmffile}{Nc_3_u1_RG_3}
        \begin{fmfgraph*}(70,52.5)
            \fmfleft{i1,i2}
            \fmfright{o1,o2}
           
            \fmf{fermion,tension=0.4,label=$\beta$,label.side=right}{v2,v1}
            \fmf{fermion,tension=0.4,label=$\alpha+\beta$,label.side=right}{v3,v2}
            \fmf{fermion,tension=0.4,label=$\beta$}{v1,v3}

            \fmf{fermion,label=$\alpha$,label.side=right}{i1,v1}
            \fmf{fermion,label=$\alpha$}{o1,v3}
            \fmf{fermion,label=$\alpha$,label.side=left}{v2,o2}
            \fmf{fermion,label=$\alpha$}{v1,i2}

            \fmfdot{v1}
            \fmfdot{v2}
            \fmfdot{v3}

             \fmfv{label=$u_{2;a}$,label.angle=180, label.dist=2mm}{v1}
             \fmfv{label=$g$,label.angle=0, label.dist=2mm}{v2}
             \fmfv{label=$g$,label.angle=0, label.dist=2mm}{v3}
        
        \end{fmfgraph*}
    \end{fmffile}}
    \caption{Representative diagrams of the contributions to the renormalisation of $u_{1}$ to single loop order under the renormalisation group.}
    \label{fig:Nc_3_u1_renormalisation}
\end{figure}

\begin{figure}[h!]
    \centering
    \subfigure[]{
    \begin{fmffile}{Nc_3_u2_RG_1}
        \begin{fmfgraph*}(70,52.5)
            \fmfleft{i1,i2}
            \fmfright{o1,o2}
           
            \fmf{fermion,left=1,tension=0.4,label=$\alpha$}{v1,v2}
            \fmf{fermion,right=1,tension=0.4,label=$\beta$}{v1,v2}
            
            \fmf{fermion,label=$\alpha$,label.side=left}{i1,v1}
            \fmf{fermion,label=$\alpha$}{v2,i2}

            \fmf{fermion,label=$\beta$}{o1,v1}
            \fmf{fermion,label=$\beta$,label.side=left}{v2,o2}

            \fmfdot{v1}
            \fmfdot{v2}

            \fmfv{label=$u_{2;a}$,label.angle=90, label.dist=2mm}{v1}
            \fmfv{label=$u_{2;a}$,label.angle=-90, label.dist=2mm}{v2}
        
        \end{fmfgraph*}
    \end{fmffile}}
    \subfigure[]{
    \begin{fmffile}{Nc_3_u2_RG_2}
        \begin{fmfgraph*}(70,52.5)
            \fmfleft{i1,i2}
            \fmfright{o1,o2}
           
            \fmf{fermion,left=1,tension=0.4,label=$\alpha_{\mu}$}{v1,v2}
            \fmf{fermion,left=1,tension=0.4,label=$\alpha_{\mu}$}{v2,v1}
            
            \fmf{fermion,label=$\alpha$}{i1,v1}
            \fmf{fermion,label=$\alpha$}{v1,i2}

            \fmf{fermion,label=$\beta$,label.side=left}{o1,v2}
            \fmf{fermion,label=$\beta$}{v2,o2}

            \fmfdot{v1}
            \fmfdot{v2}

            \fmfv{label=$u_{\mu}$,label.angle=180, label.dist=2mm}{v1}
            \fmfv{label=$u_{\mu}$,label.angle=0, label.dist=2mm}{v2}
        
        \end{fmfgraph*}
    \end{fmffile}}
    \subfigure[]{
    \begin{fmffile}{Nc_3_u2_RG_3}
        \begin{fmfgraph*}(70,52.5)
            \fmfleft{i1,i2}
            \fmfright{o1,o2}
           
            \fmf{fermion,left=1,tension=0.4,label=$\alpha$}{v1,v2}
            \fmf{fermion,left=1,tension=0.4,label=$\beta$}{v2,v1}
            
            \fmf{fermion,label=$\alpha$}{i1,v1}
            \fmf{fermion,label=$\beta$}{v1,i2}

            \fmf{fermion,label=$\beta$,label.side=left}{o1,v2}
            \fmf{fermion,label=$\alpha$}{v2,o2}

            \fmfdot{v1}
            \fmfdot{v2}

            \fmfv{label=$u_{2;a}$,label.angle=180, label.dist=2mm}{v1}
            \fmfv{label=$u_{2;a}$,label.angle=0, label.dist=2mm}{v2}
        
        \end{fmfgraph*}
    \end{fmffile}}
    \subfigure[]{
    \begin{fmffile}{Nc_3_u2_RG_4}
        \begin{fmfgraph*}(70,52.5)
            \fmfleft{i1,i2}
            \fmfright{o1,o2}
           
            \fmf{fermion,tension=0.4,label=$\alpha$,label.side=right}{v2,v1}
            \fmf{fermion,tension=0.4,label=$\alpha+\beta$,label.side=right}{v3,v2}
            \fmf{fermion,tension=0.4,label=$\alpha$}{v1,v3}

            \fmf{fermion,label=$\alpha$,label.side=right}{i1,v1}
            \fmf{fermion,label=$\beta$}{o1,v3}
            \fmf{fermion,label=$\beta$,label.side=left}{v2,o2}
            \fmf{fermion,label=$\alpha$}{v1,i2}

            \fmfdot{v1}
            \fmfdot{v2}
            \fmfdot{v3}

            \fmfv{label=$u_{1}$,label.angle=180, label.dist=2mm}{v1}
             \fmfv{label=$g$,label.angle=0, label.dist=2mm}{v2}
             \fmfv{label=$g$,label.angle=0, label.dist=2mm}{v3}
        
        \end{fmfgraph*}
    \end{fmffile}}
    \subfigure[]{
    \begin{fmffile}{Nc_3_u2_RG_5}
        \begin{fmfgraph*}(70,52.5)
            \fmfleft{i1,i2}
            \fmfright{o1,o2}
           
            \fmf{fermion,tension=0.4,label=$\beta$,label.side=right}{v2,v1}
            \fmf{fermion,tension=0.4,label=$\alpha+\beta$,label.side=right}{v3,v2}
            \fmf{fermion,tension=0.4,label=$\alpha$}{v1,v3}

            \fmf{fermion,label=$\alpha$,label.side=right}{i1,v1}
            \fmf{fermion,label=$\beta$}{o1,v3}
            \fmf{fermion,label=$\alpha$,label.side=left}{v2,o2}
            \fmf{fermion,label=$\beta$}{v1,i2}

            \fmfdot{v1}
            \fmfdot{v2}
            \fmfdot{v3}

            \fmfv{label=$u_{2;a}$,label.angle=180, label.dist=2mm}{v1}
             \fmfv{label=$g$,label.angle=0, label.dist=2mm}{v2}
             \fmfv{label=$g$,label.angle=0, label.dist=2mm}{v3}
        
        \end{fmfgraph*}
    \end{fmffile}}
    \caption{Representative diagrams of the contributions to the renormalisation of $u_{2;a}$ to single loop order under the renormalisation group.}
    \label{fig:Nc_3_u2a_renormalisation}
\end{figure}

\subsection{$N_c=4$ RG flow}
Following the same procedure as for the $N_c=3$ case, the RG flow equations for the $N_c=4$ field theory under momentum shell renormalisation are as follows:
\begin{align}
    \frac{\text{d}r_0}{dl}&=2r _0 -\frac{g^2}{4\pi^2}\frac{r_0^2}{(1+r_0)^4} +\frac{u_1+4u_{2;a}+u_{2;b}}{8\pi^2}\frac{1}{1+r_0}- \frac{2g^2}{8\pi^2}\frac{1}{(1+r_0)^2}, \\
    \frac{\text{d}g}{dl}&=g -\frac{3g^3}{8\pi^2}\frac{r_0}{(1+r_0)^4} - \frac{3}{8\pi^2} \frac{(u_{2;a}+u_4) g}{(1+r_0)^3}+\frac{g^3}{8\pi^3}\frac{1}{(1+r_0)^3}, \\
    \frac{\text{d}u_1}{dl}&= -\frac{g^2u_1}{2 \pi^2}\frac{r_0}{(1+r_0)^4}-\frac{5u_1^2+16u_{2;a}^2+4u_{2;b}^2}{16\pi^2}\frac{1}{(1+r_0)^2}+\frac{2 g^2u_{2;a}}{\pi^2}\frac{1}{(1+r_0)^3} \\ \frac{\text{d}u_{2;a}}{dl}&= -\frac{g^2u_{2;a}r_0}{2  \pi^2(1+r_0)^4}-\frac{2u_1u_{2;a}+4u_{2;a}^2+2u_{2;a}u_{2;b}+u_4^2}{8\pi^2(1+r_0)^2}+\frac{(2u_1+6u_{2;a}+2u_{2;b}+2u_4)g^2}{8\pi^2(1+r_0)^3} -\frac{2g^4}{8\pi^2(1+r_0)^4} \\ \frac{\text{d}u_{2;b}}{dl}&=-\frac{g^2u_{2;b}r_0}{2   \pi^2(1+r_0)^4} -\frac{2u_1u_{2;b}+4u_{2;a}^2+2u_{2;b}^2+2u_4^2}{8\pi^2(1+r_0)^2} +\frac{(2u_{2;a}+4u_4)g^2}{8\pi^2(1+r_0)^3} -\frac{6 g^4}{8\pi^2(1+r_0)^4}\\ \frac{\text{d}u_{4  }}{dl}&=-\frac{g^2u_{4}}{2  \pi^2}\frac{r_0}{(1+r_0)^4} -\frac{6u_{2;a}u_4}{8\pi^2(1+r_0)^2} +\frac{(4  u_{2;a}+4u_{2;b})g^2}{8\pi^2(1+r_0)^3}
\end{align}

\begin{figure}
    \centering
    \subfigure[]{
    \begin{fmffile}{Nc_4_r_RG_1}
        \begin{fmfgraph*}(70,52.5)
        \fmfleft{i}
        \fmfright{o}
        \fmf{fermion,label=$\alpha$}{i,v}
        \fmf{fermion,label=$\alpha$}{v,o}
        \fmf{fermion,right=1,tension=0.7,label=$\alpha_{\mu}$}{v,v}
        \fmfdot{v}

        \fmfv{label=$u_{\mu}$,label.angle=90, label.dist=4mm}{v}
        \end{fmfgraph*}
    \end{fmffile}}
    \subfigure[]{
    \begin{fmffile}{Nc_4_r_RG_2}
        \begin{fmfgraph*}(70,52.5)
        \fmfleft{i}
        \fmfright{o}
        \fmf{fermion,label=$\alpha$}{i,v}
        \fmf{fermion,label=$\alpha$}{v2,o}
        \fmf{fermion,left=1,tension=0.4,label=$\alpha+\beta$}{v,v2}
        \fmf{fermion,left=1,tension=0.4,label=$\beta$}{v2,v}
        \fmfdot{v}
        \fmfdot{v2}

        \fmfv{label=$g$,label.angle=0, label.dist=2mm}{v}
        \fmfv{label=$g$,label.angle=180, label.dist=2mm}{v2}
        \end{fmfgraph*}
    \end{fmffile}}
    \subfigure[]{
    \begin{fmffile}{Nc_4_r_RG_3}
        \begin{fmfgraph*}(70,52.5)
        \fmfleft{i}
        \fmfright{o}
        \fmf{fermion,label=$\alpha$}{i,v}
        \fmf{fermion,label=$\alpha$}{v2,o}
        \fmf{fermion,left=1,tension=0.4,label=$\alpha+\gamma$}{v,v2}
        \fmf{fermion,left=1,tension=0.4,label=$\gamma$}{v2,v}
        \fmfdot{v}
        \fmfdot{v2}

        \fmfv{label=$g$,label.angle=0, label.dist=2mm}{v}
        \fmfv{label=$g$,label.angle=180, label.dist=2mm}{v2}
        \end{fmfgraph*}
    \end{fmffile}}
    
    \caption{Diagrams contributing to the renormalisation of the two-point function for $N_c=4$. Each rooton two-function now has two distinct diagrams proportional to $g^2$ contributing at this order.}
    \label{fig:Nc_4_RG_diagrams}
\end{figure}

\begin{figure}
    \centering
    \subfigure[]{
    \begin{fmffile}{Nc_4_g_RG_g_cubed}
        \begin{fmfgraph*}(70,52.5)
            \fmfleft{i1}
            \fmfright{o1,o2}
           
            \fmf{fermion,tension=0.4,label=$\alpha+\gamma$,label.side=left}{v1,v2}
            \fmf{fermion,tension=0.4,label=$\beta-\gamma$}{v3,v2}
            \fmf{fermion,tension=0.4,label=$\gamma$}{v3,v1}

            \fmf{fermion,label=$\alpha$}{i1,v1}
            \fmf{fermion,label=$\beta$,label.side=left}{o1,v3}
            \fmf{fermion,label=$\alpha +\beta$,label.side=left}{v2,o2}

            \fmfdot{v1}
            \fmfdot{v2}
            \fmfdot{v3}

            \fmfv{label=$g$,label.angle=-90, label.dist=2mm}{v1}
             \fmfv{label=$g$,label.angle=0, label.dist=2mm}{v2}
             \fmfv{label=$g$,label.angle=0, label.dist=2mm}{v3}
        
        \end{fmfgraph*}
    \end{fmffile}}
    \subfigure[]{
    \begin{fmffile}{Nc_4_g_RG_2}
        \begin{fmfgraph*}(70,52.5)
            \fmfleft{i1}
            \fmfright{o1,o2}
           
            \fmf{fermion,left=1,tension=0.4,label=$\alpha+\beta$}{v1,v2}
            \fmf{fermion,left=1,tension=0.4,label=$\beta$}{v2,v1}

            \fmf{fermion,label=$\alpha$}{i1,v1}
            \fmf{fermion,label=$\beta$,label.side=right}{o1,v2}
            \fmf{fermion,label=$\alpha +\beta$,label.side=right}{v2,o2}

            \fmfdot{v1}
            \fmfdot{v2}

            \fmfv{label=$g$,label.angle=135, label.dist=2mm}{v1}
            \fmfv{label=$u_{2;a}$,label.angle=180, label.dist=2mm}{v2}
        
        \end{fmfgraph*}
    \end{fmffile}}
    \subfigure[]{
    \begin{fmffile}{Nc_4_g_RG_3}
        \begin{fmfgraph*}(70,52.5)
            \fmfleft{i1}
            \fmfright{o1,o2}
           
            \fmf{fermion,left=1,tension=0.4,label=$\alpha+\gamma$}{v1,v2}
            \fmf{fermion,left=1,tension=0.4,label=$\gamma$}{v2,v1}

            \fmf{fermion,label=$\alpha$}{i1,v1}
            \fmf{fermion,label=$\beta$}{o1,v2}
            \fmf{fermion,label=$\alpha +\beta$,label.side=right}{v2,o2}

            \fmfdot{v1}
            \fmfdot{v2}

            \fmfv{label=$g$,label.angle=135, label.dist=2mm}{v1}
            \fmfv{label=$u_4$,label.angle=180, label.dist=2mm}{v2}
        
        \end{fmfgraph*}
    \end{fmffile}}
    \caption{Diagrams contributing to the renormalisation of the three-point function to single loop order for $N_c=4$. The larger root diagram of SU$(4)$ allows for the $g^3$ diagram absent from the $N_c=3$ RG equations.}
    \label{fig:Nc_4_g_renormalisation}
\end{figure}

\begin{figure}
    \centering
    \subfigure[]{
    \begin{fmffile}{Nc_4_u1_RG_1}
        \begin{fmfgraph*}(70,52.5)
            \fmfleft{i1,i2}
            \fmfright{o1,o2}
           
            \fmf{fermion,left=1,tension=0.4,label=$\alpha$}{v1,v2}
            \fmf{fermion,right=1,tension=0.4,label=$\alpha$}{v1,v2}
            
            \fmf{fermion,label=$\alpha$,label.side=left}{i1,v1}
            \fmf{fermion,label=$\alpha$}{v2,i2}

            \fmf{fermion,label=$\alpha$,label.side=left}{o1,v1}
            \fmf{fermion,label=$\alpha$}{v2,o2}

            \fmfdot{v1}
            \fmfdot{v2}

            \fmfv{label=$u_{1}$,label.angle=90, label.dist=2mm}{v1}
            \fmfv{label=$u_{1}$,label.angle=-90, label.dist=2mm}{v2}
        
        \end{fmfgraph*}
    \end{fmffile}}
    \subfigure[]{
    \begin{fmffile}{Nc_4_u1_RG_2}
        \begin{fmfgraph*}(70,52.5)
            \fmfleft{i1,i2}
            \fmfright{o1,o2}
           
            \fmf{fermion,left=1,tension=0.4,label=$\alpha_{\mu}$}{v1,v2}
            \fmf{fermion,left=1,tension=0.4,label=$\alpha_{\mu}$}{v2,v1}
            
            \fmf{fermion,label=$\alpha$}{i1,v1}
            \fmf{fermion,label=$\alpha$}{v1,i2}

            \fmf{fermion,label=$\alpha$,label.side=left}{o1,v2}
            \fmf{fermion,label=$\alpha$}{v2,o2}

            \fmfdot{v1}
            \fmfdot{v2}

            \fmfv{label=$u_{\mu}$,label.angle=180, label.dist=2mm}{v1}
            \fmfv{label=$u_{\mu}$,label.angle=0, label.dist=2mm}{v2}

        \end{fmfgraph*}
    \end{fmffile}}
    \subfigure[]{
    \begin{fmffile}{Nc_4_u1_RG_3}
        \begin{fmfgraph*}(70,52.5)
            \fmfleft{i1,i2}
            \fmfright{o1,o2}
           
            \fmf{fermion,tension=0.4,label=$\beta$,label.side=right}{v2,v1}
            \fmf{fermion,tension=0.4,label=$\alpha+\beta$,label.side=right}{v3,v2}
            \fmf{fermion,tension=0.4,label=$\beta$}{v1,v3}

            \fmf{fermion,label=$\alpha$,label.side=right}{i1,v1}
            \fmf{fermion,label=$\alpha$}{o1,v3}
            \fmf{fermion,label=$\alpha$,label.side=left}{v2,o2}
            \fmf{fermion,label=$\alpha$}{v1,i2}

            \fmfdot{v1}
            \fmfdot{v2}
            \fmfdot{v3}

            \fmfv{label=$u_{2;a}$,label.angle=180, label.dist=2mm}{v1}
             \fmfv{label=$g$,label.angle=0, label.dist=2mm}{v2}
             \fmfv{label=$g$,label.angle=0, label.dist=2mm}{v3}
        
        \end{fmfgraph*}
    \end{fmffile}}
    \caption{Contributions to $u_1$ renormalisation at single loop order for $N_c=4$ are analogous to those for $N_c=3$.}
    \label{fig:Nc_4_u1_renormalisation}
\end{figure}

\begin{figure}
    \centering
    \subfigure[]{
    \begin{fmffile}{Nc_4_u2_RG_1}
        \begin{fmfgraph*}(60,45)
            \fmfleft{i1,i2}
            \fmfright{o1,o2}
           
            \fmf{fermion,left=1,tension=0.4,label=$\scriptstyle\alpha$}{v1,v2}
            \fmf{fermion,right=1,tension=0.4,label=$\scriptstyle\beta$}{v1,v2}
            
            \fmf{fermion,label=$\scriptstyle\alpha$,label.side=left}{i1,v1}
            \fmf{fermion,label=$\scriptstyle\alpha$}{v2,i2}

            \fmf{fermion,label=$\scriptstyle\beta$,label.side=right}{o1,v1}
            \fmf{fermion,label=$\scriptstyle\beta$,label.side=left}{v2,o2}

            \fmfdot{v1}
            \fmfdot{v2}

            \fmfv{label=$\scriptstyle u_{2;a}$,label.angle=90, label.dist=2mm}{v1}
            \fmfv{label=$\scriptstyle u_{2;a}$,label.angle=-90, label.dist=2mm}{v2}

        \end{fmfgraph*}
    \end{fmffile}}
    \subfigure[]{
    \begin{fmffile}{Nc_4_u2_RG_2}
        \begin{fmfgraph*}(60,45)
            \fmfleft{i1,i2}
            \fmfright{o1,o2}
           
            \fmf{fermion,left=1,tension=0.4,label=$\scriptstyle\alpha_{\mu}$}{v1,v2}
            \fmf{fermion,left=1,tension=0.4,label=$\scriptstyle\alpha_{\mu}$}{v2,v1}
            
            \fmf{fermion,label=$\scriptstyle\alpha$}{i1,v1}
            \fmf{fermion,label=$\scriptstyle\alpha$}{v1,i2}

            \fmf{fermion,label=$\scriptstyle\beta$,label.side=left}{o1,v2}
            \fmf{fermion,label=$\scriptstyle\beta$}{v2,o2}

            \fmfdot{v1}
            \fmfdot{v2}

            \fmfv{label=$\scriptstyle u_{\mu}$,label.angle=180, label.dist=2mm}{v1}
            \fmfv{label=$\scriptstyle u_{\mu}$,label.angle=0, label.dist=2mm}{v2}

        \end{fmfgraph*}
    \end{fmffile}}
    \subfigure[]{
    \begin{fmffile}{Nc_4_u2_RG_3}
        \begin{fmfgraph*}(60,45)
            \fmfleft{i1,i2}
            \fmfright{o1,o2}
           
            \fmf{fermion,left=1,tension=0.4,label=$\scriptstyle\alpha+\gamma$}{v1,v2}
            \fmf{fermion,right=1,tension=0.4,label=$\scriptstyle\beta-\gamma$}{v1,v2}
            
            \fmf{fermion,label=$\scriptstyle\alpha$,label.side=left}{i1,v1}
            \fmf{fermion,label=$\scriptstyle\alpha$}{v2,i2}

            \fmf{fermion,label=$\scriptstyle\beta$,label.side=right}{o1,v1}
            \fmf{fermion,label=$\scriptstyle\beta$,label.side=left}{v2,o2}

            \fmfdot{v1}
            \fmfdot{v2}

            \fmfv{label=$\scriptstyle u_{4}$,label.angle=90, label.dist=2mm}{v1}
            \fmfv{label=$\scriptstyle u_{4}$,label.angle=-90, label.dist=2mm}{v2}
        
        \end{fmfgraph*}
    \end{fmffile}}
    \subfigure[]{
    \begin{fmffile}{Nc_4_u2_RG_4}
        \begin{fmfgraph*}(60,45)
            \fmfleft{i1,i2}
            \fmfright{o1,o2}
           
            \fmf{fermion,tension=0.4,label=$\scriptstyle\beta-\gamma$,label.side=left}{v1,v2}
            \fmf{fermion,tension=0.4,label=$\scriptstyle\gamma$,label.side=right}{v3,v2}
            \fmf{fermion,tension=0.4,label=$\scriptstyle\beta-\gamma$,label.side=left}{v3,v1}

            \fmf{fermion,label=$\scriptstyle\alpha$,label.side=left}{i1,v1}
            \fmf{fermion,label=$\scriptstyle\beta$,label.side=left}{o1,v3}
            \fmf{fermion,label=$\scriptstyle\beta$,label.side=left}{v2,o2}
            \fmf{fermion,label=$\scriptstyle\alpha$,label.side=left}{v1,i2}

            \fmfdot{v1}
            \fmfdot{v2}
            \fmfdot{v3}

            \fmfv{label=$\scriptstyle u_{2;b}$,label.angle=180, label.dist=2mm}{v1}
             \fmfv{label=$\scriptstyle g$,label.angle=0, label.dist=2mm}{v2}
             \fmfv{label=$\scriptstyle g$,label.angle=0, label.dist=2mm}{v3}
        
        \end{fmfgraph*}
    \end{fmffile}}
    \subfigure[]{
    \begin{fmffile}{Nc_4_u2_RG_5}
        \begin{fmfgraph*}(60,45)
            \fmfleft{i1,i2}
            \fmfright{o1,o2}
           
            \fmf{fermion,tension=0.4,label=$\scriptstyle \alpha+\gamma$,label.side=left}{v1,v2}
            \fmf{fermion,tension=0.4,label=$\scriptstyle \gamma$,label.side=left}{v2,v3}
            \fmf{fermion,tension=0.4,label=$\scriptstyle \beta-\gamma$,label.side=right}{v1,v3}

            \fmf{fermion,label=$\scriptstyle \alpha$,label.side=left}{i1,v1}
            \fmf{fermion,label=$\scriptstyle \beta$,label.side=right}{o1,v1}
            \fmf{fermion,label=$\scriptstyle \alpha$}{v2,i2}
            \fmf{fermion,label=$\scriptstyle \beta$,label.side=left}{v3,o2}

            \fmfdot{v1}
            \fmfdot{v2}
            \fmfdot{v3}

            \fmfv{label=$\scriptstyle u_{4}$,label.angle=-90, label.dist=2mm}{v1}
             \fmfv{label=$\scriptstyle g$,label.angle=-135, label.dist=2mm}{v2}
             \fmfv{label=$\scriptstyle g$,label.angle=-45, label.dist=2mm}{v3}
        
        \end{fmfgraph*}
    \end{fmffile}}
    \subfigure[]{
    \begin{fmffile}{Nc_4_u2_RG_6}
        \begin{fmfgraph*}(60,45)
            \fmfleft{i1,i2}
            \fmfright{o1,o2}
           
            \fmf{fermion,tension=0.4,label=$\scriptstyle \gamma$,label.side=left}{v2,v1}
            \fmf{fermion,tension=0.4,label=$\scriptstyle \beta-\gamma$,label.side=right}{v2,v3}
            \fmf{fermion,tension=0.4,label=$\scriptstyle \gamma$,label.side=left}{v4,v3}

            \fmf{fermion,tension=0.4,label=$\scriptstyle \alpha+\gamma$,label.side=left}{v1,v4}

            \fmf{fermion,label=$\scriptstyle \alpha$,label.side=right}{i1,v1}
            \fmf{fermion,label=$\scriptstyle \beta$,label.side=left}{o1,v2}
            \fmf{fermion,label=$\scriptstyle \alpha$}{v4,i2}
            \fmf{fermion,label=$\scriptstyle \beta$,label.side=left}{v3,o2}

            \fmfdot{v1}
            \fmfdot{v2}
            \fmfdot{v3}
            \fmfdot{v4}

            \fmfv{label=$\scriptstyle g$,label.angle=180, label.dist=2mm}{v1}
            \fmfv{label=$\scriptstyle g$,label.angle=0, label.dist=2mm}{v2}
            \fmfv{label=$\scriptstyle g$,label.angle=-45, label.dist=2mm}{v3}
            \fmfv{label=$\scriptstyle g$,label.angle=-135, label.dist=2mm}{v4}
        
        \end{fmfgraph*}
    \end{fmffile}}
    \caption{Representative diagrams of the contributions to the renormalisation of $u_{2;a}$ to single loop order under the renormalisation group.}
    \label{fig:Nc_4_u2a_renormalisation}
\end{figure}

\begin{figure}
    \centering
    \subfigure[]{
    \begin{fmffile}{Nc_4_u2b_RG_1}
        \begin{fmfgraph*}(60,45)
            \fmfleft{i1,i2}
            \fmfright{o1,o2}
           
            \fmf{fermion,left=1,tension=0.4,label=$\scriptstyle\alpha$}{v1,v2}
            \fmf{fermion,right=1,tension=0.4,label=$\scriptstyle\beta-\gamma$}{v1,v2}
            
            \fmf{fermion,label=$\scriptstyle\alpha$,label.side=left}{i1,v1}
            \fmf{fermion,label=$\scriptstyle\alpha$,label.side=left}{v2,i2}

            \fmf{fermion,label=$\scriptstyle\beta-\gamma$,label.side=right}{o1,v1}
            \fmf{fermion,label=$\scriptstyle\beta-\gamma$,label.side=right}{v2,o2}

            \fmfdot{v1}
            \fmfdot{v2}

            \fmfv{label=$\scriptstyle u_{2;b}$,label.angle=90, label.dist=2mm}{v1}
            \fmfv{label=$\scriptstyle u_{2;b}$,label.angle=-90, label.dist=2mm}{v2}
        
        \end{fmfgraph*}
    \end{fmffile}}
    \subfigure[]{
    \begin{fmffile}{Nc_4_u2b_RG_2}
        \begin{fmfgraph*}(60,45)
            \fmfleft{i1,i2}
            \fmfright{o1,o2}
           
            \fmf{fermion,left=1,tension=0.4,label=$\scriptstyle \alpha_{\mu}$}{v1,v2}
            \fmf{fermion,left=1,tension=0.4,label=$\scriptstyle \alpha_{\mu}$}{v2,v1}
            
            \fmf{fermion,label=$\scriptstyle \alpha$,label.side=left}{i1,v1}
            \fmf{fermion,label=$\scriptstyle \alpha$,label.side=left}{v1,i2}

            \fmf{fermion,label=$\scriptstyle \beta-\gamma$,label.side=right}{o1,v2}
            \fmf{fermion,label=$\scriptstyle \beta-\gamma$,label.side=right}{v2,o2}

            \fmfdot{v1}
            \fmfdot{v2}

            \fmfv{label=$\scriptstyle u_{\mu}$,label.angle=180, label.dist=2mm}{v1}
            \fmfv{label=$\scriptstyle u_{\mu}$,label.angle=0, label.dist=2mm}{v2}
        
        \end{fmfgraph*}
    \end{fmffile}}
    \subfigure[]{
    \begin{fmffile}{Nc_4_u2b_RG_3}
        \begin{fmfgraph*}(60,45)
            \fmfleft{i1,i2}
            \fmfright{o1,o2}
           
            \fmf{fermion,left=1,tension=0.4,label=$\scriptstyle \alpha+\beta$}{v1,v2}
            \fmf{fermion,left=1,tension=0.4,label=$\scriptstyle \gamma$}{v2,v1}
            
            \fmf{fermion,label=$\scriptstyle \alpha$,label.side=left}{i1,v1}
            \fmf{fermion,label=$\scriptstyle \alpha$,label.side=left}{v2,i2}

            \fmf{fermion,label=$\scriptstyle \beta-\gamma$,label.side=right}{o1,v1}
            \fmf{fermion,label=$\scriptstyle \beta-\gamma$}{v2,o2}

            \fmfdot{v1}
            \fmfdot{v2}

            \fmfv{label=$\scriptstyle u_{4}$,label.angle=90, label.dist=2mm}{v1}
            \fmfv{label=$\scriptstyle u_{4}$,label.angle=-90, label.dist=2mm}{v2}
        
        \end{fmfgraph*}
    \end{fmffile}}
    \subfigure[]{
    \begin{fmffile}{Nc_4_u2b_RG_4}
        \begin{fmfgraph*}(60,45)
            \fmfleft{i1,i2}
            \fmfright{o1,o2}
           
            \fmf{fermion,tension=0.4,label=$\scriptstyle \gamma$,label.side=right}{v2,v1}
            \fmf{fermion,tension=0.4,label=$\scriptstyle \beta$,label.side=right}{v3,v2}
            \fmf{fermion,tension=0.4,label=$\scriptstyle \gamma$}{v1,v3}

            \fmf{fermion,label=$\scriptstyle \alpha$}{i1,v1}
            \fmf{fermion,label=$\scriptstyle \beta-\gamma$,label.side=right}{o1,v3}
            \fmf{fermion,label=$\scriptstyle \beta-\gamma$}{v2,o2}
            \fmf{fermion,label=$\scriptstyle \alpha$,label.side=left}{v1,i2}

            \fmfdot{v1}
            \fmfdot{v2}
            \fmfdot{v3}

            \fmfv{label=$\scriptstyle u_{2;a}$,label.angle=180, label.dist=2mm}{v1}
             \fmfv{label=$\scriptstyle g$,label.angle=135, label.dist=2mm}{v2}
             \fmfv{label=$\scriptstyle g$,label.angle=-135, label.dist=2mm}{v3}
        
        \end{fmfgraph*}
    \end{fmffile}}
    \subfigure[]{
    \begin{fmffile}{Nc_4_u2b_RG_5}
        \begin{fmfgraph*}(60,45)
            \fmfleft{i1,i2}
            \fmfright{o1,o2}
           
            \fmf{fermion,tension=0.4,label=$\scriptstyle \alpha+\beta$,label.side=left}{v1,v2}
            \fmf{fermion,tension=0.4,label=$\scriptstyle \beta$,label.side=left}{v2,v3}
            \fmf{fermion,tension=0.4,label=$\scriptstyle \gamma$,label.side=left}{v3,v1}

            \fmf{fermion,label=$\scriptstyle \alpha$}{i1,v1}
            \fmf{fermion,label=$\scriptstyle \beta-\gamma$,label.side=right}{o1,v1}
            \fmf{fermion,label=$\scriptstyle \alpha$,label.side=left}{v2,i2}
            \fmf{fermion,label=$\scriptstyle \beta-\gamma$,label.side=right}{v3,o2}

            \fmfdot{v1}
            \fmfdot{v2}
            \fmfdot{v3}

            \fmfv{label=$\scriptstyle u_{4}$,label.angle=-90, label.dist=2mm}{v1}
             \fmfv{label=$\scriptstyle g$,label.angle=-135, label.dist=2mm}{v2}
             \fmfv{label=$\scriptstyle g$,label.angle=-45, label.dist=2mm}{v3}
        
        \end{fmfgraph*}
    \end{fmffile}}
    \subfigure[]{
    \begin{fmffile}{Nc_4_u2b_RG_6}
        \begin{fmfgraph*}(60,45)
            \fmfleft{i1,i2}
            \fmfright{o1,o2}
           
            \fmf{fermion,tension=0.4,label=$\scriptstyle \beta$,label.side=left}{v2,v1}
            \fmf{fermion,tension=0.4,label=$\scriptstyle \gamma$,label.side=right}{v2,v3}
            \fmf{fermion,tension=0.4,label=$\scriptstyle \beta$,label.side=left}{v4,v3}

            \fmf{fermion,tension=0.4,label=$\scriptstyle \alpha+\beta$,label.side=left}{v1,v4}

            \fmf{fermion,label=$\scriptstyle \alpha$,label.side=left}{i1,v1}
            \fmf{fermion,label=$\scriptstyle \beta-\gamma$,label.side=right}{o1,v2}
            \fmf{fermion,label=$\scriptstyle \alpha$,label.side=left}{v4,i2}
            \fmf{fermion,label=$\scriptstyle \beta-\gamma$}{v3,o2}

            \fmfdot{v1}
            \fmfdot{v2}
            \fmfdot{v3}
            \fmfdot{v4}

            \fmfv{label=$\scriptstyle g$,label.angle=-45, label.dist=2mm}{v1}
            \fmfv{label=$\scriptstyle g$,label.angle=-135, label.dist=2mm}{v2}
            \fmfv{label=$\scriptstyle g$,label.angle=135, label.dist=2mm}{v3}
            \fmfv{label=$\scriptstyle g$,label.angle=45, label.dist=2mm}{v4}
        
        \end{fmfgraph*}
    \end{fmffile}}
    \caption{Representative diagrams of the contributions to the renormalisation of $u_{2;b}$ to single loop order under the renormalisation group.}
    \label{fig:Nc_4_u2b_renormalisation}
\end{figure}

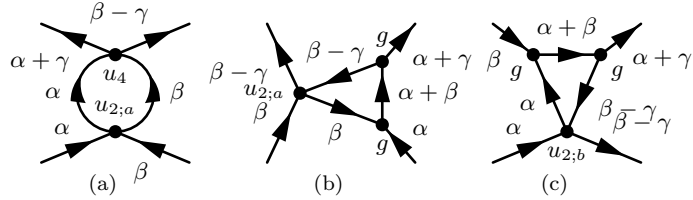
\begin{figure}[h!]
    \centering
    \subfigure[]{
    \begin{fmffile}{Nc_4_u4_RG_1}
        \begin{fmfgraph*}(70,52.5)
            \fmfleft{i1,i2}
            \fmfright{o1,o2}
           
            \fmf{fermion,left=1,tension=0.4,label=$\alpha$}{v1,v2}
            \fmf{fermion,right=1,tension=0.4,label=$\beta$}{v1,v2}
            
            \fmf{fermion,label=$\alpha$,label.side=left}{i1,v1}
            \fmf{fermion,label=$\alpha+\gamma$,label.side=left}{v2,i2}

            \fmf{fermion,label=$\beta$,label.side=left}{o1,v1}
            \fmf{fermion,label=$\beta-\gamma$,label.side=left}{v2,o2}

            \fmfdot{v1}
            \fmfdot{v2}

            \fmfv{label=$u_{2;a}$,label.angle=90, label.dist=2mm}{v1}
            \fmfv{label=$u_{4}$,label.angle=-90, label.dist=2mm}{v2}
        
        \end{fmfgraph*}
    \end{fmffile}}
    \subfigure[]{
    \begin{fmffile}{Nc_4_u4_RG_2}
        \begin{fmfgraph*}(70,52.5)
            \fmfleft{i1,i2}
            \fmfright{o1,o2}
           
            \fmf{fermion,tension=0.4,label=$\beta-\gamma$,label.side=right}{v2,v1}
            \fmf{fermion,tension=0.4,label=$\alpha+\beta$,label.side=right}{v3,v2}
            \fmf{fermion,tension=0.4,label=$\beta$}{v1,v3}

            \fmf{fermion,label=$\beta$}{i1,v1}
            \fmf{fermion,label=$\alpha$,label.side=right}{o1,v3}
            \fmf{fermion,label=$\alpha+\gamma$}{v2,o2}
            \fmf{fermion,label=$\beta-\gamma$,label.side=left}{v1,i2}

            \fmfdot{v1}
            \fmfdot{v2}
            \fmfdot{v3}

            \fmfv{label=$u_{2;a}$,label.angle=180, label.dist=2mm}{v1}
             \fmfv{label=$g$,label.angle=90, label.dist=2mm}{v2}
             \fmfv{label=$g$,label.angle=-90, label.dist=2mm}{v3}
        
        \end{fmfgraph*}
    \end{fmffile}}
    \subfigure[]{
    \begin{fmffile}{Nc_4_u4_RG_3}
        \begin{fmfgraph*}(70,52.5)
            \fmfleft{i1,i2}
            \fmfright{o1,o2}
           
            \fmf{fermion,tension=0.4,label=$\alpha$,label.side=left}{v1,v2}
            \fmf{fermion,tension=0.4,label=$\alpha+ \beta$,label.side=left}{v2,v3}
            \fmf{fermion,tension=0.4,label=$\beta-\gamma$,label.side=left}{v3,v1}

            \fmf{fermion,label=$\alpha$,label.side=left}{i1,v1}
            \fmf{fermion,label=$\beta-\gamma$,label.side=left}{v1,o1}
            \fmf{fermion,label=$\beta$}{i2,v2}
            \fmf{fermion,label=$\alpha+\gamma$}{v3,o2}

            \fmfdot{v1}
            \fmfdot{v2}
            \fmfdot{v3}

            \fmfv{label=$u_{2;b}$,label.angle=-90, label.dist=2mm}{v1}
             \fmfv{label=$g$,label.angle=-135, label.dist=2mm}{v2}
             \fmfv{label=$g$,label.angle=-45, label.dist=2mm}{v3}
        
        \end{fmfgraph*}
    \end{fmffile}}
    \caption{Representative diagrams of the contributions to the renormalisation of $u_{4 }$ to single loop order under the renormalisation group.}
    \label{fig:Nc_4_u4_renormalisation}
\end{figure}
      
The relevant Feynman diagrams for each parameter are presented in Figs. \ref{fig:Nc_4_RG_diagrams},\ref{fig:Nc_4_g_renormalisation},\ref{fig:Nc_4_u1_renormalisation},\ref{fig:Nc_4_u2a_renormalisation},\ref{fig:Nc_4_u2b_renormalisation}, and \ref{fig:Nc_4_u4_renormalisation}. As for $N_c=3$, the quantum corrections vanish at late RG flow times as $r_0$ becomes large, and so - provided there is no mechanism fine tuning the original color ice model to the critical point - a mean field treatment is expected to be sufficient to describe the most probable first order phase transition to an ordered phase out of the liquid phase upon tuning $J_{\pm}/J_z$.

\newpage

\section{Survey of effective Hamiltonians arising from flux frustration degeneracy breaking}

\begin{figure}
    \centering
    
    \includegraphics[width=0.4\linewidth]{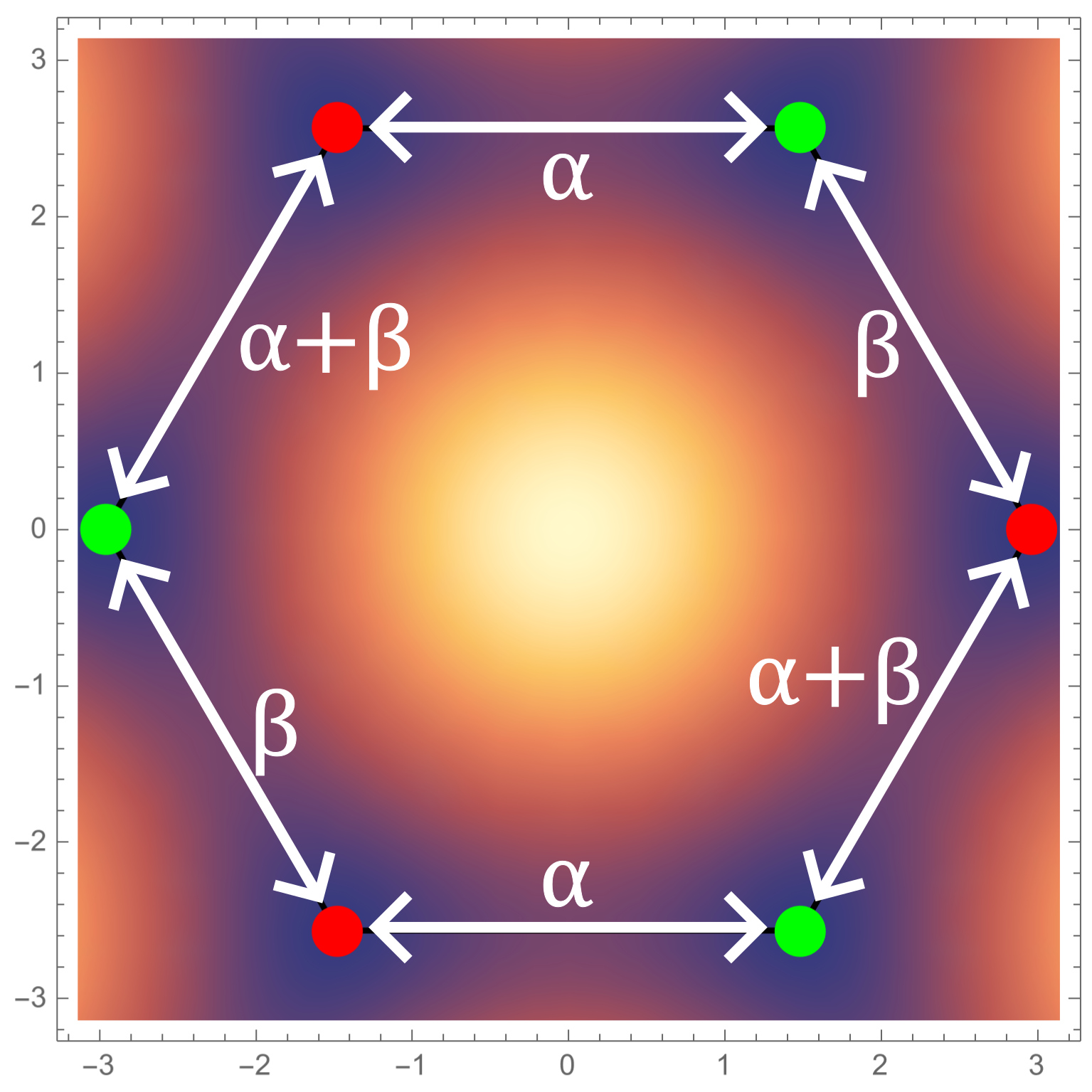}
    
    \caption{Action of $e^{i\theta\alpha^iE^i_l}$ perturbation on degenerate flux states for the $\text{U}(1)$ and $\mathbb{Z}_{3m}$  $N_c=3$ models for $V<0$. For these theories, the energy per plaquette in the $V<0$ phase is minimized at the $K$ and $K'$ points. As in the main text, the background color plot demonstrates the variation of $\sum_{\alpha}\cos(\alpha^iB^i)$ across the full $U(1)$ BZ. Electric field perturbations act to shift the flux eigenvalues of neighboring plaquettes by vectors shown by the arrows. Amongst these plaquettes, the total flux is conserved modulo a reciprocal lattice vector. For $d=2$, the perturbation acts on two plaquettes, `raising' the flux of one plaquette, and `lowering' the other, such that the effective flux Hamiltonian is a projected quantum XY model.}
    \label{fig:Supp_Nc3_flux_diagrams_1}
\end{figure}

\begin{figure}
    \centering
    \subfigure[]{
    \includegraphics[width=0.4\linewidth]{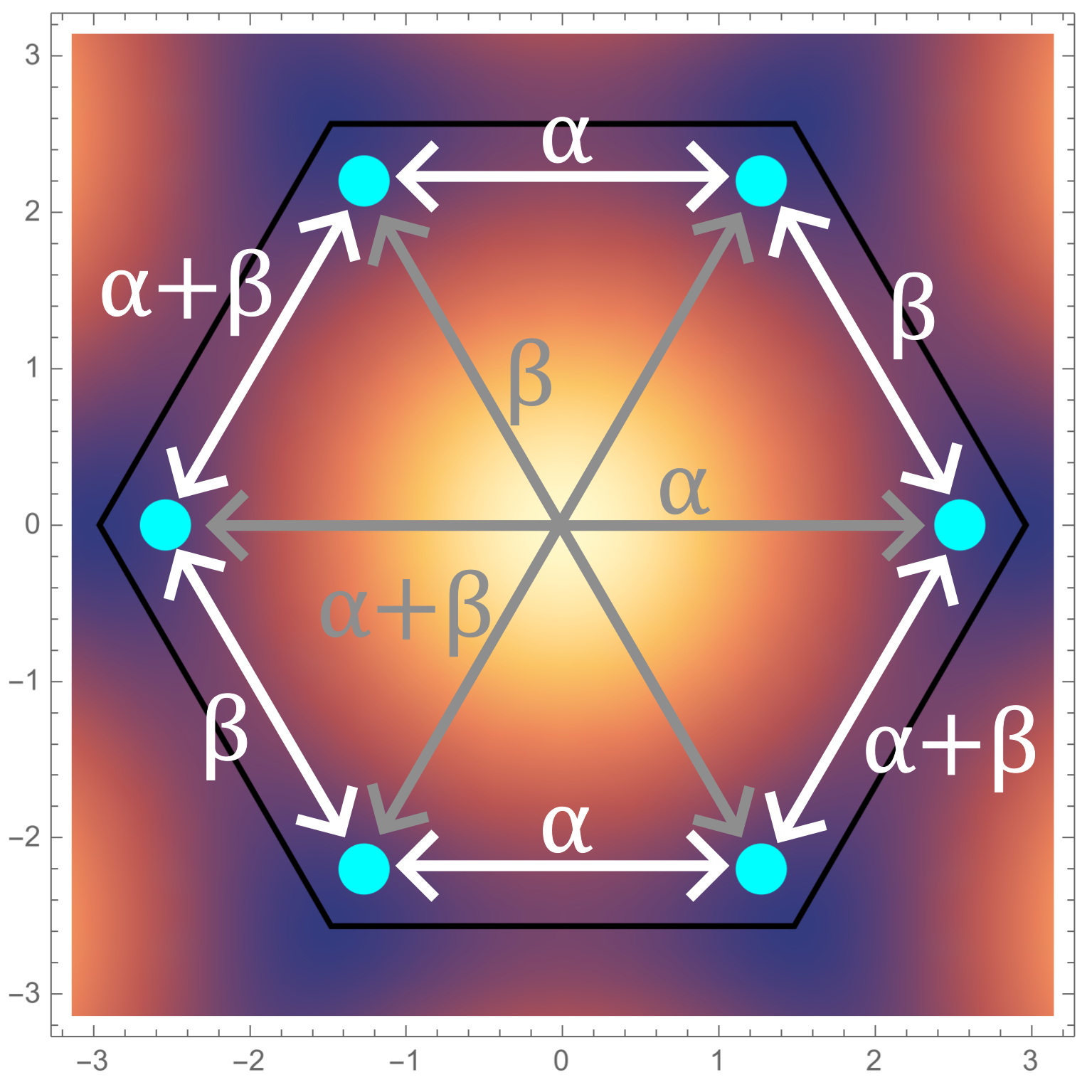}
    }
    \subfigure[]{
    \includegraphics[width=0.4\linewidth]{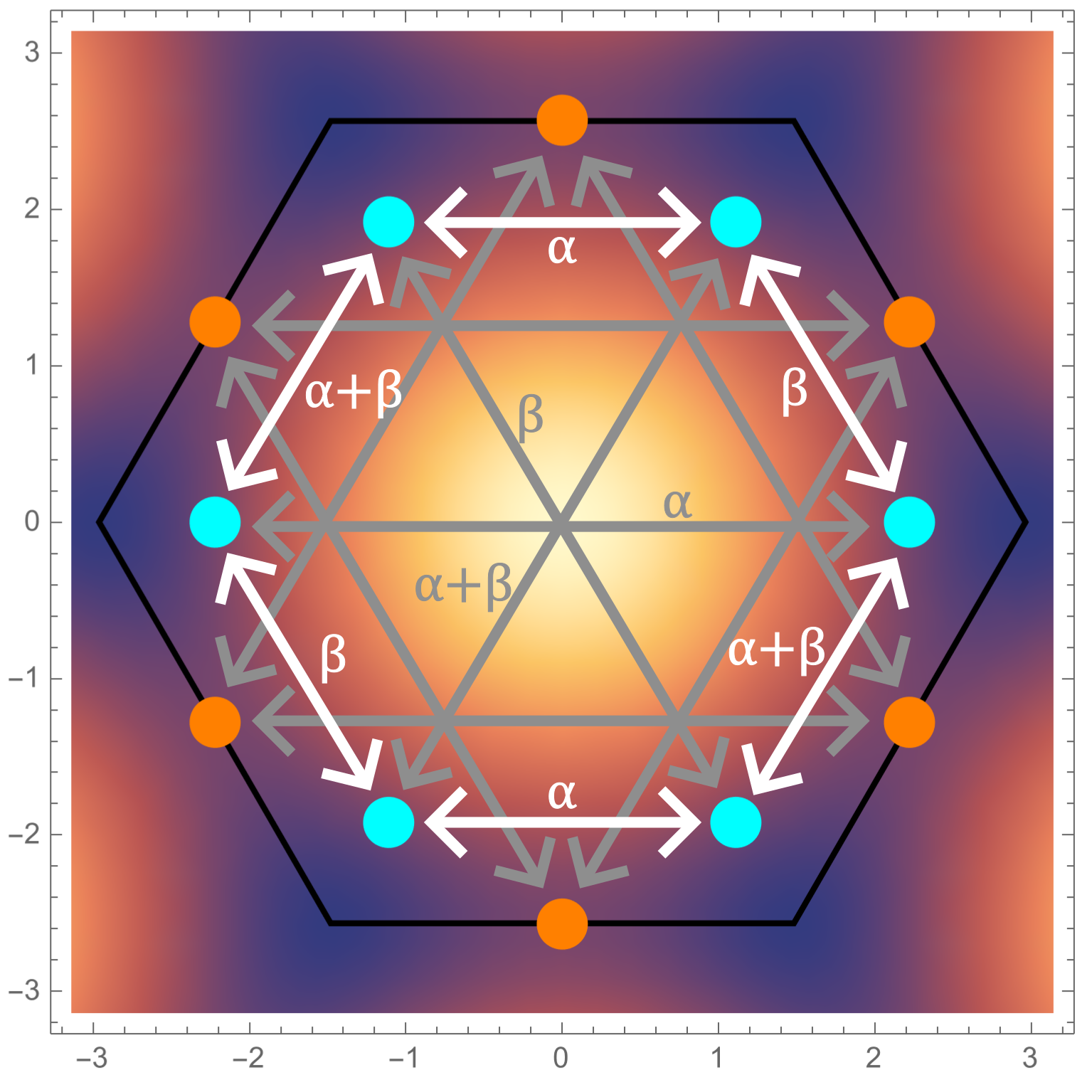}
    }
    \caption{Action of $e^{i\theta\alpha^iE^i_l}$ operators on the minimizing flux states in the $V<0$ phase for (a) the $\mathbb{Z}_7$ and (b) $\mathbb{Z}_4$ theories. The $\mathbb{Z}_{3m\pm1}$ models for $m\ge2$, of which $\mathbb{Z}_7$ is a representative, possess six flux minima, and can be coupled by two distinct families of flux shift operators, here shown as white and gray arrows. The $\mathbb{Z}_4$ model, shown in (b), is unique in possessing 9 flux minima. Just as in the $\textbf{6}$ representation model, there are two families of field operators (two flux shift magnitudes). The larger flux shifts between the cyan $\textbf{6}$ representation states coincide with the moves connecting the orange $\textbf{3}$ representation states, coupling these sectors together.}
    \label{fig:Supp_Nc3_flux_diagrams_2}
\end{figure}

We discuss here a few more particular details concerning flux frustration, and the breaking of the resulting extensive degeneracy, in exactly solvable quantum color liquid models. As in the main text, we examine toric-code inspired Hamiltonians with $\mathbb{Z}_N^{\otimes N_c-1}|_{\mathbb{Z}_2\times S_{N_c}}$ gauge symmetry, defined on the edges of bipartite lattices in $d$ dimensions of the following form:
\begin{equation}
    H = J_{Q} \sum_{v} \sum_i \left(Q^i_v\right)^2 -V \sum_p \sum_{\alpha} \cos\left(\sum_i\alpha^iB^i_p\right).
    \label{eq:Supp_Toric_Hamiltonian}
\end{equation}
Operators $Q^i_v$ and $B^i_p$ are as defined in the main text. These models are extensively degenerate for $V<0$ and $N_c>2$.For $N_c=2$ and gauge group $\mathbb{Z }_{2n}$ or $U(1)$, there is a unique flux eigenvalue that minimizes the energy per plaquette in the $V<0$ phase, and the system remains topologically ordered, now with a finite flux per plaquette (assuming a total even number of plaquettes to satisfy the constraint that the product of all plaquette operators is unity). Lattice symmetries must be implemented projectively, and spinon dynamics are altered by the presence of fluxes, but critically the ground state remains a quantum liquid.

Except for these special cases, generically there are multiple flux states, connected by the $\mathbb{Z}_2\times S_{N_c}$ point group symmetry of the $\mathfrak{su}(N_c)$ root lattice, that minimize the energy per plaquette for $V<0$, leading to an extensive degeneracy of ground states in the unperturbed model.

The number of such points, which representation of $\mathbb{Z}_2\times S_{N_c}$ they transform under, and the effective Hamiltonian resulting from perturbation by $e^{i\alpha^iE^i_l}$ operators is highly detail dependent. In this section, we exhibit some of the wide variety of behaviors that can be encountered.

Flux frustration is avoided for $N_c=2$ only if the gauge group is $\mathbb{Z}_{2m}$ or $U(1)$. For odd order $N$, there is a two-fold flux degeneracy per plaquette. There remains the global flux constraint, and additional local constraints in $d=3$ such that the product of plaquette operators over any closed surface (or the entire system in $d=2$) is unity.

Breaking the extensive degeneracy via $-U\sum_l\cos(E_l)$, one obtains the following effective Hamiltonian on the dual lattice in two dimensions:
\begin{equation}
    H_{\text{eff}} = P \left(-U \sum_{\langle p,q\rangle} \sigma^+_p\sigma^-_q+\sigma^+_q\sigma^-_p\right)P.
\end{equation}
The effective Hamiltonain is the quantum XY model, with the modification that the system is projected into the zero flux sector (modulo $2\pi$), though as the Hamiltonian conserves flux absolutely, this does not change the physics bar excluding certain total flux sectors.

In the main text, the case of a $\mathbb{Z}_2^{\otimes 2}|_{\mathbb{Z}_2\times S_{3}}$ gauge theory was discussed in two dimensions. It was found that there were three flux minimizing states in this case, with an effective nearest-neighbor exchange Hamiltonian promoting entanglement between neighboring fluxes. By varying $N$, the number of flux minima, and hence the effective theory for the degenerate sector, can be altered.

We can classify different scenarios by the irrep under which the flux minima transform with the $\mathbb{Z}_2\times S_{3}$ point group. Varying the size of the group $N$, one finds the following scenarios given in Table \ref{tabel:Nc3_effective_theories}.

\begin{table}
\begin{center}
\begin{tabular}{|c||c|c|c|c|}
\hline Group & Representation & Hamiltonian (2D) \\
\hline
$\mathbb{Z}_2$ & $\mathbf{3}$ & $P\left(-U \sum_{\langle p,q\rangle}\sum_{\alpha}X^{\alpha}_pX^{\alpha}_q\right)P$ \\
$\mathbb{Z}_{3m}$; $U(1)$ & $\mathbf{2}$ & $P\left(-U \sum_{\langle p,q\rangle}\sigma^+_p\sigma^{-}_q +\sigma^-_p\sigma^{+}_q\right)P$  \\
$\mathbb{Z}_4$ & $\mathbf{3} \oplus \mathbf{6}$  & Eq. (\ref{eq:Supp_Nc3_3_6_rep_Hamiltonian})   \\
$\mathbb{Z}_{3m-1}$;  $\mathbb{Z}_{3m+1}$ \ ($m\ge2$) & $\mathbf{6}$ & Eq. (\ref{eq:Supp_Nc3_6_rep_Hamiltonian})  \\
\hline
\end{tabular}
\caption{Representations of the point group $\mathbb{Z}_2\times S_3$ realized by the energy minimizing fluxes for the different $\mathbb{Z}_N$ gauge theories with $N_c=3$. The representation of the point group realized determines the effective flux Hamiltonian.}
\label{tabel:Nc3_effective_theories}
\end{center}
\end{table}

The next most simple case to consider for $N_c=3$ in two dimensions are the $\mathbb{Z}_{3m}$ and $U(1)$ models. Here, there are two flux states that minimize the energy per plaquette for $V<0$, these being the $K$ and $K'$ points of the flux Brillouin zone ($\pm B_K$). These fluxes satisfy the algebra $3B_K=0$ (up to a reciprocal lattice vector). The $e^{i\alpha^i E^i_l}$ flux shifts behave as $\sigma^+$ and $\sigma^{-}$ operators, just as in the odd-$N$ $N_c=2$ models.

The two-dimensional effective Hamiltonian is thus a ferromagnetic quantum $XY$ model, projected into the net zero flux sector. Whilst the ground state is highly non-trivial, heuristically the flux through a given plaquette will be strongly entangled with the rest of the system, and equally likely to be $+B_K$ or $-B_K$. The motion of rooton excitations on top of this state will again incur additional internal phase shifts in the wavefunction, raising the expected energy, and indicating that rootons may hybridize with excitations of the flux vacuum. 

The $\mathbb{Z}_{3m-1}$ and $\mathbb{Z}_{3m+1}$ cases both possess six flux minima which transform under the same representation of $\mathbb{Z}_2\times S_{3}$. The effective Hamiltonian induced by $e^{i\alpha^i E^i_l}$ flux shifts is somewhat more complicated. Denoting the six flux minima states with roman indices $a,b,...,e$, and defining operators $R_{ab}$ which change an initially type $a$ flux to type $b$, and otherwise annihilate the state, the effective Hamiltonian is:
\begin{multline}
    H_{\text{eff}}= P\Biggl(-U_1 \sum_{p,q} \bigl( [R_{ab;p}+R_{ed;q}][R_{ba;p}+R_{de;q}] +[R_{bc;p}+R_{fe;q}][R_{cb;p}+R_{ef;q}] + [R_{cd;p}+R_{af;q}][R_{dc;p}+R_{fa;q}] + \text{H.c.} \bigr)\\  -U_2 \sum_{p,q} \bigl( R_{cf;p}R_{fc;q}+R_{ad;p}R_{da;q}+R_{eb;p}R_{be;q} + \text{H.c.}\bigr)\Biggr)P. \label{eq:Supp_Nc3_6_rep_Hamiltonian}
\end{multline}
Unlike in the previous scenarios considered, there are two distinct flux swap processes not related by symmetry, and hence the effective Hamiltonian allows two independent perturbation parameters $U_1$ and $U_2$. Figure \ref{fig:Supp_Nc3_flux_diagrams_2} (a) presents the flux minima and the action of flux shift operators upon them for the $\mathbb{Z}_7$ theory.

The $\mathbb{Z}_4$ is a special case, where the energies of a $\mathbf{3}$ representation state set, and a $\mathbf{6}$ representation set coincide. The resulting Hamiltonian is the sum of the $\mathbb{Z}_2$ and $\mathbb{Z}_{3m\pm1}$ Hamiltonians, with some additional terms coupling the two cycles $a,b,...,e$ and $A,B,C$. The allowed swaps are shown diagrammatically in Fig. \ref{fig:Supp_Nc3_flux_diagrams_2} (b). The $\mathbb{Z}_4$ model Hamiltonian is:
\begin{multline}
    H_{\text{eff}}= P\Biggl(-U_1 \sum_{p,q} \bigl( [R_{ab;p}+R_{ed;q}][R_{ba;p}+R_{de;q}] +[R_{bc;p}+R_{fe;q}][R_{cb;p}+R_{ef;q}] + [R_{cd;p}+R_{af;q}][R_{dc;p}+R_{fa;q}] + \text{H.c.} \bigr)\\  -U_2 \sum_{p,q} \bigl( [R_{cf;p}+X_{BC;p}][R_{fc;q}+X_{BC;q}]+[R_{ad;p}+X_{AC;p}][R_{da;q}+X_{AC;p}]+[R_{eb;p}+X_{AB;p}][R_{be;q}+X_{AB;q}] + \text{H.c.}\bigr)\Biggr)P. \label{eq:Supp_Nc3_3_6_rep_Hamiltonian}
\end{multline}

Whilst the effective Hamiltonians upon breaking the extensive degeneracy of the $V<0$ phase varying greatly with $N$, we note that the values of the flux minima for cases $\mathbb{Z}_{3m\pm1}$ continuously approach the $\pm B_K$ values as $N$ increases. 

\begin{figure}
    \centering
    
    \includegraphics[width=0.4\linewidth]{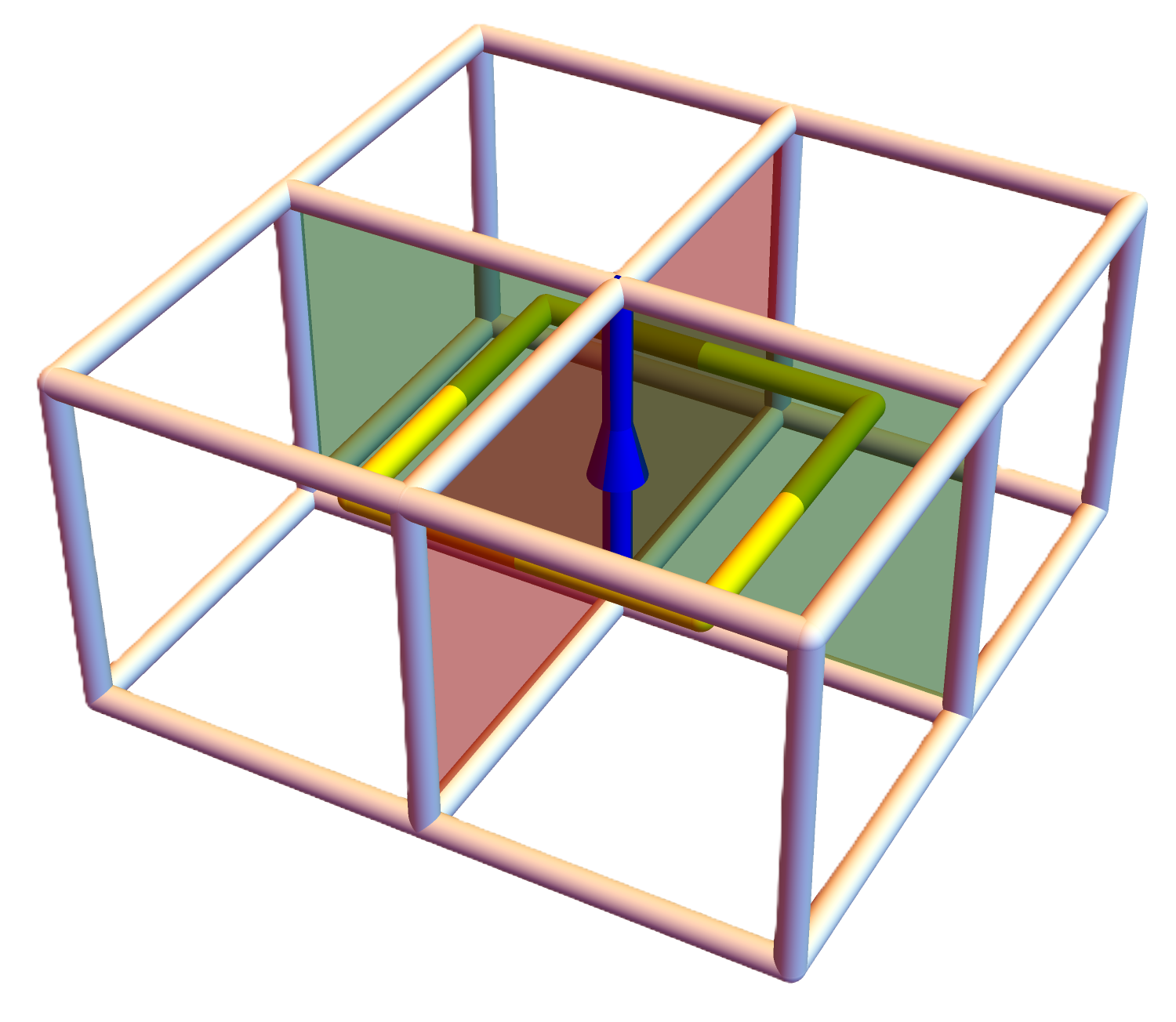}
    
    \caption{Demonstration of the action of $e^{i\alpha^iE^i_l}$ operators in three dimensions. Acting on the link shown in blue, the the operator shifts the flux of all connected plaquettes, with a sign to this shift given by the relative orientation of that plaquette and the positive circulation sense about the dual plaquette - shown in yellow - to the blue link.}
    \label{fig:Supp_3d_E_perturbation_action}

\end{figure}

Let us now extend this discussion to three dimensions. The effective flux theories motivated above can be understood as theories defined on the Hodge dual lattice to the original. In two dimensions, this means plaquette operators are mapped to the vertices of the new model, and operators defined on edges, $e^{i\alpha^i E^i_l}$, are defined on dual edges on the dual lattice. In three dimensions however, plaquettes are dual to edges, and edges to plaquettes, and so the effective Hamiltonians generated from the above procedure will have degrees of freedom defined on the links of a dual lattice, and the interactions generated by the $e^{i\alpha^i E^i_l}$ operators will be ring exchange operators around primitive plaquettes (see Fig. \ref{fig:Supp_3d_E_perturbation_action}).

Another consideration that must be made in three dimensions is that the total flux through any closed surface must be a reciprocal lattice vector. Deviations of the flux through a closed surface from zero correspond to quantized vison excitations $V^i_{\nu}$ that are defined on each primitive volume $\nu$.

Taking as an example the $U(1)$ $N_c=3$ model on a cubic lattice, the ground state sector for $V<0$ consists of all states with all plaquettes having flux $\pm B_K$, and total flux through any closed surface some reciprocal lattice vector. Treating these fluxes again as Ising variables, this requires the flux configuration out of each cubic volume to be 3-in-3-out, all in, or or out. Adding perturbations introduces ring exchange operators to the Hamiltonian, suggesting that the effective flux theory could be described by a quantum `flux' liquid coexisting with the color liquid.

\begin{figure}
    \centering
    \subfigure[]{
    \includegraphics[width=0.4\linewidth]{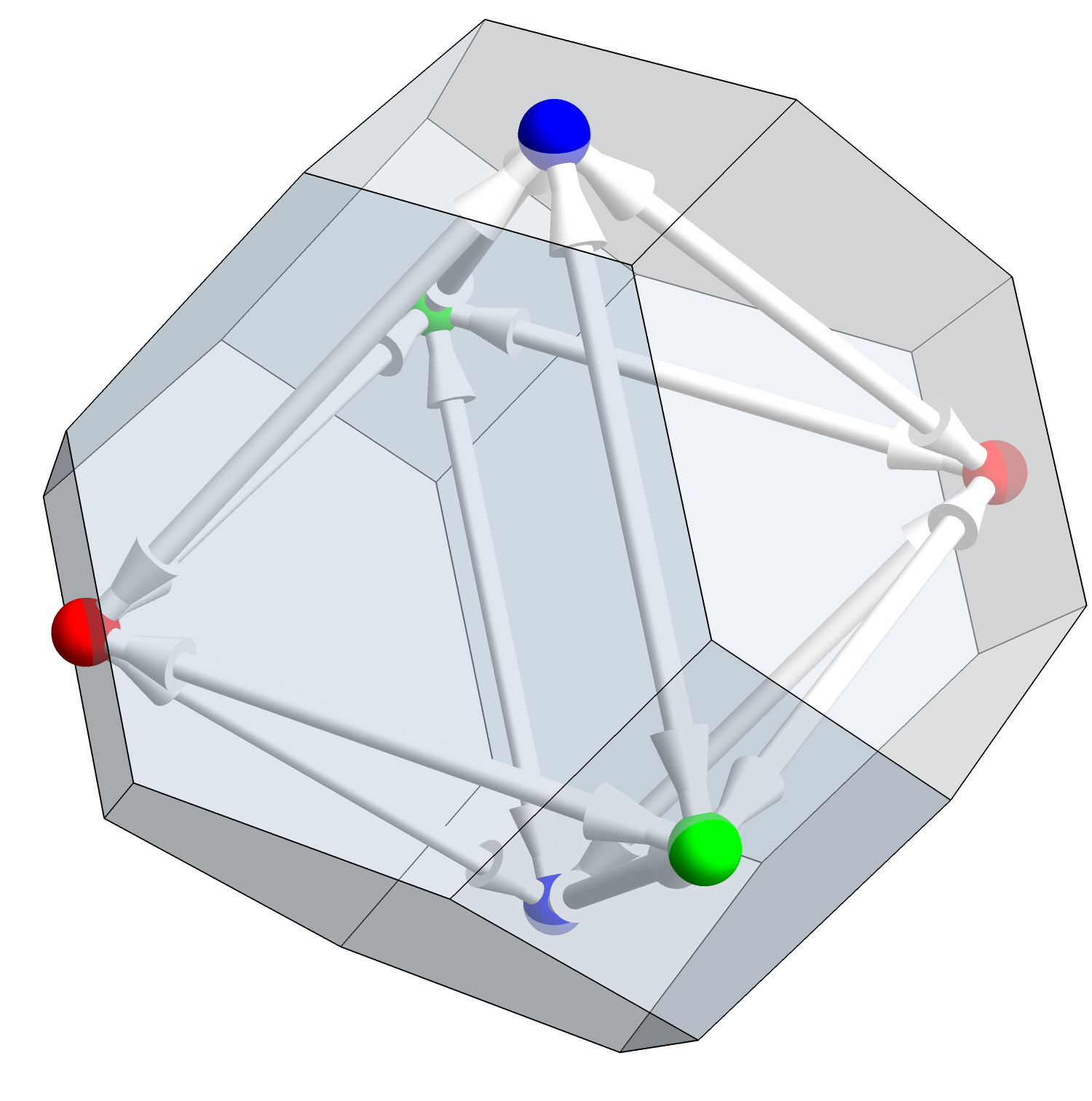}
    }
    \subfigure[]{
    \includegraphics[width=0.4\linewidth]{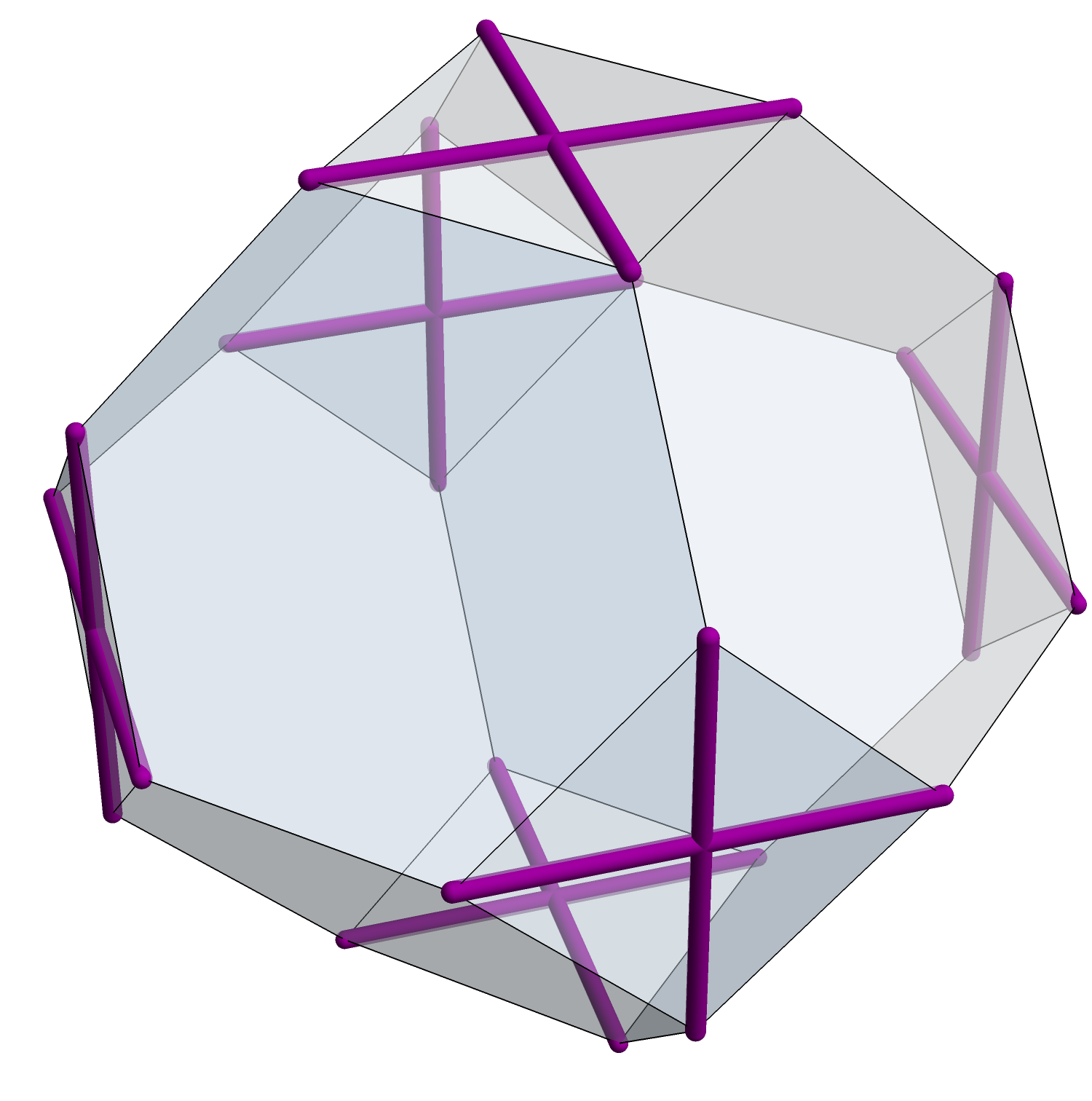}
    }
    \caption{System of flux states minimizing the energy per plaquette in the $V<0$ phases of the $\mathbb{Z}_2$ ((a)) and $\text{U}(1)$ $N_c=4$ theories. The minimizing flux states are presented on the $d=3$ Brillouin zone of the $N_c=4$ root lattice. The $\mathbb{Z}_2$ theory exhibits the same behavior for $N_c=4$ as for $N_c=3$. The three minimizing flux states obey the same algebra in both models, and are acted upon by the $e^{i\alpha^iE^i_l}$ operators in the same fashion. The $N_C=4$ $\text{U}(1)$ model meanwhile possesses a continuum of minimizing fluxes, shown as purple lines in panel (b).}
    \label{fig:Supp_Nc4_flux_diagrams}

\end{figure}

Finally, we briefly turn to $N_c=4$ models. Fig. \ref{fig:Supp_Nc4_flux_diagrams} (a) presents a diagram of the flux swap operators acting on the minimizing flux states of the $V<0$ phase for a $\mathbb{Z}_2$ model. It turns out that the effective flux Hamiltonian is entirely equivalent to the $N_c=3$ $\mathbb{Z}_2$ theory, with flux minima obeying the same algebra, and an identical effective Hamiltonian. 

The $U(1)$ gauge theory exhibits especially interesting behavior. The flux states that minimize the energy plaquette in the $V<0$ phase are not a set of discrete points, but a continuous set, as shown in Fig. \ref{fig:Supp_Nc4_flux_diagrams} (b). 

